\documentclass{article}
\usepackage[a4paper, total={7in, 9in}]{geometry}
\usepackage{hyperref}
\hypersetup{
    colorlinks=true,
    linkcolor=red,
    filecolor=magenta,      
    urlcolor=red,
    citecolor=blue,
}
\usepackage{amsmath,amsfonts,mathtools,amsthm,amssymb,amstext}

\usepackage{lipsum}
\usepackage{algorithmic}
\usepackage{algorithm}
\usepackage{array}
\usepackage[caption=false,font=normalsize,labelfont=sf,textfont=sf]{subfig}
\usepackage{textcomp}
\usepackage{stfloats}
\usepackage{url}
\usepackage{verbatim}
\usepackage{graphicx}
\usepackage{xcolor}
\usepackage{cases}
\usepackage{hyperref}
\usepackage[normalem]{ulem}
\usepackage{makecell} 

\usepackage{tabularx}

\newtheorem*{remark}{Remark}

\newtheorem{theorem}{Theorem}
\newtheorem{corollary}{Corollary}[theorem]
\newtheorem{lemma}{Lemma}

\usepackage[style=ieee,backend=bibtex]{biblatex}
\title{\textbf{Point process simulation of generalised hyperbolic L\'{e}vy processes}}

\author{\textbf{Yaman~K{\i}ndap and Simon~Godsill} \\
        {\small Signal Processing and Communications Laboratory} \\
        {\small University of Cambridge, UK}
       }
       
\date{\today}

\begin{document}
\maketitle

\begin{abstract}
Generalised hyperbolic (GH) processes are a class of stochastic processes that are used to model the dynamics of a wide range of complex systems that exhibit heavy-tailed behavior, including systems in finance, economics, biology, and physics. In this paper, we present novel simulation methods based on subordination with a generalised inverse Gaussian (GIG) process and using a generalised shot-noise representation that involves random thinning of infinite series of decreasing jump sizes. Compared with our previous work on GIG processes, we provide tighter bounds for the construction of rejection sampling ratios, leading to improved acceptance probabilities in simulation. Furthermore, we derive methods for the adaptive determination of the number of points required in the associated random series using concentration inequalities. Residual small jumps are then approximated using an appropriately scaled Brownian motion term with drift. Finally the rejection sampling steps are made significantly more computationally efficient through the use of squeezing functions based on lower and upper bounds on the L\'{e}vy density. Experimental results are presented illustrating the strong performance under various parameter settings and comparing the marginal distribution of the GH paths with exact simulations of GH random variates. The new simulation methodology is made available to researchers through the publication of a Python code repository.
\end{abstract}

\section{Introduction}
\label{sec:intro}

The statistical properties of randomly evolving phenomena in many real-world applications can be studied using stochastic differential equations. The random behaviour in such systems are typically characterised through a Brownian motion term which implicitly assumes that some version of the Central-limit theorem is valid for the observed random behaviour. However, numerous dynamical systems exhibit more heavy-tailed characteristics than the Gaussian, see for example applications in financial modelling (\cite{Mandelbrot1963,Fama1965a,Cont:2003}), communications (\cite{Azzaoui2010,Fahs2012,FreitasEganClavierEtAl2017,LiebeherrBurchardCiucu2012,ShevlyakovKim2006,WarrenThomas1991}), signal processing \cite{Nikias1995}, image analysis (\cite{Achim2001,Achim2006}), audio processing (\cite{Godsill_Rayner_1996,Lombardi2006}), climatological sciences 
(\cite{Katz1992,Katz2002}), in the medical sciences
\cite{ChenWangMcKeown2010} and for the understanding of sparse modelling/compressive sensing (\cite{UnserTaftiAminiEtAl2014,UnserTaftiSun2014,Unser2014,AminiUnser2014,CarrilloRamirezArceEtAl2016,Lopes2016,ZhouYu2017,Tzagkarakis2009,AchimBuxtonTzagkarakisEtAl2010}). In such cases the stochastic driving term can be characterised by a L\'{e}vy process which generalises the sample paths and the marginal distributions of the term to include other parametric families such as the Poisson process and the $\alpha$-stable process.

In general a continuous-time randomly evolving system may possess a continuous random Brownian motion component as well as sudden discrete random changes at random times (`jumps'). General L\'{e}vy processes encompass both of these classes of random evolution such that the increments of the process are independent and stationary (\cite{Bertoin_1997,kenlevy}). 
Here we focus on a broad class of such processes that are composed purely of jumps.

While there is substantial theoretical and applied interest in simulation of L\'{e}vy processes {\em per se\/}, in our work we are ultimately concerned with modelling and inference for systems driven by non-Gaussian L\'{e}vy processes, such as the linear stochastic differential equation (SDE) model \cite{oksendalSDE}
\[
dX(t)=AX(t)dt+hdW(t)
\]
where the more standard Brownian motion is replaced by a non-Gaussian L\'{e}vy process $\{W(t)\}$ (the so-called Background-Driving L\'{e}vy process (BDLP) \cite{Barndorffnielsen1977InfiniteDO}), see for example our earlier work with Stable law L\'{e}vy processes (\cite{Lemke2015,Godsill_Riabiz_Kont_2019,RiabizArdeshiriGodsill2017}). The work presented here though is focussed purely on the generation of the underlying L\'{e}vy processes, with the extension of our point process methods to the SDE case being in principle straightforward, as demonstrated in (\cite{Lemke2015,Godsill_Riabiz_Kont_2019}) for the Stable law case. A publication in preparation will present convergence results from extending this paper to the linear SDE case.

In this paper, we study simulation methods for a very broad class of L\'{e}vy processes, the generalised hyperbolic (GH) process (\cite{Barndorff-NielsenMikoschResnick2001,Eberlein2003}) (also known as generalised hyperbolic L\'{e}vy motion \cite{Eberlein_2001}), which captures various degrees of semi-heavy- or heavy-tailed behaviour such that the tails may be designed to be lighter than non-Gaussian Stable laws (which possess infinite variance \cite{Samorodnitsky_Taqqu_1994}), but heavier than a Gaussian \cite{Borak2011}. Some important special cases include the hyperbolic process \cite{EberleinKeller1995}, the normal inverse-Gaussian process (NIG) \cite{BarndorffNielsen1978} and the variance-gamma process \cite{MadanSeneta1990}, which were introduced in the context of modelling empirical financial returns, and the Student-t process (see also \cite{ShahWilsonGhahramani2014,pmlr-v38-solin15} which was introduced as an extension to Gaussian processes in Machine Learning, although such processes are not equivalent to the Student-t L\'{e}vy processes simulated here). The GH distribution is defined as a normal variance-mean mixture where the required mixing distribution is the generalised inverse-Gaussian (GIG) distribution. Our current work improves a point process simulation framework for GIG processes \cite{GodsillKindap2021}, extending it to the variance-mean mixture representation of the GH process, and in addition providing substantial modifications and improvements to the original methods. 

The simulation of the sample paths of L\'{e}vy processes is a key area of research that enables the use of L\'{e}vy processes in inference and decision-making. In \cite{Rosinski_2001}, Rosi{\'n}ski surveys generalised shot-noise series representations of L\'{e}vy processes and their relation with point processes, and this is the general framework adopted for the current paper  (see \cite{GodsillKindap2021,Lemke_Godsill_2015,Riabiz2020,Godsill_Riabiz_Kont_2019} and references therein, for our previous studies using this methodology). Other relevant developments include \cite{B_N_1997} which present the theory of NIG processes, \cite{Rydberg_1997} which present approximate sampling methods for the NIG case, and \cite{B_N_Shephard_2001} which give applications of shot -noise series based methods for non-Gaussian Ornstein-Uhlenbeck (OU) processes. Exact simulation methods for the class of tempered stable (TS) processes are studied in (\cite{Zhang_2011a,Qu_etal_2020,Grabchak2019RejectionSF,Sabino2022}). In addition, approximate simulation methods for GH L\'{e}vy fields, which are infinite-dimensional GH L\'{e}vy processes, are studied in \cite{barth2017approximation}. 

It is shown in \cite{Barndorffnielsen1977InfiniteDO} that the GH distribution is infinitely divisible and hence can be the distribution of a L\'{e}vy process at time $t=1$. The GH distribution possesses a five parameter probability density function defined for random variables on the real line as follows (\cite{Eberlein_2001,Eberlein2003})

\begin{align}
    f_{GH}(x) =  a(\lambda, \alpha, \beta, \delta) \left( \delta^2+(x-\mu)^2 \right)^{(\lambda - \frac{1}{2})/2} \times K_{\lambda-\frac{1}{2}} \left( \alpha \sqrt{\delta^2 + (x-\mu)^2} \right) \mathrm{exp}(\beta(x-\mu))\label{eqn:GH_pdf}
\end{align}

\noindent where

\begin{equation*}
    a(\lambda, \alpha, \beta, \delta) = \frac{(\alpha^2 - \beta^2)^{\lambda/2}}{\sqrt{2\pi} \alpha^{\lambda-\frac{1}{2}} \delta^{\lambda} K_{\lambda} (\delta \sqrt{\alpha^2 - \beta^2})}
\end{equation*}

\noindent $K_{\nu}(\cdot)$ is the modified Bessel function of the second kind with index $\nu$. The parameter $\lambda \in \mathbb{R}$ characterises the tail behaviour, $\alpha > 0$ determines the shape, $0 \leq |\beta| < \alpha$ controls the skewness, $\mu \in \mathbb{R}$ is a location parameter and $\delta > 0$ is the scale parameter. Alternative parametrisations of the probability density function in the limiting parameter settings are discussed in \cite{Eberlein2003}. 

The three parameter probability density function $f_{GIG}(\lambda, \delta, \gamma)$ of the GIG distribution may be linked to Eq. (\ref{eqn:GH_pdf}) via a variance-mean mixture of Gaussians \cite{Eberlein_2001}. Using the parameterisation $\gamma = \sqrt{\alpha^2-\beta^2}$, the variance-mean mixture for the GH distribution may be expressed as

\begin{equation}
    f_{GH}(x) = \int_{0}^{\infty} \mathcal{N} (x; \mu + \beta u, u ) f_{GIG} \left(u; \lambda, \delta, \sqrt{\alpha^2 - \beta^2} \right) du \label{mv_mixture}
\end{equation}

\noindent where $u$ is a GIG distributed random variable. Random variate generation algorithms for a GIG variable are studied in \cite{Devroye_2014,Hormann2013}, and their extension to GH distributed random variables are then obtained through the normal variance-mean construction shown in Eq. (\ref{mv_mixture}).

GH processes are generally intractable for simulation since the L\'{e}vy density associated with the GIG process is expressed as an integral involving certain Bessel functions. Simulation methods based on generalised shot-noise representation of the GIG L\'{e}vy process are given in \cite{GodsillKindap2021}. These methods rely on the construction of dominating point processes that are tractable for simulation, followed by thinning methods derived from upper bounds on the intractable integrand. In the earlier work the series generated must be truncated to a finite number of terms which needs to be tuned by the user, and hence may be inefficient in some parameter regimes. In this paper we show for the first time a practical method for simulation of paths of the GH process, based on subordination with a GIG process. The paper provides several significant contributions. The first contribution  is to  provide improved simulation methods for the underlying GIG process based on tighter bounds for the construction of dominating processes and the corresponding thinning method, and a proof of convergence is provided for the first time, based on an earlier result by \cite{Rosinski2001}. Secondly, we derive \textit{adaptive} truncation methods for approximating the infinite series involved in our representation which allow for the first time an automatic choice of the truncation level for the jumps of the GIG process. Once truncation has occurred we then approximate the residual error committed by adding an appropriately scaled Brownian motion term with drift, motivated by Central-limit theorem-style results for the residual error. Furthermore, the thinning (rejection sampling)  methods are made significantly more computationally efficient through the introduction of `squeezing' functions that both upper- and lower-bound the acceptance probabilities. Finally, acceptance probability bounds are derived and convergence properties of the novel simulation methods are compared against the methods introduced in \cite{GodsillKindap2021}. The simulation methodology is made available to researchers through the publication of a Python code repository\footnote{https://github.com/yamankindap/gh-levy-simulation}.

The paper is organised as follows. Section \ref{sec:representations} presents the necessary preliminaries for simulation of L\'{e}vy processes and their corresponding point processes, using a generalised shot-noise approach. Section \ref{sec:GIG_process} introduces the specific form of the GIG L\'{e}vy density and derives various bounds on these densities as well as constructing dominating L\'{e}vy densities from the related bounds. Section \ref{sec:simulation_algorithms} gives simulation algorithms for the GH L\'{e}vy process based on the simulation of the previously discussed dominating L\'{e}vy processes and associated thinning methods. Section \ref{sec:adaptive_truncation}, presents an adaptive truncation method for the infinite series involved in generalised shot-noise representations and a method of approximating the residual series. Section \ref{sec:squeezing_functions} gives a practical sampling algorithm based on squeezing functions for increasing the efficiency of simulation. Section \ref{sec:simulations} presents example simulations, comparing the distribution of the paths generated with exact simulations of GH random variates.

\section{Generalised shot-noise representations}
\label{sec:representations}

In this section series representations of L\'{e}vy processes given in (\cite{Rosinski_2001, Kallenberg_2002}) that enable their simulation are reviewed. Let $W(t)$ be a L\'{e}vy process on some time interval of interest $t\in[0,T]$ having no drift or Brownian motion part, and hence containing purely jumps; then the characteristic function (CF) is given by (\cite{Kallenberg_2002}, Corollary 13.8), as
\begin{align*}
    E \left[ \exp(iuW(t)) \right] = \exp \left( t \left[\int_{\mathbb{R}\setminus \{ 0 \}} (e^{iuw} -1-iw{\cal I}(|w|<1))Q(dw) \right] \right)
\end{align*}

\noindent where $Q$ is a L\'{e}vy measure on $\mathbb{R}_0 := \mathbb{R}\setminus \{ 0 \}$ satisfying $\int_{\mathbb{R}_0}\min(1,w^2)Q(dw)<\infty$. Under this definition $W(T)$ is a random variable whose distribution is infinitely divisible.

We will also require a more restricted class of non-negative, non-decreasing L\'{e}vy processes $X(t)$, the {\em subordinator\/} process, whose 
CF is given by:
\begin{align*}
    E \left[ \exp(iuX(t)) \right] = \exp \left( t \left[\int_{0}^\infty (e^{iux} -1)Q_X(dx) \right] \right)
\end{align*}
and which has the more restrictive requirement that \begin{equation}
\int_{0}^\infty\min(1,x)Q_X(dx) < \infty\label{sub_cond}\end{equation}

 $Q_X(dx)$ defines the {\em density\/} of jumps for $\{X(t)\}$ such that the expected number of jumps of size $x\in[a,b]$ is $\mu_{[a,b]}=\int_{a}^bQ_X(dx)$ and the number of jumps is a Poisson random variable with mean $\mu_{[a,b]}$. We will be dealing with {\em infinite activity\/} processes for which $\int_{0}^\infty Q_X(dx)\rightarrow \infty$ and hence there are almost surely an infinite number of jumps in time interval $[0,T]$.

In order to generate sample paths from the GH process, we will use the so-called variance-mean mixture representation of its L\'{e}vy measure,
\begin{equation}
Q_{GH}(dw)=\int_{0}^\infty {\cal N}(dw;\mu+\beta x, x)Q_{GIG}(dx)\label{eq:Q_GH_mv_mixture}
\end{equation}
which is the normal mixture representation of the GH L\'{e}vy measure, analogous to the normal mixture representation of its probability density (\ref{mv_mixture}), and where $Q_{GIG}$ is the L\'{e}vy measure of a Generalised inverse Gaussian (GIG) subordinator process (see \cite{Barndorff-Nielsen1997,WolpertIckstadt1998} and the following section for further detail).

It is first required to simulate a realisation $\{x_i\}_{i=1}^\infty$ of the jumps  from the underlying GIG subordinator process and to use a generalised shot-noise representation \cite{Rosinski_2001} to simulate from $Q_{GH}$:
\begin{equation}
\label{eqn:shot_noise_gen}
    W(t)=\sum_{i=1}^\infty W_i{\cal I}_{V_i\leq t}- t c_i
\end{equation}
\noindent where $\{ V_i \in [0,T] \}_{i=1}^{\infty}$ are i.i.d. uniform random variables representing the arrival time of jumps, and independent of the jump sizes  $\{ W_i \}_{i=1}^{\infty}$, which are independently distributed as:
\[
W_i\sim{\cal N}(\mu+\beta x_i, x_i)
\]

Note that \cite{Rosinski_2001} proves the almost sure convergence of such series to $W(t)$ for $x_i$ non-increasing, i.e. jumps of $X(t)$ are generated in order of decreasing size. The terms $c_i$ are centering terms which we may take as zero for the GH class of processes as a result of the condition in Eq. (\ref{sub_cond}).

The task remaining is to generate ordered realisations of the jumps in the subordinator, $\{x_i\}_{i=1}^\infty$.
Here the L\'{e}vy-It\^{o} integral representation of $X(t)$ may be invoked:

\begin{align}
\label{eqn:levyito}
    X(t) &= \int_{(0,\infty)} x N([0, t], dx)
\end{align}

\noindent where $N$ is a bivariate point process with mean measure $Leb. \times Q$ on $[0,T] \times \mathbb{R}_0$ which may be represented with Dirac functions as
\begin{equation}
\label{eqn:original_point_process}
    N = \sum_{i=1}^{\infty} \delta_{V_i, X_i}
\end{equation}

\noindent where again $\{ V_i \in [0,T] \}_{i=1}^{\infty}$ are i.i.d. uniform random variables independent of $\{X_i\}$ which represent the arrival time of jumps, $\{ X_i \}_{i=1}^{\infty}$ are the jump sizes, and $T$ is the duration of the time interval considered. If we substitute $N$ into Eq. (\ref{eqn:levyito}) we obtain a series representation of $X(t)$ as:

\begin{equation}
X(t)=\sum_{i=1}^\infty X_i{\cal I}_{V_i\leq t}\label{x_series}
\end{equation}

The classical method to generate such a subordinator process, with L\'{e}vy measure $Q$ (\cite{Ferguson_Klass,Rosinski_2001, Wolpert_Ickstadt_1998, WolpertIckstadt1998}) simulates jumps of decreasing size by an appropriate transformation of the epochs of a unit rate Poisson process. Briefly, an arbitrarily large number of epochs $\{\Gamma_i\}_{i=1,2,...}$ is randomly simulated from a unit rate Poisson process. These terms may be transformed into the jump magnitudes of the corresponding subordinator process by calculating  the upper tail probability of the L\'{e}vy measure $Q^+(x)=Q ([x,\infty))<\infty$. A corresponding non-increasing function $h(\cdot)$ is then defined as the inverse tail probability, $h(\gamma)=\inf_x\{x;\,Q^+(x)=\gamma\}$ which assigns a non-increasing jump value to each of the ordered Poisson epochs, $\{X_i=h(\Gamma_i)\}$. Thus, small $\Gamma_i$ values correspond to large jumps $h(\Gamma_i)$ and vice versa. It can be seen from this definition that  $\mathbb{E}[\#\{ X_i ; X_i \geq x \}]=Q^+(x)$: this procedure is essentially following an analogous formulation to the standard inverse CDF method for random variate generation, but applied here to a point process intensity function instead of a probability distribution. Formally, the mapping theorem \cite{Kingman1992} ensures that the resulting transformed process is a Poisson process having the correct L\'{e}vy density $Q(x)$. 

Since there is an infinite number of jumps in the series representation (\ref{x_series}), the simulation is in practice truncated at a finite number of terms and the remaining small jumps are ignored or approximated somehow \cite{Asmussen2001}, a topic that is addressed in Section \ref{sec:adaptive_truncation} of this paper. 

The generic method reviewed here requires the explicit evaluation of the inverse tail measure  $h(\gamma)$ which is not tractable for the GIG process. An alternative approach was devised in  \cite{GodsillKindap2021},  simulating from a tractable dominating point process $N_0$ having L\'{e}vy measure $Q_0$ such that $dQ_0(x)/dQ(x) \geq 1, \,\, \forall x \in (0, \infty)$ for which $h_{0}(\gamma)$ is directly available. The resulting samples from $N_0$ are then thinned with probability $dQ(x)/dQ_0(x)$ as in  (\cite{Lewis_Shedler_1979,Rosinski_2001}) to obtain the desired jump magnitudes $\{ x_i \}$ of the subordinator process. The generic procedure is given in Alg. \ref{alg:simulate_Q} for a point process $Q(x)$ having dominating density $Q_0(x)\geq Q(x)$ and $h_0(\gamma)=\inf_x\{x;\,Q_0^+(x)=\gamma\}$.

\begin{algorithm}[H]
\caption{Generation of the jumps of a point process having L\'{e}vy density $Q(x)$ and dominating process $Q_0(x)\geq Q(x)$.}
\label{alg:simulate_Q}
\begin{enumerate}
    \item Assign $N=\emptyset$,
    \item Generate the epochs of a unit rate Poisson process, $\{\Gamma_i;\,i=1,2,3...\}$,
    \item For $i=1,2,3...$,
        \begin{itemize}
            \item Compute $x_i=h_0(\Gamma_i)$
            \item With probability $Q(x_i)/Q_0(x_i)$, accept $x_i$ and assign $N=N\cup x_i$.
        \end{itemize}
\end{enumerate}
\end{algorithm}

Note that our work here will later  require partial simulation of such point processes on measurable sets $A$ on jump magnitudes, i.e. 
$Q_{A}(x)=\mathbb{I}_A(x)Q(x)$, and typically $A$ will simply be an interval $(a,b]$, $b\leq \infty$. This partial simulation is straightforwardly achieved by replacing Step 2) in Alg. 
\ref{alg:simulate_Q} with the steps provided in Alg. \ref{alg:Poisson_gen}.

\begin{algorithm}[H]
\caption{Generation of Poisson process epochs corresponding to jump magnitudes $ x_i \in (a,b]$ where $a>b$.}
\label{alg:Poisson_gen}
\begin{itemize}
\item $i=1$, $\Gamma_0=Q_0^+(a)$
\item While $\Gamma_{i-1}<Q_0^+(b)$
\begin{enumerate}
    \item $E_i\sim Exp(1)$
    \item $\Gamma_i=\Gamma_{i-1}+E_i$
    \item $i=i+1$
\end{enumerate}
\item Return $\{\Gamma_j\}_{j=1}^{i-1}$
\end{itemize}
\end{algorithm}

As before $Q_0^+(x)=Q_0 ([x,\infty))$ and $Exp(1)$ is the unit mean exponential distribution, and noting that the While loop     in Alg. \ref{alg:Poisson_gen} may in practice be replaced with a draw from $M\sim Poisson(Q_0^+(b)-Q_0^+(a))$ followed by $M$ iid draws for the (unordered) $\Gamma_i$ terms from a uniform distribution $U(Q_0^+(b)-Q_0^+(a))$.

A rejection sampling  procedure such as Alg. \ref{alg:simulate_Q} may be viewed within the generalised shot-noise framework of \cite{Rosinski2001} in which the process is expressed as a random function of the underlying Poisson epochs $\{\Gamma_i\}$ as follows
\[
X(t)=\sum_i H(\Gamma_i,e_i){\cal I}(V_i\leq t)
\]
where $H(\gamma,.)$ is a non-increasing function of $\gamma$, and $e_i$ are random variables or vectors drawn independently across $i$.
\cite{Rosinski2001} Th. 4.1 proves the a.s. convergence of such series under mild conditions. In particular the conditions of Th. 4.1 (A) are satisified.  First the distribution of 
$H(.,.)$ is expressed as a probability kernel $\sigma(\gamma,A)$ for measurable sets $A$:
\[
\mathbb{P}\{H(\gamma,e)\in A\}=\sigma(\gamma,A)
\]
and it follows from the Marking Theorem \cite{Kingman1992} that the resulting point process has L\'{e}vy measure
\[
Q(A)=\int_0^\infty \sigma(\gamma,A) d\gamma
\]

Applying this to verify Alg. \ref{alg:simulate_Q}, take $H(\gamma,e)=h(\gamma)e$ and 
$e\in\{0,1\}$ binomial with $\mathbb{P}\{e=1\}=Q(h_0(\gamma))/Q_0(h_0(\gamma))$. We will consider only non-zero jumps, since `jumps' of size zero do not impact the point process, and indeed L\'{e}vy measures $Q(dx)$ are not defined for $x=0$.  Then it follows for measurable sets $A_0=A\backslash 0$ that
\[
\sigma(\gamma,A_0)={\cal I}(h_0(\gamma)\in A_0) \frac{Q(h_0(\gamma))}{Q_0(h_0(\gamma))}
\]
and hence the resulting L\'{e}vy measure is

\begin{align*}
Q_1(A_0) &= \int_0^\infty {\cal I}(h_0(\gamma)\in A_0) \frac{Q(h_0(\gamma))}{Q_0(h_0(\gamma))} d\gamma \\
&= \int_{x\in A_0} \frac{Q(x)}{Q_0(x)} (Q_0(x)dx) \\ &= \int_{x\in A_0} Q(x) dx, 
\end{align*}

\noindent as required. Here we have made the substitution $x=h_0(\gamma)$, so $\gamma=Q_0^+(x)$ and $d\gamma$=$Q_0(x)dx$.  
While the procedure of Alg. \ref{alg:simulate_Q} is well known to be valid, see e.g. \cite{Rosinski2001}, we include the sketch proof here since we will use more sophisticated versions of it to prove validity of our own algorithms for GIG and GH process simulation in subsequent sections of the paper.

Simple and well known examples of the procedures in Algs. \ref{alg:simulate_Q} and \ref{alg:Poisson_gen} are  the {\em tempered stable\/} and {\em Gamma\/} processes, which we will  require as part our sampling procedures for the GIG process later in the paper. 
 The corresponding L\'{e}vy densities and thinning probabilities for these cases are given in \cite{GodsillKindap2021} (Section 2.1 and 2.2). The associated sampling algorithms are repeated here for reference purposes in Algorithms \ref{alg:tempered_stable} and \ref{alg:gamma_gen}. 

\begin{algorithm}[H]
\caption{Generation of the jumps of a tempered stable process with L\'{e}vy density $Q_{TS}(x) = Cx^{-1-\alpha} e^{-\beta x}$ ($x\geq 0$) where $0<\alpha<1$ is the tail parameter and $\beta\geq 0$ is the tempering parameter.}
\label{alg:tempered_stable}
\begin{enumerate}
    \item Assign $N_{TS}=\emptyset$,
    \item Generate the epochs of a unit rate Poisson process, $\{\Gamma_i;\,i=1,2,3...\}$,
    \item For $i=1,2,3...$,
        \begin{itemize}
            \item Compute $x_i=\left(\frac{\alpha\Gamma_i}{C}\right)^{-1/\alpha}$,
            \item With probability $e^{-\beta x_i}$, accept $x_i$ and assign $N_{TS}=N_{TS}\cup x_i$.
        \end{itemize}
\end{enumerate}
\end{algorithm}

\begin{algorithm}[H]
\caption{Generation of the jumps of a gamma process with L\'{e}vy density $Q_{Ga}(x) = {C}{x^{-1}}e^{-\beta x}$ ($x\geq 0$) where $C>0$ is the shape parameter and $\beta>0$ is the rate parameter.}
\label{alg:gamma_gen}
\begin{enumerate}
    \item Assign $N_{Ga}=\emptyset$,
    \item Generate the epochs of a unit rate Poisson process, $\{\Gamma_i;\,i=1,2,3...\}$,
    \item For $i=1,2,3...$,
        \begin{itemize}
            \item Compute $x_i=\frac{1}{\beta \left( \exp(\Gamma_i / C) - 1 \right)}$,
            \item With probability $(1+\beta x) \exp(-\beta x_i)$, accept $x_i$ and assign $N_{Ga}=N_{Ga}\cup x_i$.
        \end{itemize}
\end{enumerate}
\end{algorithm}

\section{The generalised inverse Gaussian L\'{e}vy process}
\label{sec:GIG_process}

In this section, the  GIG L\'{e}vy process and its L\'{e}vy measure are defined. Tractable bounds on this L\'{e}vy measure are required in order to simulate the GIG (and hence the GH) process, and in this section we provide improved bounds compared with those in \cite{GodsillKindap2021}. These improved bounds are proven in the following section to have higher acceptance rates for the rejection sampling procedures that underlie the algorithms.

The density of the L\'{e}vy measure of a GIG process (\cite{Eberlein2003}, Eq. 74), following a change of variables as in \cite{GodsillKindap2021}, is given by

\begin{equation*}
\label{eqn:gig_levy_density_full}
    \frac{e^{-x\gamma^2/2}}{x} \left[ \frac{2}{\pi^2} \int_{0}^{\infty} \frac{e^{-\frac{z^2x}{2\delta^2}}}{z|H_{|\lambda|}(z)|^2}dz + \text{max}(0,\lambda)  \right], \quad x>0
\end{equation*}

\noindent where $H_{\lambda}(z)=J_{\lambda}(z) + iY_{\lambda}(z)$ is the Bessel function of the third kind, also known as the Hankel function of the first kind, which is defined in terms of $J_{\lambda}(z)$, the Bessel function of the first kind, and  $Y_{\lambda}(z)$, the  Bessel function of the second kind. The presence of an integral involving the Bessel function makes the simulation of such processes intractable except for certain edge cases.

Naturally, the GIG L\'{e}vy density can be divided into two terms as

\begin{equation*}
    Q_{GIG}(x) =\frac{2 e^{-x\gamma^2/2}}{\pi^2x}\int_0^\infty\frac{e^{-\frac{z^2x}{2\delta^2}}}{z|H_{|\lambda|}(z)|^2}dz
\end{equation*}

\noindent and a second term, present only for $\lambda>0$ as

\begin{equation}
    \frac{\lambda e^{-x\gamma^2/2}}{x},\,\,x>0 \label{gamma_process_positive_lambda}
\end{equation}

\noindent which is the L\'{e}vy density of a gamma process with shape parameter $\lambda$ and rate $\gamma^2/2$. It is straightforward to simulate from this second term using Alg. \ref{alg:gamma_gen}, thus our attention is directed towards simulation of the point process with L\'{e}vy density $Q_{GIG}(x)$. 

In order to avoid any direct calculation of the integral in $Q_{GIG}(x)$, the general approach proposed in \cite{GodsillKindap2021} is to consider a bivariate point process $Q_{GIG}(x, z)$ on $(0,\infty) \times (0,\infty)$ which has, by construction, the GIG L\'{e}vy density as its marginal, i.e. $Q_{GIG}(x)=\int_0^\infty Q_{GIG}(x,z)dz$ such that

\begin{equation}
    Q_{GIG}(x, z) = \frac{2 e^{-x\gamma^2/2}}{\pi^2x}\frac{e^{-\frac{z^2x}{2\delta^2}}}{z|H_{|\lambda|}(z)|^2} \label{Q_GIG_def}
\end{equation}

\noindent Thus, joint samples $\{ x_i, z_i \}$ are simulated from the point process with intensity function $Q_{GIG}(x, z)$, from which the samples $\{x_i\}$ are retained as samples from $Q_{GIG}(x)$. However, simulation from $Q_{GIG}(x, z)$ is still intractable because of the presence of the Bessel function. This is overcome by constructing tractable bivariate dominating point processes with intensity function $Q^{0}_{GIG}(x, z)$ and thinning with probability $Q_{GIG}(x,z) / Q^0_{GIG}(x,z)$ to yield samples from the desired process $Q_{GIG}$. 

The generic approach proposed here will involve a marginal-conditional factorisation of both point processes:
\[
Q^{0}_{GIG}(x, z)=Q^0_{GIG}(x)Q^0_{GIG}(z|x),
\]
\[
Q_{GIG}(x, z)=Q_{GIG}(x)Q_{GIG}(z|x),
\]
where $Q^0_{GIG}(z|x)$ and $Q_{GIG}(z|x)$ are proper probability densities, i.e. $\int_0^\infty Q^0_{GIG}(z|x)dz=1$
 and $\int_0^\infty Q_{GIG}(z|x)dz=1$. Thus $z$ may be interpreted as a {\em marking} variable 
 and $(x,z) \in (0,\infty)\times(0\infty)$ form a bivariate Poisson process \cite{Kingman1992}. The generic algorithm for sampling $Q_{GIG}(x)$ is then given below, followed by its proof of validity under the generalised shot noise approach. 
\begin{algorithm}[H]
\caption{Generation of the jumps of a point process having L\'{e}vy density $Q(x)=\int_0^\infty Q(x)Q(z|x)dz$ and dominating process $Q_0(x,z)=Q_0(x)Q_0(z|x)$ such that $Q_0(x,z)\geq Q(x,z)$.}
\label{alg:simulate_Q_xz}
\begin{enumerate}
    \item Assign $N=\emptyset$,
    \item Generate Poisson epochs $\{\Gamma_i\}$ and corresponding ordered jump sizes $\{x_i=h_0(\Gamma_i)\}$ from the marginal process $Q_0(x)$ using Alg. \ref{alg:simulate_Q}
    \item For $i=1,2,3...$,
        \begin{itemize}
            \item Simulate $z_i\sim Q_0(z|x_i)$
            \item With probability $Q(x_i,z_i)/Q_0(x_i,z_i)$, accept $x_i$ and assign $N=N\cup x_i$.
        \end{itemize}
\end{enumerate}
\end{algorithm}
We now proceed to prove the convergence of Alg. 5 using \cite{Rosinski2001} Th. 4.1 (A). Note that the proof is here presented for the first time. 
\begin{lemma}
\label{lemma:1}
The Generalised Shot noise process 
$X(t)=\sum_i H(\Gamma_i,e_i){\cal I}(V_i\leq t)$ generated according to Alg. \ref{alg:simulate_Q_xz} converges a.s. to the Poisson point process with L\'{e}vy density $Q(x)$.  
\end{lemma}
\begin{proof}
Alg. \ref{alg:simulate_Q_xz} generates, for each point $x_i$, an auxiliary marking variable $z_i\sim Q_0(z_i|x_i)$, and an acceptance variable $a_i\in \{0,1\}$ that is binomial with $\mathbb{P}\{a_i=1\}=Q(x_i,z_i)/Q_0(x_i,z_i)$. Thus set $e_i=(z_i,a_i)\in(0,\infty)\times \{0,1\}$ and hence
\[
H(\gamma,(z_i,a_i))=h_0(\gamma_i)a_i
\]
with resulting probability kernel
\begin{align*} 
\sigma(\gamma,A) &= \int_0^\infty {\cal I}(h_0(\gamma)\in A_0)\left(Q(h_0(\gamma),z)/Q_0(h_0(\gamma),z)\right)Q_0(z|h_0(\gamma))dz \\ &= {\cal I}(h_0(\gamma)\in A_0)Q(h_0(\gamma))/Q_0(h_0(\gamma))
\end{align*} 

\noindent where once again $A_0$ are all measurable sets excluding 0.
Hence  the resulting L\'{e}vy measure is
\begin{align*}
Q_1(A_0)&= \int_0^\infty {\cal I}(h_0(\gamma)\in A_0)Q(h_0(\gamma))/Q_0(h_0(\gamma)) d\gamma\\
&=\int_{x\in A_0} Q(x) dx
\end{align*}
as required. The remaining conditions in \cite{Rosinski2001} Th. 4.1 (A) are simply that $Q()$ is a L\'{e}vy measure, which is true by construction (it is a subordinator and hence satisfies (\ref{sub_cond})), and a second technical condition that is always satisfied by subordinators, see \cite{Rosinski2001} Remark 4.1. 
\end{proof}

\begin{figure}[!t]
\centering
\includegraphics[width=0.7\textwidth]{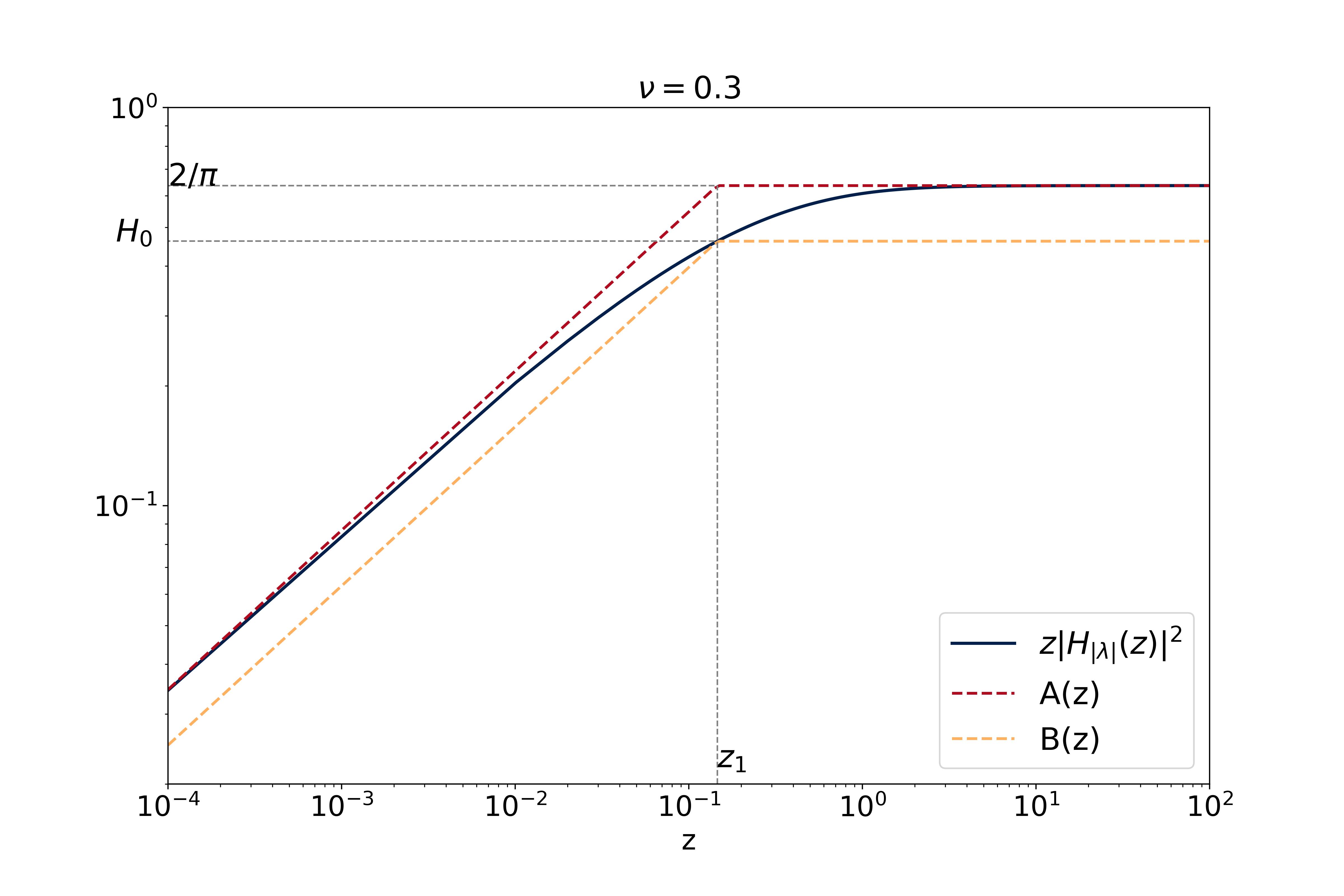}
\caption{Plot of Bessel function bounds, $\nu=0.3$. $z_0$ set equal to $z_1$ and $z_1=\left(\frac{ 2^{1-2\nu}\pi}{\Gamma^2(\nu)}\right)^{1/(1-2\nu)}$.}
\label{fig:theorem_bounds2}
\end{figure}

A new set of bounds on $z|H_{\nu}(z)|^2$ is now given in Theorem \ref{theorem:1} below, which will be used in bounding the overall function (\ref{Q_GIG_def}).
The bounds are graphically illustrated for the two distinct parameter ranges in Figs. \ref{fig:theorem_bounds} and \ref{fig:theorem_bounds2}. 
The proof of the theorem follows a similar scheme to Theorem 2 in \cite{GodsillKindap2021} and is hence only briefly stated:  

\begin{figure}[!t]
\centering
\includegraphics[width=0.7\textwidth]{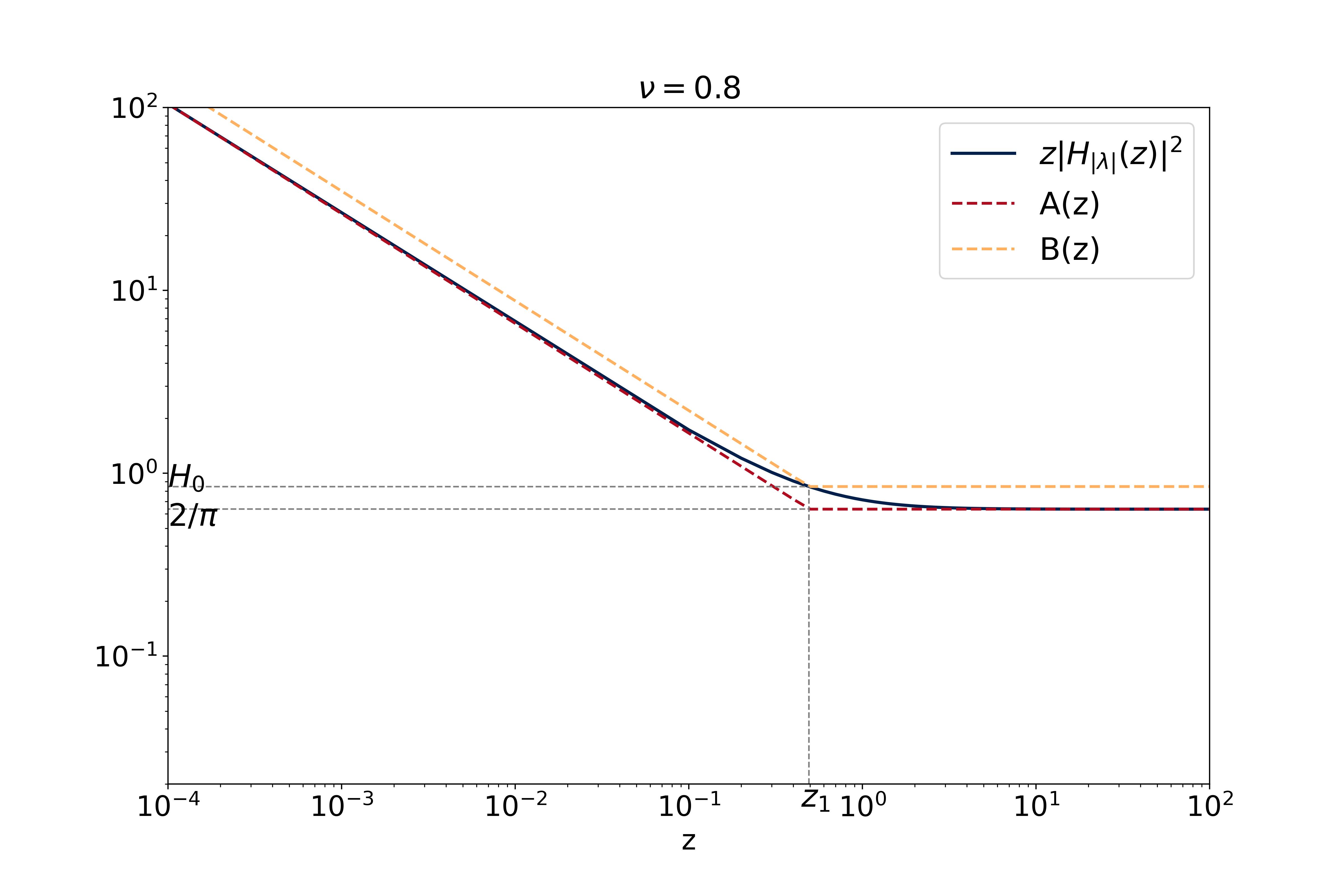}
\caption{Plot of Bessel function bounds, $\nu=0.8$. $z_0$ set equal to $z_1$ and $z_1=\left(\frac{ 2^{1-2\nu}\pi}{\Gamma^2(\nu)}\right)^{1/(1-2\nu)}$.}
\label{fig:theorem_bounds}
\end{figure}

\begin{theorem}
\label{theorem:1}
Choose a point $z_0\in(0,\infty)$ and compute $H_0=z_0|H_{\nu}(z_0)|^2$. This will define the corner point on a  piecewise lower or upper bound. Choose now any $0\leq z_1 \leq \left(\frac{ 2^{1-2\nu}\pi}{\Gamma^2(\nu)}\right)^{1/(1-2\nu)}$ and define the following functions:
\[
    A(z)=\begin{cases}\frac{2}{\pi}\left(\frac{z_1}{z}\right)^{2\nu-1},&z < z_1 \\ \frac{2}{\pi},&z\geq z_1 \end{cases}
\]
and
\[
    B(z)=\begin{cases}H_0\left(\frac{z_0}{z}\right)^{2\nu-1},&z<z_0\\H_0,&z\geq z_0\end{cases}
\]
Then, for $0<\nu\leq 0.5$,
\begin{equation}
    A(z)\geq z|H_{\nu}(z)|^2\geq B(z) \label{bound_lam_leq_0_5}
\end{equation}
and for $\nu\geq 0.5$,
\begin{equation}
    A(z)\leq z|H_{\nu}(z)|^2\leq B(z) \label{bound_lam_geq_0_5}
\end{equation}
with all inequalities becoming equalities when $\nu=0.5$, and both $A(z)$ bounds (left side inequalities) becoming tight at $z=0$ and $z=\infty$. 
\end{theorem}
\begin{proof}
As in Theorem 2 of \cite{GodsillKindap2021}, but now we allow for a range of values  $0\leq z_1 \leq \left(\frac{ 2^{1-2\nu}\pi}{\Gamma^2(\nu)}\right)^{1/(1-2\nu)}$. The proof is an obvious extension of the original theorem since for any value of $z_1$ less than $\left(\frac{ 2^{1-2\nu}\pi}{\Gamma^2(\nu)}\right)^{1/(1-2\nu)}$, the function $A(z)$ is shifted to the left and lies below of those plotted in Fig. \ref{fig:theorem_bounds} and above of those plotted in Fig. \ref{fig:theorem_bounds2}, hence providing a valid but less tight bounding function.
\end{proof}

\begin{remark}
\label{new_corollary}
Choice of any $z_1<\left(\frac{ 2^{1-2\nu}\pi}{\Gamma^2(\nu)}\right)^{1/(1-2\nu)}$ leads to a poorer (less tight) bound on the true function $z|H_{\nu}(z)|^2$. In particular, as $z_1\rightarrow 0$ we obtain the crude and well known (\cite{Watson1944}, Section 13.75) bound $A(z)=\frac{2}{\pi}$, which was employed in a first version of our method for $|\lambda|>0.5$ (\cite{GodsillKindap2021}, Theorem 1).  This asymptotic bound forms the loosest bound $A(z)$ and all other valid choices of $z_1$ yield increasingly tight bounds on the required L\'{e}vy density  $Q_{GIG}(x,z)$, which we employ in our new and improved sampling algorithms for $|\lambda|>0.5$. 
\end{remark}

\begin{corollary}
\label{cor:theorem1}
For any positive $z$ and fixed $|\lambda|$, the following bounds are obtained by replacing $z|H_{\nu}(z)|^2$ with $A(z)$  and $B(z)$ in the definition of $Q_{GIG}(x,z)$ (\ref{Q_GIG_def}). For the case $|\lambda|\geq 0.5$ we have:
\begin{equation}
Q_{GIG}^{B}(x,z)\leq    Q_{GIG}(x,z) \leq Q_{GIG}^{A}(x,z)\label{eq:bound_geq_05}
\end{equation}
and for $0<|\lambda|\leq 0.5$ we have:
\begin{equation}
Q_{GIG}^{A}(x,z)\leq    Q_{GIG}(x,z) \leq Q_{GIG}^{B}(x,z)\label{eq:bound_leq_05}
\end{equation}
with equality being achieved in both cases for $|\lambda|=0.5$. 
Here $Q_{GIG}^{A}(x,z)$ is defined as:
\begin{subnumcases}{\label{Q_GIG_bound_A} Q_{GIG}^{A}(x,z) =}
     \frac{e^{-x \gamma^2/2}}{\pi x}\frac{z^{2|\lambda|-1} e^{-\frac{z^2 x}{2 \delta^2}}}{z_1^{2|\lambda|-1}},\,\,\, z < z_1 \label{Q_GIG_bound_A_case1} \\ \frac{e^{-x \gamma^2/2}}{\pi x} e^{-\frac{z^2 x}{2 \delta^2}}, \quad \,\,z \geq z_1 \label{Q_GIG_bound_A_case2}
\end{subnumcases} and $Q_{GIG}^{B}(x,z)$ defined as:
\begin{subnumcases}{\label{Q_GIG_bound_B}
Q_{GIG}^{B}(x,z)=}
\frac{ 2 e^{-x\gamma^2/2}}{\pi^2x}\frac{z^{2|\lambda|-1}e^{-\frac{z^2x}{2\delta^2}}}{H_0z_0^{2|\lambda|-1}},\,\,\,z<z_0 \label{Q_GIG_bound_B_case1}\\
\frac{ 2 e^{-x\gamma^2/2}}{\pi^2x}\frac{e^{-\frac{z^2x}{2\delta^2}}}{H_0},\,\,\,z\geq z_0 \label{Q_GIG_bound_B_case2}
\end{subnumcases}
\end{corollary}
\begin{remark}
\label{rem:piH_0/2} Setting $z_0=z_1$, it can be clearly seen that the ratio $Q_{GIG}^{A}(x,z)/Q_{GIG}^{B}(x,z)=\pi H_0/2$, a constant independent of the value of $x$ and $z$. This fact, which can be clearly visualised in Figs. \ref{fig:theorem_bounds} and \ref{fig:theorem_bounds2} (note the log-scale), is utilised later in  
our development of a retrospective `squeezed' sampler, see Section \ref{sec:squeezing_functions}.\end{remark}

\begin{corollary}
\label{cor_GIG_bound21}
The bound in Eq. (\ref{Q_GIG_bound_A_case1}) can be rewritten in factorised form as
\begin{align*}
\label{A_bound_decomposed1}
   Q^A_{N_1}(x,z) &=\frac{e^{-x \gamma^2/2}}{\pi x} \frac{z^{2|\lambda|-1} e^{-\frac{z^2 x}{2 \delta^2}}}{z_1^{2|\lambda|-1}}{\cal I}_{0<z<z_1}\\& = \frac{ e^{-x\gamma^2/2}}{\pi x^{1+|\lambda|}}\frac{(2\delta^2)^{|\lambda|}\gamma(|\lambda|,z_1^2x/(2\delta^2))}{2z_1^{2|\lambda|-1}} \frac{\Gamma(|\lambda|)\sqrt{\text{Ga}}
(z||\lambda|,x/(2\delta^2))}{\gamma(|\lambda|,z_1^2x/(2\delta^2))} {\cal I}_{0<z<z_1}\\
&=Q^A_{N_1}(x)Q^A_{N_1}(z|x)
\end{align*}
where 
\[
Q^A_{N_1}(z|x)=\frac{\Gamma(|\lambda|)\sqrt{\text{Ga}}
(z||\lambda|,x/(2\delta^2))}{\gamma(|\lambda|,z_1^2x/(2\delta^2))}{\cal I}_{0<z<z_1}
\]
is a conditional right-truncated square-root gamma density\footnote{The square-root gamma density $\sqrt{Ga}$ is the density of $X^{0.5}$ when $X\sim Ga(x|\alpha,\beta)$, which has the probability density function $
    \sqrt{\text{Ga}}(x|\alpha, \beta) = \frac{2 \beta^{\alpha}}{\Gamma(\alpha)} x^{2\alpha-1} e^{-\beta x^2} 
$} with its associated normalising constant. The marginal term $Q^A_{N_1}(x)$ is  a modified tempered $|\lambda|$-stable process\footnote{The lower/upper incomplete gamma functions are defined, for $\text{Re}(s) > 0$, as:
\begin{align*}
    \gamma(s,x) = \int_{0}^{x} t^{s-1} e^{-t} dt,\,\,\,\,
    \Gamma(s,x) = \int_{x}^{\infty} t^{s-1} e^{-t} dt
\end{align*}} .
\end{corollary}

\begin{corollary}
\label{cor_GIG_bound22}
The bound in Eq. (\ref{Q_GIG_bound_A_case2}) can be rewritten in a similar way as
\begin{align*}
\label{A_bound_decomposed2}
   Q^A_{N_2}(x,z)& =\frac{e^{-x \gamma^2/2}}{\pi x} e^{-\frac{z^2 x}{2 \delta^2}}{\cal I}_{z\geq z_1}\\ &=\frac{ e^{-x\gamma^2/2}}{\pi x^{3/2}}\frac{(2\delta^2)^{0.5}\Gamma(0.5,z_1^2x/(2\delta^2))}{2} \frac{\Gamma(0.5)\sqrt{\text{Ga}}
(z|0.5,x/(2\delta^2))}{\Gamma(0.5,z_1^2x/(2\delta^2))} {\cal I}_{z\geq z_1}\\
&= Q^A_{N_2}(x) Q^A_{N_2}(z|x)
\end{align*}
where 
\[
Q^A_{N_2}(z|x)=\frac{\Gamma(0.5)\sqrt{\text{Ga}}
(z|0.5,x/(2\delta^2))}{\Gamma(0.5,z_1^2x/(2\delta^2))} {\cal I}_{z\geq z_1}
\]
is a conditional left-truncated square-root gamma density with its associated normalising constant. The marginal term $Q^A_{N_2}(x)$ is a modified tempered $0.5$-stable process.
\end{corollary}

In Corollaries \ref{cor_GIG_bound21} and \ref{cor_GIG_bound22}, the point process intensities correspond marginally to a (modified) tempered stable process in $x$, and conditionally to a truncated $\sqrt{Ga}$ density for $z$. This feature enables sampling from the dominating bivariate point process $Q^A_{GIG}(x,z)$ by first sampling $x$ and then, conditional on the value of $x$, sampling a corresponding $z$ value. Here, for the parameter range $|\lambda|\geq 0.5$, we are extending the approach previously derived only for the parameter range $0<|\lambda|<0.5$  \cite{GodsillKindap2021}, (and here summarised in Section \ref{sec:lambda_leq_0.5}). 
Notice that allowing $z_1 \rightarrow 0$, as discussed in the Remark following Theorem \ref{theorem:1}, will result in our previous crude bounding function for the parameter range $|\lambda|>0.5$  (\cite{GodsillKindap2021}, Corollary 1).

Since the functions $Q_{GIG}^{A}(x, z)$ and $Q_{GIG}^{B}(x, z)$ form both upper and lower bounds on the target point process $Q_{GIG}(x, z)$, see (\ref{eq:bound_geq_05}) and (\ref{eq:bound_leq_05}), we are able to construct effective sampling algorithms for all parameter ranges $|\lambda|>0$ based upon rejection sampling ideas, see sections \ref{sec:lambda_geq_0.5} and \ref{sec:lambda_leq_0.5}. Furthermore, in Section \ref{sec:squeezing_functions} we use the corresponding lower bounds to create efficient `squeezed' versions of these algorithms. In the case of  $0 < |\lambda| \leq 0.5$, the new algorithm is a direct development of that presented in \cite{GodsillKindap2021}, while in the case $|\lambda|>0.5$ the algorithm now follows the same structure as for the other parameter range, in contrast with the previous approach from \cite{GodsillKindap2021} that uses a cruder bound. In all parameter ranges we propose significant improvements over the previous work, including better bounds for simulation of the marginal process, adaptive truncation, simulation of a Gaussian residual term and squeezed rejection sampling. 

In the next section we show how to move from simulations of the underlying GIG process towards our ultimate aim here, which is simulation of the GH process.

\section{Simulating GH processes}
\label{sec:simulation_algorithms}

In this section simulation algorithms for the GH L\'{e}vy process are presented. Our approach relies on the definition of a GH process as a subordinated Brownian motion where the subordinator is a GIG process (\cite{Barndorffnielsen1977InfiniteDO,BarndorffNielsen1978}). In this approach, jumps sizes $\{x_i\}$ are first generated from the underlying GIG process $Q_{GIG}$, see previous sections for full details. Then, jumps for the corresponding GH process are obtained as
 \begin{equation}
    \label{eqn:normal_variance_mean_mapping}
        w_i = \mu + \beta x_i + \sigma \sqrt{x_i} u_i, \,\,\,u_i\overset{iid}{\sim}{\cal N}(0,1)
    \end{equation}
for some $\beta\in\Re$ and $\sigma>0$.

The conditional simulation of the GH process is common to all parameter regimes and is presented in Alg. \ref{alg:Q_GH_gen}. Given jump times and magnitudes $(V_i, W_i)$ the corresponding value of the GH L{\'e}vy process at $t$ is given in Eq. (\ref{eqn:shot_noise_gen}).

\begin{algorithm}
\caption{Simulation of GH process.}
\label{alg:Q_GH_gen}
\begin{enumerate}
    \item Generate a large number of points $x_i$ from the GIG L\'{e}vy process $Q_{GIG}$, see Algorithms \ref{gen_N_1} and \ref{gen_N_2} for $|\lambda|\geq 0.5$, or Algorithms \ref{alg:Bgen_N_1} and \ref{alg:Bgen_N_2} for $0< |\lambda| \leq 0.5$,
    \item If $\lambda>0$, draw an additional set of points $\{x_j'\}$ from the gamma process   $\frac{\lambda e^{-x\gamma^2/2}}{x},\,\,x>0$, see  (\ref{gamma_process_positive_lambda})
    and take the union $\{x_i\}\leftarrow\{x_i\}\cup\{x_j'\}$,
    \item For each point $x_i$, draw an independent and identically distributed random variate $u_i \sim \mathcal{N}(0,1)$,
    \item The corresponding GH jump sizes $w_i$ are obtained as:
    \begin{equation*}
        w_i = \mu + \beta x_i + \sigma \sqrt{x_i} u_i
    \end{equation*}
    \item For each jump $w_i$, draw an independent and identically distributed jump time $v_i \sim \mathcal{U}(0,T)$.
    \item The process $W(t)$ is then obtained as:
    \[
    W(t)=\sum_{i=1}^\infty w_i{\cal I}(v_i\leq t)
    \]
\end{enumerate}    
\end{algorithm}

\noindent We now detail the methods for generation of the  underlying GIG process.

\subsection{The case for $|\lambda| \geq 0.5$}
\label{sec:lambda_geq_0.5}
Here a new algorithm is presented for simulation in the  parameter range $|\lambda| \geq 0.5$, based on the bound $Q_{GIG}^{A}(x,z)$ derived in previous sections, which  is an improved bound compared with that in \cite{GodsillKindap2021} Algorithm 3.
The process associated with the L{\'e}vy density $Q_{GIG}^{A}(x,z)$ can be considered as a marked point process split into two independent point processes $N_1$ and $N_2$ having factorised (marginal-conditional) intensity functions as given in Corollaries \ref{cor_GIG_bound21} and \ref{cor_GIG_bound22}, respectively.
    
Both $N_1$ and $N_2$ correspond to a marginal modified tempered stable process for $x$ and a conditional truncated $\sqrt{Ga}$ density for $z$. Each simulated pair $(x,z)$ is accepted with probability equal to the ratio $Q_{GIG}(x,z)/Q_{GIG}^{A}(x,z)$. As a result of the piecewise form of $Q_{GIG}^{A}(x,z)$, the accept/reject steps for $N_1$ and $N_2$ may be treated independently and the union of points from the two processes forms the final set of GIG points. The thinning probabilities for points drawn from $Q^A_{N_1}$ and $Q^A_{N_2}$ are then:

\begin{equation}
\frac{Q_{GIG}(x,z)}{Q^A_{N_1}(x, z)} = \frac{2}{\pi |H_{|\lambda|}(z)|^2 \left( \frac{z^{2|\lambda|}}{z_1^{2|\lambda|-1}} \right)} \label{eq:Q_N1_ratio}
\end{equation}

\begin{equation}
\frac{Q_{GIG}(x,z)}{Q^A_{N_2}(x, z)} = \frac{2}{\pi z|H_{|\lambda|}(z)|^2} \label{eq:Q_N2_ratio}
\end{equation}

Due to the presence of upper and lower incomplete gamma functions in the marginal point process envelopes $Q^A_{N_1}(x)$ and $Q^A_{N_2}(x)$ defined as

\begin{equation}
    Q^A_{N_1}(x) = \frac{ e^{-x\gamma^2/2}}{\pi x^{1+|\lambda|}}\frac{(2\delta^2)^{|\lambda|}\gamma(|\lambda|,z_1^2x/(2\delta^2))}{2z_1^{2|\lambda|-1}}\label{eq:QAN1}
\end{equation}

\noindent and

\begin{equation}
    Q^A_{N_2}(x) = \frac{ e^{-x\gamma^2/2}}{\pi x^{3/2}}\frac{(2\delta^2)^{0.5}\Gamma(0.5,z_1^2x/(2\delta^2))}{2} \label{eq:QAN2}
\end{equation}

\noindent direct simulation from these L\'{e}vy densities are still intractable and hence dominating processes and associated thinning methods are required.

For $Q^A_{N_1}(x)$ the following bound is used to formulate a tractable dominating process (\cite{Neuman2013}, Theorem 4.1):

\begin{equation}
\label{eqn:lower_gamma_ineq}
    \frac{a \gamma(a,x)}{x^a} \leq \frac{(1+ae^{-x})}{(1+a)}
\end{equation}

\noindent so that the dominating process can be expressed as:

\begin{align}
\label{eqn:dom_N1_A}
    Q^A_{N_1}(x) &\leq \frac{e^{-x\gamma^2/2}}{\pi x} \frac{z_1 (1+|\lambda|e^{-(z_1^2 x)/(2 \delta^2)})}{2 |\lambda| (1+|\lambda|)} \nonumber\\
    &=\frac{z_1}{2 \pi (1+|\lambda|)}\left(\frac{e^{-x\gamma^2/2}}{|\lambda|x} +\frac{e^{-x(\gamma^2/2+z_1^2/(2 \delta^2)})}{x}\right)\nonumber \\ 
    &= Q_{N_1}^{A,d}(x)
\end{align}

Notice that the point process associated with $Q_{N_1}^{A,d}(x)$ may be considered as the union of two independent gamma processes. Points are then independently accepted with probability $Q_{N_1}^{A}(x)/Q_{N_1}^{A,d}(x)$. The corresponding algorithm is given in Alg. \ref{alg:phase1_N1}.

\begin{algorithm}[H]
\caption{Sampling from $Q^{A}_{N_1}(x)$.}
\label{alg:phase1_N1}
\begin{enumerate}
    \item $N=\emptyset$,
    \item Generate a gamma process $N_{Ga}^{1}$ having parameters $a_1 = \frac{z_1}{2 \pi |\lambda| (1+|\lambda|)}$ and $\beta_1=\gamma^2/2$,
    \item Generate a gamma process $N_{Ga}^{2}$ having parameters $a_2 = \frac{z_1}{2 \pi (1+|\lambda|)}$ and $\beta_2= \gamma^2/2 + z_1^{2}/(2 \delta^2)$,
    \item $N = N_{Ga}^{1} \cup N_{Ga}^{2}$,
    \item For each $x \in N$ accept with probability
    \[
        \frac{(2\delta^2)^{|\lambda|} \gamma(|\lambda|, (z_1^2 x)/(2\delta^2)) |\lambda| (1+|\lambda|)}{x^{|\lambda|} z_1^{2|\lambda|} (1+|\lambda| \text{exp}\left( - z_1^2 x / (2\delta^2) \right) ) }
    \]
    \noindent otherwise reject and delete $x$ from $N$.
\end{enumerate}
\end{algorithm}

Having simulated the $x$ values from the marginal point process associated with $Q^{A}_{N_1}(x)$, the corresponding $z$ values are simulated from a right-truncated square-root gamma density as in Corollary \ref{cor_GIG_bound21} and accept-reject steps are carried out to obtain samples from the $N_1$ point process. The complete procedure is outlined in Alg. \ref{gen_N_1}.

\begin{algorithm}[H]
\caption{Generation of $N_1$ for $|\lambda| \geq 0.5$}
\label{gen_N_1}
\begin{enumerate}
    \item $N_1={\emptyset}$,
    \item Simulate $x_i$ from the marginal point process associated with $Q^{A}_{N_1}(x)$ as given in Alg. \ref{alg:phase1_N1},
    \item For each $x_i$, simulate a $z_i$ from a truncated square-root gamma density
    \[
        \frac{\Gamma(|\lambda|)\sqrt{\text{Ga}} (z||\lambda|,x_i/(2\delta^2))}{\gamma(|\lambda|,z_1^2x_i/(2\delta^2))} {\cal I}_{0<z<z_1}
    \]
    \item With probability
    \[
        \frac{Q_{GIG}(x_i,z_i)}{Q^A_{N_1}(x_i, z_i)} = \frac{2}{\pi |H_{|\lambda|}(z_i)|^2 \left( \frac{z_i^{2|\lambda|}}{z_1^{2|\lambda|-1}} \right)}
    \]
    \noindent accept $x_i$, i.e. set $N_1=N_1\cup x_i$, otherwise discard $x_i$.
\end{enumerate}
\end{algorithm}
 
For the simulation of $Q^A_{N_2}(x)$ in (\ref{eq:QAN2}), a bound on the term $\Gamma(0.5, z_{1}^{2}x/(2\delta^{2}))$ is established by using the well-known equivalence $\Gamma(0.5, x) = \sqrt{\pi} \, \mathrm{erfc}(\sqrt{x})$ where $\mathrm{erfc}(\cdot)$ is the complementary error function. Two valid upper bounds on the gamma function are then obtained directly from \cite{ChianiDardariSimon2003} as

\begin{equation}
\label{eqn:upper_gamma_ineq}
    \Gamma(0.5, x) \leq \sqrt{\pi} \left[ \frac{1}{2} e^{-2x} + \frac{1}{2} e^{-x} \right] \leq \sqrt{\pi} e^{-x}
\end{equation}

While the first inequality is tighter, using it requires the simulation of two TS processes instead of a single process. Hence in the current implementation the right hand bound in Eq. (\ref{eqn:upper_gamma_ineq}) is chosen  and the associated dominating point process envelope is then given by

\begin{align}
\label{eqn:dom_N2_A}
    Q^A_{N_2}(x) &\leq \frac{ \delta e^{-\left[ \frac{z_1^2}{2\delta^2} + \frac{\gamma^2}{2} \right]x}}{ \sqrt{2\pi}x^{3/2}} \nonumber\\
    &= Q_{N_2}^{A,d}(x)
\end{align}

\noindent which can be simulated as a TS process and for each $x_i$ the probability of acceptance is $\Gamma(0.5,z_1^2 x_i/(2\delta^2))/(\sqrt{\pi} e^{-z_1^2 x /(2\delta^2)})$. The corresponding algorithm is given in Alg. \ref{alg:phase1_N2}.

\begin{algorithm}[H]
\caption{Sampling from $Q^{A}_{N_2}(x)$.}
\label{alg:phase1_N2}
\begin{enumerate}
    \item Generate a tempered stable process $N_{MTS}$ with parameters $C=\frac{\delta}{\sqrt{2\pi}}$, $\alpha=0.5$ and $\beta=\frac{z_1^2}{2\delta^2} + \frac{\gamma^2}{2}$ using Algorithm \ref{alg:tempered_stable}.
    \item For each point $x\in N_{MTS}$, accept with probability $\Gamma(0.5,z_1^2 x/(2\delta^2))/(\sqrt{\pi} e^{-z_1^2 x /(2\delta^2)})$, otherwise reject and delete $x$ from $N_{MTS}$.
\end{enumerate}
\end{algorithm}

Using the simulated values $x_i$, the corresponding $z_i$ values are generated from the conditional left-truncated square-root gamma density $Q^A_{N_2}(z|x)$ and the whole procedure is outlined in Alg. \ref{gen_N_2}.

\begin{algorithm}[H]
\caption{Generation of $N_2$ for $|\lambda| \geq 0.5$}
\label{gen_N_2}
\begin{enumerate}
    \item $N_2={\emptyset}$,
    \item Simulate $x_i$ from the marginal point process associated with $Q^{A}_{N_2}(x)$ as shown in Algorithm \ref{alg:phase1_N2},
    \item For each $x_i$, simulate a $z_i$ from a truncated square-root gamma density
    \[
        \frac{\Gamma(0.5)\sqrt{\text{Ga}} (z|0.5,x_i/(2\delta^2))}{\Gamma(0.5,z_1^2x_i/(2\delta^2))} {\cal I}_{z\geq z_1}
    \]
    \item With probability 
    \[
        \frac{Q_{GIG}(x_i,z_i)}{Q^A_{N_2}(x_i, z_i)} = \frac{2}{\pi z_i|H_{|\lambda|}(z_i)|^2}
    \]
    \noindent accept $x_i$, i.e. set $N_2=N_2\cup x_i$, otherwise discard $x_i$.
\end{enumerate}
\end{algorithm}

Note that the bound shown in Eq. (\ref{eqn:upper_gamma_ineq}) is a significant improvement over the bound based on the complete gamma function as used in Alg. 7 of \cite{GodsillKindap2021}. 

Finally, the set of points $N=N_1\cup N_2$ is a realisation of jump magnitudes corresponding to a GIG process having intensity function $Q_{GIG}(x)$. The associated GH process may be obtained using Alg. \ref{alg:Q_GH_gen}.
\begin{remark}
Note that whenever $\lambda > 0$, the set of points generated from the process with intensity function in Eq. (\ref{gamma_process_positive_lambda}) is added (by taking a union operation) to the set of points coming from $Q_{GIG}(x)$ in Algorithms \ref{gen_N_1} and \ref{gen_N_2} for $|\lambda|\geq 0.5$, or Algorithms \ref{alg:Bgen_N_1} and \ref{alg:Bgen_N_2} for $0< |\lambda| \leq 0.5$,  to obtain the full set of jumps $\{ x_i\}$ from the required GIG process. 
\end{remark}

\begin{remark}
    Notice that for $\gamma = 0$, the gamma process $N_{Ga}^{1}$ in Alg. \ref{alg:phase1_N1} is not well-defined since its rate parameter is zero and hence $N_1$ cannot be simulated. In this case, set $z_1=0$ and then only samples from $N_2$ are required. In this case the conditional density for $z$ becomes the complete square-root gamma density instead of a truncated one and the marginal modified tempered stable process for $x$ reduces to a standard TS process that does not require any further thinning. The simulation algorithm then becomes equivalent to Alg. 3 in \cite{GodsillKindap2021}. For all other values of $\gamma$ we set $z_1 \leq \left(\frac{2^{1-2\nu}\pi}{\Gamma^2(\nu)}\right)^{1/(1-2\nu)}$, in order to achieve the improved bound according to Theorem \ref{theorem:1}.
\end{remark}

\subsection{Acceptance rates for simulation from $Q_{GIG}^{A}(x,z)$}
We now analyse the acceptance rates for the new procedure. This will enable a quantitative comparison with our previous methods in \cite{GodsillKindap2021}. 
The acceptance probabilities for the two point processes $N_1$ and $N_2$ are obtained from (\ref{eq:Q_N1_ratio}) and (\ref{eq:Q_N2_ratio}) as

\begin{align*}
 \rho_1(x,z)& =  \frac{2}{\pi |H_{|\lambda|}(z)|^2 \left( \frac{z^{2|\lambda|}}{z_1^{2|\lambda|-1}} \right)}\\  \rho_2(x,z) &= \frac{2}{\pi z|H_{|\lambda|}(z)|^2}\,.
\end{align*}
 The expected value of the acceptance rates for fixed $x$ may be evaluated w.r.t. the sampling densities for $z$, i.e. $Q^A_{N_1}(z|x)$ (\ref{eq:QAN1}) and  $Q^A_{N_2}(z|x)$ (\ref{eq:QAN2}):

\begin{align*}
    \mathbb{E}\left[ \rho_1(x,z) \right] =& \int_{0}^{z_1} \frac{2}{\pi z |H_{|\lambda|}(z)|^2 \left( \frac{z}{z_1} \right)^{2|\lambda|-1}} \frac{\Gamma(|\lambda|)\sqrt{\text{Ga}}
(z||\lambda|,x/(2\delta^2))}{\gamma(|\lambda|,z_1^2x/(2\delta^2))} dz
\end{align*}

\begin{align*}
    \mathbb{E}\left[ \rho_2(x,z) \right] =& \int_{z_1}^{\infty} \frac{2}{\pi z|H_{|\lambda|}(z)|^2} \frac{\Gamma(0.5)\sqrt{\text{Ga}}
(z|0.5,x/(2\delta^2))}{\Gamma(0.5,z_1^2x/(2\delta^2))} dz
\end{align*}

However, the presence of the term $z|H_{|\lambda|}(z)|^2$ in both integrals makes them intractable. The expected acceptance rates may then be bounded by using the same functions $A(z)$ and $B(z)$, introduced in Theorem \ref{theorem:1}, to replace $z|H_{|\lambda|}(z)|^2$. Note that only the lower bound on the acceptance rates are of interest here since upper bounding both expectations using $A(z)$ leads to a trivial  upper bound of $1$ on the acceptance rates.

The acceptance rates associated with the $N_1$ and $N_2$ processes can be lower bounded using Theorem \ref{theorem4} and the proof is provided in the Appendix.

\begin{theorem}
\label{theorem4}
Choose a point $z_0 \in [0,\infty)$ and compute $H_0=z_0|H_{|\lambda|}(z_0)|^2$. For any fixed $x$ and $|\lambda| \geq 0.5$, the following lower bounds on $\mathbb{E}\left[ \rho_1(x,z) \right]$ and $\mathbb{E}\left[ \rho_2(x,z) \right]$  apply:

\begin{align}
 \mathbb{E}\left[ \rho_1(x,z) \right] \geq \left\{
\begin{array}{ll}
        \frac{2}{\pi H_0} \Bigg[ \left( \frac{z_1}{z_0} \right)^{2|\lambda|-1} \frac{\gamma(|\lambda|, \frac{z_0^2 x}{2 \delta^2})}{\gamma(|\lambda|,\frac{z_1^2 x}{2 \delta^2} )}
        + \left( \frac{z_1^2 x}{2 \delta^2} \right)^{|\lambda|-0.5} \frac{\left( \gamma(0.5, \frac{z_1^2 x}{2 \delta^2}) - \gamma(0.5, \frac{z_0^2 x}{2 \delta^2}) \right)}{\gamma(|\lambda|, \frac{z_1^2 x}{2 \delta^2})} \Bigg], \, z_0 \in [0, z_1) \\
       \frac{2}{\pi H_0} \left( \frac{z_1}{z_0} \right)^{2|\lambda|-1} \, \, , \,  z_0 \in [z_1,\infty) \\
\end{array} 
\right. 
\end{align}

\begin{align}
 \mathbb{E}\left[ \rho_2(x,z) \right] \geq \left\{
\begin{array}{ll}
        \frac{2}{\pi H_0} , \, z_0 \in [0, z_1) \\
        \frac{2}{\pi H_0} \Bigg[ \frac{\Gamma(0.5, \frac{z_0^2 x}{2 \delta^2})}{\Gamma(0.5, \frac{z_1^2 x}{2 \delta^2})} + \left( \frac{z_0^2 x}{2 \delta^2} \right)^{0.5-|\lambda|} \frac{\left( \gamma(|\lambda|, \frac{z_0^2 x}{2 \delta^2}) - \gamma(|\lambda|, \frac{z_1^2 x}{2 \delta^2}) \right)}{\Gamma(0.5, \frac{z_1^2 x}{2 \delta^2})} \Bigg], \,  z_0 \in [z_1,\infty)
\end{array} 
\right. 
\end{align}

\end{theorem}

Note that the corner point $z_0\in (0,\infty)$ may be chosen arbitrarily, while $z_1$ is considered to be fixed. Therefore the lower bound may be optimised for both $N_1$ and $N_2$ w.r.t. $z_0$ for each $x$ value. For $N_1$ the optimised lower bound may be expressed as

\begin{align*}
    \mathbb{E}\left[ \rho_1(x,z) \right] \geq \underset{z_0}{\text{max}} \begin{cases} & \frac{2}{\pi H_0} \left( \frac{z_1}{z_0} \right)^{2|\lambda|-1}, \quad z_0 \in (z_1, \infty) \\ & \frac{2}{\pi H_0} \bigg( \left( \frac{z_1}{z_0} \right)^{2|\lambda|-1} \frac{\gamma(|\lambda|, \frac{z_0^2 x}{2 \delta^2})}{\gamma(|\lambda|, \frac{z_1^2 x}{2 \delta^2})} + \left( \frac{z_1^2 x}{2 \delta^2} \right)^{|\lambda|-0.5} \frac{\left[ \gamma(0.5, \frac{z_1^2 x}{2 \delta^2}) - \gamma(0.5, \frac{z_0^2 x}{2 \delta^2}) \right]}{\gamma(|\lambda|, \frac{z_1^2 x}{2 \delta^2})} \bigg)  \, , \quad z_0 \in (0, z_1) \end{cases}
\end{align*}

Similarly for $N_2$ the optimised lower bound may be expressed as

\begin{align*}
    \mathbb{E}\left[ \rho_2(x,z) \right] \geq \underset{z_0}{\text{max}} \begin{cases} &  \frac{2}{\pi H_0} \Bigg( \frac{\Gamma(0.5, \frac{z_0^2 x}{2 \delta^2})}{\Gamma(0.5, \frac{z_1^2 x}{2 \delta^2})} + \left( \frac{z_0^2 x}{2 \delta^2} \right)^{0.5-|\lambda|} \frac{\left[ \gamma(|\lambda|, \frac{z_0^2 x}{2 \delta^2}) - \gamma(|\lambda|, \frac{z_1^2 x}{2 \delta^2}) \right]}{\Gamma(0.5, \frac{z_1^2 x}{2 \delta^2})} \Bigg), \, \, z_0 \in (z_1, \infty) \\
    & \frac{2}{\pi H_0}, \quad z_0 \in (0, z_1) \end{cases}
\end{align*}

\begin{figure}[!t]
\centering
\includegraphics[width=0.7\textwidth]{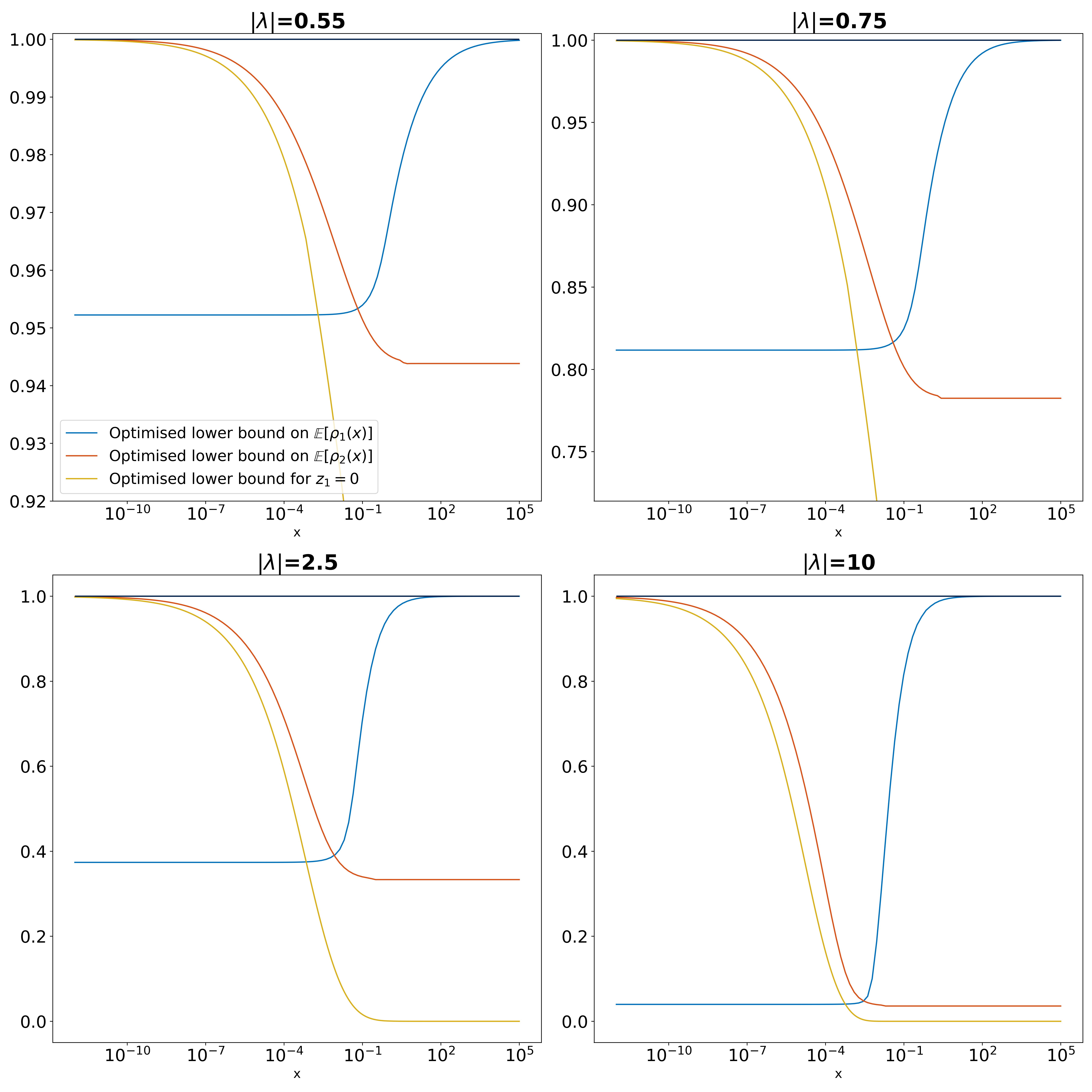}
\caption{Plot of optimised lower bounds on $\mathbb{E}\left[ \rho_1(x) \right]$ and $\mathbb{E}\left[ \rho_2(x) \right]$, for various $|\lambda| > 0.5$. $\delta = 0.1$ in all cases and bounds do not depend on $\gamma$.}
\label{fig:acceptance_rates}
\end{figure}

Note that the acceptance rates for our previous algorithm Section 3.1.1 of \cite{GodsillKindap2021} can be considered as a limiting case of the new procedure corresponding to $z_1=0$, and hence the new acceptance rates are at least as large as the previous rates for each $x>0$. In Fig. \ref{fig:acceptance_rates}, the optimised lower and upper bounds on the acceptance rates for the new procedure are shown. Additionally for comparison, the lower bounds on the previous algorithm ($z_1=0$) are plotted in Fig. \ref{fig:acceptance_rates}. This illustrates that the new procedure is a significant improvement in terms of the acceptance rate for $N_2$ (i.e. $E[\rho_2(x)]$) for small $|\lambda|$ values, and a slight improvement for larger $|\lambda|$ values.

\subsection{The case of $0<|\lambda| \leq 0.5$}
\label{sec:lambda_leq_0.5}
This parameter range was covered in  \cite{GodsillKindap2021} and we review the basics here for completeness.  We provide several improvements to this approach, including the more efficient sampling of point process $N_1$ as two gamma processes, improved bounds on the incomplete gamma functions, as well as the adaptive truncation method, residual approximation and the squeezed sampling methods detailed in subsequent sections. 

The process associated with the upper bounding L{\'e}vy density $Q_{GIG}^{B}(x,z)$ for this parameter range, see (\ref{Q_GIG_bound_B_case1}) and (\ref{Q_GIG_bound_B_case2}), can be considered once again as a marked point process split into two independent point processes $N_1$ and $N_2$ having factorised intensity functions as given in Corollary 2 of \cite{GodsillKindap2021} as

\begin{align*}
    N_1:\,\,\,\,\, &\frac{ e^{-x\gamma^2/2}}{\pi^2x^{1+|\lambda|}}\frac{(2\delta^2)^{|\lambda|}\gamma(|\lambda|,z_0^2x/(2\delta^2))}{H_0z_0^{2|\lambda|-1}} \frac{\Gamma(|\lambda|)\sqrt{\text{Ga}}
(z||\lambda|,x/(2\delta^2))} {\gamma(|\lambda|,z_0^2x/(2\delta^2))}{\cal I}_{z<z_0} \\
&=Q^B_{N_1}(x)Q^B_{N_1}(z|x)\\
    N_2:\,\,\,\,\, &\frac{ e^{-x\gamma^2/2}}{\pi^2x^{3/2}}\frac{(2\delta^2)^{0.5}\Gamma(0.5,z_0^2x/(2\delta^2))}{H_0} \frac{\Gamma(0.5)\sqrt{\text{Ga}}
(z|0.5,x/(2\delta^2))}{\Gamma(0.5,z_0^2x/(2\delta^2))}{\cal I}_{z\geq z_0}\\
&=Q^B_{N_2}(x)Q^B_{N_2}(z|x)
\end{align*}
with 
\begin{equation}
Q^B_{N_1}(x)=\frac{ e^{-x\gamma^2/2}}{\pi^2x^{1+|\lambda|}}\frac{(2\delta^2)^{|\lambda|}\gamma(|\lambda|,z_0^2x/(2\delta^2))}{H_0z_0^{2|\lambda|-1}}\label{eq:QBN1}
\end{equation}
and
\begin{equation}
Q^B_{N_2}(x)= 
\frac{ e^{-x\gamma^2/2}}{\pi^2x^{3/2}}\frac{(2\delta^2)^{0.5}\Gamma(0.5,z_0^2x/(2\delta^2))}{H_0} \label{eq:QBN2}
\end{equation}

Again, $N_1$ and $N_2$ correspond to a marginal modified tempered stable process for $x$ and a conditional truncated $\sqrt{Ga}$ density for $z$. The upper and lower incomplete gamma functions in the marginal point process envelopes $Q_{N_1}^{B}(x)$ and $Q_{N_2}^{B}(x)$ require the use of dominating processes and thinning methods similar to section \ref{sec:lambda_geq_0.5}. 

Using the bound in Eq. (\ref{eqn:lower_gamma_ineq}) the density $Q_{N_1}^{B}(x)$ can be transformed into two independent gamma processes and the methodology is summarised in Alg. \ref{alg:phase2_N1}. This simulation algorithm improves upon the method shown in Algorithm 4 of \cite{GodsillKindap2021} by transforming the problem of simulating a tempered stable process into that of simulating two gamma processes, which are known to converge rapidly in terms of the number of points required in Eq. (\ref{eqn:shot_noise_gen}). The convergence of these sums are discussed further in Section \ref{sec:adaptive_truncation}. Having simulated points $x_i$ from the marginal density $Q_{N_1}^{B}(x)$, $z_i$ values are simulated from a right-truncated square-root gamma density and the accept-reject step for each $x_i$ is performed as shown in Alg. \ref{alg:Bgen_N_1}.

\begin{algorithm}[H]
\caption{Sampling from $Q^{B}_{N_1}(x)$.}
\label{alg:phase2_N1}
\begin{enumerate}
    \item $N=\emptyset$,
    \item Generate a gamma process $N_{Ga}^{1}$ having parameters $a_1 = \frac{z_0}{\pi^2 H_0 |\lambda| (1+|\lambda|)}$ and $\beta_1=\gamma^2/2$,
    \item Generate a gamma process $N_{Ga}^{2}$ having parameters $a_2 = \frac{z_0}{\pi^2 H_0 (1+|\lambda|)}$ and $\beta_2=\gamma^2/2 + z_0^2/(2 \delta^2)$,
    \item $N = N_{Ga}^{1} \cup N_{Ga}^{2}$,
    \item For each $x \in N$ accept with probability
    \[
        \frac{(2\delta^2)^{|\lambda|} \gamma(|\lambda|, (z_0^2 x)/(2\delta^2)) |\lambda| (1+|\lambda|)}{x^{|\lambda|} z_0^{2|\lambda|} (1+|\lambda| \text{exp}\left( - z_0^2 x / (2\delta^2) \right) ) }
    \]
    \noindent otherwise reject and delete $x$ from $N$.
\end{enumerate}
\end{algorithm}

\begin{algorithm}[H]
\caption{Generation of $N_1$ for $0< |\lambda| \leq 0.5$}
\label{alg:Bgen_N_1}
\begin{enumerate}
    \item $N_1={\emptyset}$,
    \item Simulate $x_i$ from the marginal point process associated with $Q^{B}_{N_1}(x)$ as shown in Algorithm \ref{alg:phase2_N1},
    \item For each $x_i$, simulate a $z_i$ from a truncated square-root gamma density
    \[
        \frac{\Gamma(|\lambda|)\sqrt{\text{Ga}} (z||\lambda|,x_i/(2\delta^2))}{\gamma(|\lambda|,z_0^2x_i/(2\delta^2))} {\cal I}_{0<z<z_0}
    \]
    \item With probability
    \[
        \frac{Q_{GIG}(x,z)}{Q_{N_1}^{B}(x,z)} = \frac{H_0}{|H_{|\lambda|}(z_i)|^2\left(\frac{z_i^{2|\lambda|}}{z_0^{2|\lambda|-1}}\right)}
    \]
    \noindent accept $x_i$, i.e. set $N_1=N_1\cup x_i$, otherwise discard $x_i$.
\end{enumerate}
\end{algorithm}

For $Q^B_{N_2}(x)$, the incomplete gamma function is upper bounded by the complete gamma function to produce a process that is tractable for simulation and the corresponding algorithm is given in Alg. \ref{alg:phase2_N2}. For each $x_i$ value simulated from the marginal density, a corresponding $z$ value is sampled from the conditional left-truncated square-root gamma density and the whole methodology is outlined in Alg. \ref{alg:Bgen_N_2}. 

Note that the bound defined in Eq. (\ref{eqn:upper_gamma_ineq}) could also be used in this parameter regime to obtain a more efficient sampling algorithm; however, the acceptance rates using the simpler bound are found to work well. 

\begin{algorithm}[H]
\caption{Sampling from $Q^{B}_{N_2}(x)$.}
\label{alg:phase2_N2}
\begin{enumerate}
    \item Generate a tempered stable process $N_{MTS}$ with parameters $C=\frac{(2\delta^2)^{0.5} \Gamma(0.5)}{\pi^2H_0}$, $\alpha=0.5$ and $\beta=\gamma^2/2$,
    \item For each point $x\in N_{MTS}$, accept with probability $\Gamma(0.5, z_0^2 x/(2\delta^2))/\Gamma(0.5)$, otherwise reject and delete $x$ from $N_{MTS}$.
\end{enumerate}
\end{algorithm}

\begin{algorithm}[H]
\caption{Generation of $N_2$ for $0< |\lambda| \leq 0.5$}
\label{alg:Bgen_N_2}
\begin{enumerate}
    \item $N_2={\emptyset}$,
    \item Simulate $x_i$ from the marginal point process associated with $Q^{B}_{N_2}(x)$ as shown in Algorithm \ref{alg:phase2_N2},
    \item For each $x_i$, simulate a $z_i$ from a truncated square-root gamma density
    \[
        \frac{\Gamma(0.5)\sqrt{\text{Ga}} (z|0.5,x_i/(2\delta^2))}{\Gamma(0.5,z_0^2x_i/(2\delta^2))} {\cal I}_{z\geq z_0}
    \]
    \item With probability
    \[
        \frac{Q_{GIG}(x,z)}{Q_{N_2}^{B}(x,z)} = \frac{H_0}{z_i|H_{|\lambda|}(z_i)|^2}
    \]
    \noindent accept $x_i$, i.e. set $N_2=N_2\cup x_i$, otherwise discard $x_i$.
\end{enumerate}
\end{algorithm}

Lastly, the set of points $N=N_1 \cup N_2$ is a realisation of jump magnitudes corresponding to a GIG process having intensity function $Q_{GIG}(x)$ and once again the corresponding GH process may be obtained using Alg. \ref{alg:Q_GH_gen}.

\section{Adaptive truncation and Gaussian approximation of residuals}
\label{sec:adaptive_truncation}

The shot noise methods discussed in previous sections involve infinite series of decreasing random variables and in practice must be truncated after a finite number of terms. In this section we propose novel methods for adaptive determination of the number of terms required in the truncated series. This adaptive truncation can both save substantially on the computational burden of generating very long shot noise series, and also ensure (probabilistically) a specified error tolerance. Furthermore, we provide lower bounds on the mean and variance of the GIG and GH residual sequences which will be used in order to approximate the residual term as Brownian motion once the adaptive truncation procedure is terminated.

The adaptive truncation and residual approximation methods studied in this section are designed to match the moments of a realisation from a GH process at time $t$ to its theoretical moments. For a subordinator L\'{e}vy process $X(t)$ with L{\'e}vy measure $Q(dx)$ and finite first and second moments, the mean and variance of the subordinator may be expressed as \cite{Barndorff-Nielsen-Shephard}

\begin{equation}
\label{eqn:subordinator_mean}
    \mathbb{E}[X(t)] = \frac{t}{T} \int_{0}^{\infty} x Q(dx)
\end{equation}
\noindent and
\begin{equation}
\label{eqn:subordinator_var}
    \text{Var}[X(t)] = \frac{t}{T} \int_{0}^{\infty} x^{2} Q(dx)
\end{equation}

\noindent where the distribution of the process at time $T$, denoted as $X(T)$, defines the associated random variable. These integrals are intractable for the GIG case. Hence upper and lower bounds on these integrals are studied.

For normal variance-mean processes $W(t)$, such as the GH process, the associated L\'{e}vy measure can be expressed as a function of the subordinator L{\'e}vy measure $Q_{GIG}(dx)$ as in (\ref{eq:Q_GH_mv_mixture}). Hence the mean and variance of the GH process can be found in terms of the moments of a GIG process as

\begin{equation}
\label{eqn:normalvarmean_mean}
    \mathbb{E}[W(t)] = \frac{t}{T} \beta \mathbb{E}[X(T)]
\end{equation}
\noindent and
\begin{equation}
\label{eqn:normalvarmean_var}
    \text{Var}[W(t)] = \frac{t}{T} \left( \beta^2 \text{Var}[X(T)] + \sigma^{2} \mathbb{E}[X(T)] \right)
\end{equation}

\noindent where $\beta$ and $\sigma$ are the skewness and scale parameters as defined in (\ref{eqn:normal_variance_mean_mapping}).

In Section \ref{subsec:adaptive_trunc} we study adaptive truncation of infinite series for subordinator processes and provide associated algorithms for the GIG case. In principle the adaptive truncation scheme can also be described in terms of the GH moments using (\ref{eqn:normalvarmean_mean}) and (\ref{eqn:normalvarmean_var}). This transformation to the GH moments is explicitly required for the Gaussian approximation of residual moments of the GH process and lower bounds on these residuals are studied in Section \ref{subsec:lower_bounds_residual_approx}. The Gaussian approximation we use is motivated by proofs of the convergence of normal variance-mean mixture residuals to a Brownian motion in all cases of the GH process except for the normal-gamma process, and these results will be presented in a forthcoming publication.

\subsection{Adaptive truncation of shot noise series}
\label{subsec:adaptive_trunc}

The shot noise series for a subordinator $X(t)$ with its jumps truncated at $\varepsilon$ may be defined as

\begin{equation}
    X^{\varepsilon}(t) = \sum_{ \{ i : x_{i} \geq \varepsilon  \} } x_i{\cal I}_{V_i\leq t} \label{eq:Wc}
\end{equation}

\noindent The difference of $X(t)$ and the truncated series $X^{\varepsilon}(t)$ characterises the residual error caused by truncation and may be expressed as a random process $R^{\varepsilon}(t)$ such that

\begin{align}
\label{eqn:residual_error}
    R^{\varepsilon}(t) &= X(t) - X^{\varepsilon}(t) \nonumber \\
    &= \sum_{ \{ i : x_i < \varepsilon  \} } x_i {\cal I}_{V_i\leq t}
\end{align}

\noindent where $\varepsilon$ is the value at which the jump magnitude sequence $\{ x_i \}$ are stochastically truncated (i.e. the truncated series $X^\varepsilon$ has a random number of terms with $x_i$ greater than or equal to $\varepsilon$).

The statistical properties of the residual error $R^{\varepsilon}(t)$ as a function of $\varepsilon$ can be used to study the convergence of the truncated series to the L{\'e}vy process $X(t)$. The number of terms used in the approximation of $X(t)$ may be dynamically adjusted depending on the particular realisations of $\{ x_i \}$ and the required precision of approximation. Theorem \ref{thm:probabilistic_bound} below and its Corollary describes the construction of a probabilistic bound on the residual error caused by truncation of a subordinator process in terms of upper bounds on its residual moments and provide the residual mean and variance for the tempered stable and gamma processes which are used as dominating processes for sampling the GIG process.

Note that the bound in the Theorem below is a pointwise bound at a particular time $t$, whereas ideally a {\em pathwise\/} bound might be desired that applies across all times. Martingale inequalities can be used in principle to achieve this, see \cite{wolpert_2021}, although these may not be directly applicable here as we do not in general have an exact characterisation of the residual mean and variance for the GIG process, only upper and lower bounds on these. 
\begin{theorem}
\label{thm:probabilistic_bound}
    For the residual error $R^{\varepsilon}(t)$ associated with truncation of a subordinator process, the following probabilistic bound applies for any $E>\bar{\mu_{\varepsilon}}$ and truncation level $\varepsilon>0$:
    
    \begin{equation}
        \text{Pr} \left( R^{\varepsilon}(t) \geq E\right) \leq \frac{\bar{\sigma}_{\varepsilon}^2}{(E-\bar{\mu}_{\varepsilon})^2}
    \end{equation}
    
    \noindent where $\bar{\mu}_{\varepsilon}\geq\mu_{\varepsilon}$ and $\bar{\sigma}_{\varepsilon}\geq \sigma_{\varepsilon}$ are upper bounds on $\mu_{\varepsilon}=\mathbb{E}[R^{\varepsilon}(t)]$ and $\sigma_{\varepsilon}^2=\text{var}(R^{\varepsilon}(t))$, and $E$ is a threshold that may depend on the random realisation $X^{\varepsilon}(t)$. 
\end{theorem}

\begin{proof}
    The mean and variance of a subordinator process $X(t)$ are given in (\ref{eqn:subordinator_mean}) and (\ref{eqn:subordinator_var}). Similarly, the mean and variance of the truncated process residual $R^{\varepsilon}(T)$ can be found as 
    
    \begin{equation}
        \mu_{\varepsilon}=\mathbb{E}[R^{\varepsilon}(t)] = \frac{t}{T} \int_{0}^{\varepsilon} x Q(dx)\label{mu_c}
    \end{equation}
    \noindent and
    \begin{equation}
        \sigma_{\varepsilon}^2=\text{Var}(R^{\varepsilon}(t)) = \frac{t}{T} \int_{0}^{\varepsilon} x^{2} Q(dx)\label{sigma_c}
    \end{equation}
    where both of these integrals are well-defined and finite for any valid subordinator and $0<\varepsilon<\infty$, by condition (\ref{sub_cond}).
    
    Now, using the expected value and standard deviation of $R^{\varepsilon}(t)$ we may bound the residual error using concentration inequalities. Specifically, Chebyshev's inequality states that for a random variable $R^{\varepsilon}(t)$ with finite expected value $\mu_{\varepsilon}$ and finite non-zero variance $\sigma_{\varepsilon}^2$
    \[
        \text{Pr} \left( |R^{\varepsilon}(t) - \mu_{\varepsilon}| \geq k\sigma_{\varepsilon} \right) \leq \frac{1}{k^2}
    \]
    \noindent We require here only the right tail probability mass corresponding to the event $R^{\varepsilon}(t) - \mu_{\varepsilon} \geq k\sigma_{\varepsilon}$, and this is clearly less than or equal to the probability of the event $|R^{\varepsilon}(t) - \mu_{\varepsilon}| \geq k\sigma_{\varepsilon}$;  hence rearranging we arrive at
    \begin{equation*}
        \text{Pr} \left( R^{\varepsilon}(t) \geq \mu_{\varepsilon} + k \sigma_{\varepsilon} \right) \leq \frac{1}{k^2}
    \end{equation*}
    Now, if we have instead  upper bounds $\mu_{\varepsilon}\leq \bar{\mu}_{\varepsilon}$ and $\sigma_{\varepsilon}\leq \bar{\sigma}_{\varepsilon}$, it is clear that $\mu_{\varepsilon} + k \sigma_{\varepsilon}\leq \bar{\mu}_{\varepsilon} + k \bar{\sigma}_{\varepsilon} $
    and so
    \begin{equation}
        \text{Pr} \left( R^{\varepsilon}(t) \geq \bar{\mu}_{\varepsilon} + k \bar{\sigma}_{\varepsilon} \right) \leq \frac{1}{k^2} \label{eq:Pr_R_eps}
    \end{equation}
    Finally,  a simple rearrangement with $E=\bar{\mu}_{\varepsilon}+k\bar{\sigma}_{\varepsilon}$ leads to the theorem as stated.
\end{proof}









\begin{corollary}
\label{cor:upper_bounds_on_GIG}
Our simulation algorithms for the GIG process involve thinning/rejection sampling operations in order to generate point processes $N_1$ and $N_2$ from gamma and tempered stable dominating processes, see Algs. \ref{alg:phase1_N1}, \ref{alg:phase1_N2}, \ref{alg:phase2_N1} and \ref{alg:phase2_N2}. By construction, the resulting thinned processes have L\'{e}vy density $Q(x)$ strictly less than or equal to that of the dominating process in each case, say $Q^0(x)$, i.e. $Q(x)\leq Q^0(x)$. Hence the means and variances of the truncation error, calculated using (\ref{mu_c}) and (\ref{sigma_c}), are strictly less than or equal to those of the corresponding dominating gamma and tempered stable processes, and we may thus take the means and variances of the underlying TS or gamma processes as the upper bounds $\bar{\mu}_{\varepsilon}$ and $\bar{\sigma}_{\varepsilon}$ required in Theorem \ref{thm:probabilistic_bound}.  

For the TS process the expected value and variance of the residual process is found as
    
    \begin{align}
        \mu_{TS}(t) &= \frac{t}{T} \int_{0}^{\varepsilon} C x^{-\alpha} e^{-\beta x} dx \nonumber\\
        & = \frac{t C \beta^{\alpha-1}}{T} \gamma \left(1-\alpha, \beta \varepsilon \right)\label{mu_TS}
    \end{align}
    \noindent and
    \begin{align}
        \sigma^{2}_{TS}(t) &= \frac{t}{T} \int_{0}^{\varepsilon} C x^{1-\alpha} e^{-\beta x} dx \nonumber \\
        & = \frac{t C \beta^{\alpha-2}}{T}  \gamma \left(2-\alpha, \beta \varepsilon \right) \label{sigma_TS}
    \end{align}
    In limit as $\beta\rightarrow 0$ we obtain the stable subordinator whose moments can be obtained either directly or as limits of the above TS case using $
    \frac{\gamma(s,x)}{x^s} \to \frac{1}{s} \quad \text{as} \quad x \to 0
$, giving:

\begin{equation}
   \mu_{S}(t) = \frac{t C \varepsilon^{1-\alpha}}{T (1-\alpha)}
\end{equation}
\noindent and
\begin{equation}
    \sigma^{2}_{S}(t) = \frac{t C \varepsilon^{2-\alpha}}{T (2-\alpha)}
\end{equation}

    Similarly for the gamma process the expected value and variance of the residual process can be found as
    
    \begin{align}
        \mu_{Ga}(t) &= \frac{t}{T} \int_{0}^{\varepsilon} C e^{-\beta x} dx \nonumber \\
        & = \frac{tC}{T\beta} \gamma \left(1, \beta \varepsilon \right) \label{gamma_residual_mean}
    \end{align}
    \noindent and
    \begin{align}
        \sigma^{2}_{Ga}(t) &= \frac{t}{T} \int_{0}^{\varepsilon} C x e^{-\beta x} dx \nonumber \\
        & = \frac{tC}{T\beta^{2}} \gamma \left(2, \beta \varepsilon \right) \label{gamma_residual_variance}
    \end{align}

Take, for example, generation of $N_2$ in Algs. \ref{alg:phase1_N2} and \ref{gen_N_2}. The starting point is generation of a TS process with parameters $C=\frac{\delta}{\sqrt{2\pi}}$, $\alpha=0.5$ and $\beta=\frac{z_1^2}{2\delta^2} + \frac{\gamma^2}{2}$, implemented using Algorithm \ref{alg:tempered_stable}. The mean and variance for the truncated residual of  this process are obtained from (\ref{mu_TS}) and (\ref{sigma_TS}). Algs. \ref{alg:phase1_N2} and \ref{gen_N_2} then perform random thinning on the TS points. Hence the resulting process $N_2$ has truncated residual  with mean and variance no larger than those of the corresponding TS process. Thus Theorem \ref{thm:probabilistic_bound} applies, using the TS mean and variance as the upper bounds ${\bar{\mu}}_{\varepsilon}$ and ${ \bar{\sigma} }_{\varepsilon}^2$. The other cases of $N_1$ and $N_2$ simulation follow a similar argument, using the appropriate gamma or TS process to generate upper bounds on the moments required for Theorem \ref{thm:probabilistic_bound}. 

Finally, a probabilistic upper bound on the GIG residual may be obtained by adding the upper bounds on the means and variances for $N_1$ and $N_2$, since the two point processes are independent.
\end{corollary}

\begin{corollary}
\label{cor:residual_imp}
Improved residual errors and corresponding bounds on these are available if the mean $\mu_\epsilon$ is available, as then the improved estimate $\hat{X}^\epsilon(t)={X}^\epsilon(t)+\mu_\epsilon$ may be formed as proposed in \cite{Asmussen2001}. In our GH case however we only have upper and lower bounds $\underline{\mu}_\epsilon\leq\mu_\epsilon\leq\bar{\mu}_\epsilon$ established in Theorems \ref{thm:probabilistic_bound} and \ref{thm:lower_bounds_1} (see below). In this case it would seem appropriate to take a conservative line and substitute the lower bound $\underline{\mu}_\varepsilon$ in place of ${\mu}_\epsilon$. Then Theorem \ref{thm:probabilistic_bound} can be modified in step (\ref{eq:Pr_R_eps}) as follows,
\begin{equation}
        \text{Pr} \left( |R^{\varepsilon}(t)-\underline{\mu}_\varepsilon | \geq \bar{\mu}_{\varepsilon}-\underline{\mu}_\varepsilon + k \bar{\sigma}_{\varepsilon} \right) \leq \frac{1}{k^2}\label{new_equal}
\end{equation}

\noindent 
To justify this, use Chebyshev directly to give 
\[
\text{Pr} ( | R^{\varepsilon}(t)- \mu_\varepsilon | \geq k \sigma_\varepsilon ) \leq 1/k^2
\]
But we have 
\begin{align*}
A\coloneqq\{|R^{\varepsilon}(t)-\underline{\mu}_\varepsilon &| \geq \bar{\mu}_{\varepsilon}-\underline{\mu}_\varepsilon + k {\sigma}_{\varepsilon}\}\\&\subseteq\{ | R^{\varepsilon}(t)- \mu_\varepsilon | \geq k \sigma_\varepsilon \}\coloneqq B
\end{align*}
and hence $\text{Pr}(A)\leq \text{Pr} (B) $ from which (\ref{new_equal}) follows.

Rearranging this expression with $E=\bar{\mu}_{\varepsilon}-\underline{\mu}_\varepsilon + k \bar{\sigma}_{\varepsilon}$ a new expression is obtained, valid for $E+\underline{\mu}_\varepsilon-\bar{\mu}_{\varepsilon}>0$:
\begin{equation*}
        \text{Pr} \left( |R^{\varepsilon}(t)-\underline{\mu}_\varepsilon| \geq E\right) \leq \frac{\bar{\sigma}_{\varepsilon}^2}{(E+\underline{\mu}_\varepsilon-\bar{\mu}_{\varepsilon})^2 }
\end{equation*}
This expression would then be recommended for practical use in adaptive truncation schemes, yielding always smaller probabilities of exceedance than Theorem \ref{thm:probabilistic_bound} for fixed threshold $E$ since $E+\underline{\mu}_\epsilon-\bar{\mu}_{\varepsilon}\geq E-\bar{\mu}_{\varepsilon}$.
\end{corollary}

Using the probabilistic bounds given in Theorem \ref{thm:probabilistic_bound} and its Corollaries, an adaptive truncation scheme can be devised to determine a suitable value for $\varepsilon$ for {\em each\/} generated realisation of the process. As $\varepsilon$ decreases, so we accumulate sequentially the realised value of $X^{\varepsilon}(t)$ (or its mean-adjusted version from Corollary \ref{cor:residual_imp}) according to Eq. (\ref{eq:Wc}). A tolerance $E=\tau X^{\varepsilon}(t)$ is chosen, where $0 < \tau \ll 1$, which is designed to truncate the series once the predicted residual has become very small in comparison with the series realised to level $\varepsilon$. Then a probability threshold, $p_T \ll 1$  is chosen for comparison with $Pr(R^{\varepsilon}(t)\geq E)$, in order to decide when to terminate the simulation.  A generic adaptive truncation scheme is outlined in Alg. \ref{alg:adaptive_simulation} for a point process $N$ associated with a subordinator L{\'e}vy process $X(t)$ having L\'{e}vy density $Q(x)$.

\begin{algorithm}[H]
\caption{Simulation of $N$ with adaptive truncation using tolerance $\tau$ and probability threshold $p_T$ for a point process with L\'{e}vy density $Q()$ having moment bounds $\bar{\mu}_{\varepsilon_n}$ and  $\bar{\sigma}_{\varepsilon_n}$.}
\label{alg:adaptive_simulation}
\begin{enumerate}
\item Define a decreasing truncation schedule $\varepsilon_0=\infty>\varepsilon_1>\varepsilon_2>...>\varepsilon_n>...$ and initialise $E_0=0$, $N=\emptyset$. Set $n=1$, $B=\text{False}$.
\item While $B=\text{False}$,
\begin{itemize}\item If $\frac{\bar{\sigma}_{\varepsilon_n}^2}{(\tau E_n-\bar{\mu}_{\varepsilon_n})^2} > p_T$
\begin{itemize}
\item[] Simulate points $\{x_i\}$ from $Q(x){\cal I}_{x\in (\varepsilon_{n},\varepsilon_{n-1}]}$ using  Algs. \ref{alg:simulate_Q} and \ref{alg:Poisson_gen} (with $a=\varepsilon_{n}$ and $b=\varepsilon_{n-1}$),
\item[] Set $N=N\cup \{x_i\}$,
\item[] Set $E_n=E_{n-1}+\sum_i x_i$,
\end{itemize}
\item Else $B=\text{True}$.
\item $n=n+1$.
\end{itemize}
\item Return $N$. 
\end{enumerate}
\end{algorithm}

\begin{algorithm}[H]
\caption{Simulation of $N=\cup_{k=1}^K N_k$ with L\'{e}vy density $Q(x)=\sum_{k=1}^KQ_k(x)$, $x>0$, having moment bounds $\bar{\mu}^k_{\varepsilon_n}$ and  $\bar{\sigma}^k_{\varepsilon_n}$ with  adaptive truncation using tolerance $\tau$ and probability threshold $p_T$.}
\label{alg:adaptive_simulation1}
\begin{enumerate}
\item Define a decreasing truncation schedule $\varepsilon_0=\infty>\varepsilon_1>\varepsilon_2>...>\varepsilon_n>...$ and initialise $E=0$, $N_k=\emptyset$. Set $n=1$, $B_k=\text{False}$, $B=\text{False}$.
\item While $B=\text{False}$
\begin{itemize}
    \item For $k=1$ to $K$,
    \begin{enumerate}
        \item[] If $\frac{{\bar{\sigma}^{k ^{2}}}_{\varepsilon_n}}{(\tau E_n-{\bar{\mu}^k}_{\varepsilon_n})^2} > p_T$
        \begin{itemize}
            \item[] Simulate points $\{x_i\}$ from $Q_k(x){\cal I}_{x\in (\varepsilon_{n},\varepsilon_{n-1}]}$ using e.g. Algs. \ref{alg:simulate_Q} and \ref{alg:Poisson_gen} (with $a=\varepsilon_{n}$ and $b=\varepsilon_{n-1}$)
            \item[] Set $N=N\cup \{x_i\}$.
            \item[] Set $E=E+\sum_i x_i$
        \end{itemize}
        \item[] Else $B_k=\text{True}$.
    \end{enumerate}
    \item $B=\wedge_kB_k$, Logical AND operation
    \item $n=n+1$
\end{itemize}
\item Return
$N$. 
\end{enumerate}
\end{algorithm}


The GH simulation algorithms studied in this work and \cite{GodsillKindap2021} are made up of two independent point processes $N_1$ and $N_2$. An adaptive truncation algorithm such as Alg. \ref{alg:adaptive_simulation} can be applied separately to each process to obtain the resulting jumps. This misses a trick however, since either series could in principle be truncated even earlier once its residual error is very small relative to the accumulated sum of both $N_1$ and $N_2$. There are many workable schemes based around this idea and one possible such approach is presented in Alg. \ref{alg:adaptive_simulation1}. It is presented in a general form that can apply to the parallel simulation and adaptive truncation of $K$ independent subordinator point processes $N_k$ having L\'{e}vy densities $Q_k(x)$ and overall L\'{e}vy density  $Q(x)=\sum_{k=1}^KQ_k(x)$. 

For $N_1$ in the most general settings of Alg. \ref{gen_N_1} and \ref{alg:Bgen_N_1}, the dominating process is made up of two independent gamma processes $N_{Ga}^{1}$ and $N_{Ga}^{2}$. In the case of $N_2$ in both settings, Alg. \ref{gen_N_2} and \ref{alg:Bgen_N_2}, a single dominating tempered stable process is required. Hence an efficient method of simulation is running Alg. \ref{alg:adaptive_simulation1} on these 3 independent dominating processes. The residual means and variances of the tempered stable and gamma processes required by Alg. \ref{alg:adaptive_simulation} and Alg. \ref{alg:adaptive_simulation1} are shown in Corollary \ref{cor:upper_bounds_on_GIG}. It is worth noting that the convergence of a gamma process is typically significantly faster than a tempered stable process and so the $N_1$ process tends to terminate much sooner than $N_2$.

\begin{remark}
In the edge parameter setting $\lambda < 0$ and $\gamma = 0$, the marginal point process simulation methods shown in Alg. \ref{alg:phase1_N1} and \ref{alg:phase2_N1} are not valid as a result of $N_{Ga}^{1}$ becoming undefined for $\gamma = 0$. For this setting, Alg. 3 of \cite{GodsillKindap2021} may be used together with the adaptive truncation and residual approximation methods introduced in this section.

For this parameter setting the tempered stable process defined in Alg. 3 of \cite{GodsillKindap2021} becomes a stable process since the tempering parameter $\beta$ is equal to $0$. The associated residual moments of a stable process are presented in Corollary \ref{cor:upper_bounds_on_GIG} which should be used to implement Step 2) of Alg. \ref{alg:adaptive_simulation}. 
\end{remark}

\subsection{Gaussian approximation of residual errors}
\label{subsec:lower_bounds_residual_approx}

Here we present a Brownian motion approximation method for the residual error $R^{\varepsilon}(t)$ of a GIG or GH process caused by the truncation of a shot noise series as defined in Eq. (\ref{eqn:residual_error}). Such an approach is well known from previous work, see e.g. \cite{Asmussen2001}, but here we propose an intermediate solution in which a Brownian motion is injected whose drift and variance are lower bounds compared with the exact result, which is intractable in general for the GIG and GH processes.

The theoretical mean and variance of the residual error for the GH case can be found as a function of the mean and variance of an associated GIG residual error $R^{\varepsilon}(t)$ using Eqs. (\ref{eqn:normalvarmean_mean}) and (\ref{eqn:normalvarmean_var}). Hence similar to Section \ref{subsec:adaptive_trunc}, we provide lower bounds $\underline{\mu_{\varepsilon}}$, $\underline{\sigma_{\varepsilon}^2}$ on the mean and variance of a residual GIG process as a function of the truncation level $\varepsilon$. Together with the upper bounds discussed in Corollary \ref{cor:upper_bounds_on_GIG}, these lower bounds characterise the residual error of truncating the infinite shot noise series for the GIG and GH cases. Using the residual approximation module the series representation of the GIG L\'{e}vy process $X(t)$ can be expressed as
\begin{equation*}
    X(t) \approx \frac{t \underline{\mu_{\varepsilon}}}{T}  + \frac{\underline{\sigma_{\varepsilon}}}{\sqrt{T}} \mathcal{B}(t) + X^{\varepsilon}(t)
\end{equation*}
where $X^{\varepsilon}(t)$ is computed in the usual way as $\sum_{ \{ i : x_{i} \geq \varepsilon  \} } x_i{\cal I}_{V_i\leq t}$, $\mathcal{B}(t)$ is an independent standard Brownian motion term and $\underline{\mu_{\varepsilon}}$, $\underline{\sigma_{\varepsilon}^2}$ are lower bounds on the mean and variance of the residual error $R^{\varepsilon}(T)$ given a truncation level $\varepsilon$.

According to Eqs. (\ref{eqn:normalvarmean_mean}) and (\ref{eqn:normalvarmean_var}), the lower bounds on the moments of the residual error $R_{W}^{\varepsilon}(t)$ of the associated GH process $W(t)$ can be obtained as
\begin{equation}
\label{eqn:gh_residual_mean}
    \mathbb{E}[R_{W}^{\varepsilon}(t)] \geq \frac{t}{T} \beta \underline{\mu_{\varepsilon}}
\end{equation}
\begin{equation}
\label{eqn:gh_residual_variance}
    \text{Var}[R_{W}^{\varepsilon}(t)] \geq \frac{t}{T} \left( \beta^2 \underline{\sigma_{\varepsilon}^2} + \sigma^{2} \underline{\mu_{\varepsilon}} \right)
\end{equation}

\noindent Hence for the GH process the same procedure is adopted to obtain an approximation as
\begin{equation}
    W(t) \approx \frac{t}{T} \beta \underline{\mu_{\varepsilon}} + \frac{\sqrt{ \beta^2 \underline{\sigma_{\varepsilon}^2} + \sigma^{2} \underline{\mu_{\varepsilon}} }}{\sqrt{T}}  \mathcal{B}(t) + W^{\varepsilon}(t) \label{eq:residual_series_repr}
\end{equation}

The approximation of the residual error is in addition to the adaptive truncation methods described in Section \ref{subsec:adaptive_trunc} that provide a specified level of truncation $\varepsilon$ for the simulation of sample paths from a GH process. In Theorem \ref{thm:lower_bounds_1} below we provide the required lower bounds on the mean and variance of a residual GIG process with L\'{e}vy density $Q_{GIG}(x)$ as a function of $\varepsilon$. The derivation of these bounds are presented for a specific $T$ as it is straightforward to scale the moments according to time. The lower bounds are then used to evaluate the mean and variance of the residual error in the GH process simulation and approximate the contribution of the residual small jumps according to Eq. (\ref{eq:residual_series_repr}). A Brownian motion approximation to the residual is known to hold for many shot noise series, see \cite{Asmussen2001}, with the gamma process being a well-known exception that does not converge to a Gaussian as $\varepsilon\rightarrow 0$. In our own work we have proven convergence of the shot noise series for the GH process to a Brownian motion in all cases except the normal-gamma, and these results will be presented in a future publication.

\begin{theorem}
\label{thm:lower_bounds_1}
Given a truncation level $\varepsilon$, a residual sequence $R^{\varepsilon}(T)$ of GIG jumps with mean $\mu_{\varepsilon}=E[R^{\varepsilon}(T)]$ and variance $\sigma_{\varepsilon}^2=\text{Var}[R^{\varepsilon}(T)]$ may be lower bounded by:
\begin{equation*}
    \underline{\mu_{\varepsilon}} \leq \mu_{\varepsilon}
\end{equation*}
\begin{equation*}
   \underline{\sigma_{\varepsilon}^2} \leq {\sigma_{\varepsilon}^2}
\end{equation*}
\noindent where the bounds are defined as:

\begin{subnumcases}{\nonumber \underline{\mu_{\varepsilon}} =}
     \frac{C_{Ga}^{B} \gamma \left( 1, \beta_{Ga}^{B} \varepsilon \right)}{\beta_{Ga}^{B}} + \frac{C_{TS}^{B} \gamma \left( 0.5, \beta_{TS}^{B} \varepsilon \right)}{{\beta_{TS}^{B}}^{0.5}}, \,\,\,\, |\lambda| \geq 0.5 \nonumber\\ 
     \frac{C_{Ga}^{A} \gamma \left( 1, \beta_{Ga}^{A} \varepsilon \right)}{\beta_{Ga}^{A}} + \frac{C_{TS}^{A} \gamma \left( 0.5, \beta_{TS}^{A} \varepsilon \right)}{{\beta_{TS}^{A}}^{0.5}}, \,\,\,\, |\lambda| < 0.5 \nonumber
\end{subnumcases}

\begin{subnumcases}{\nonumber \underline{\sigma_{\varepsilon}^2} =}
     \frac{C_{Ga}^{B} \gamma \left( 2, \beta_{Ga}^{B} \varepsilon \right)}{{\beta_{Ga}^{B}}^2} + \frac{ C_{TS}^{B} \gamma \left( 1.5, \beta_{TS}^{B} \varepsilon \right) }{{\beta_{TS}^{B}}^{1.5}}, \,\,\,\, |\lambda| \geq 0.5 \nonumber\\ 
     \frac{C_{Ga}^{A} \gamma \left( 2, \beta_{Ga}^{A} \varepsilon \right)}{{\beta_{Ga}^{A}}^2} + \frac{ C_{TS}^{A} \gamma \left( 1.5, \beta_{TS}^{A} \varepsilon \right) }{{\beta_{TS}^{A}}^{1.5}}, \,\,\,\, |\lambda| < 0.5 \nonumber
\end{subnumcases}

\noindent where

\[
C_{Ga}^{A} = \frac{z_{1}}{2 \pi|\lambda|} \quad \text{and} \quad \beta_{Ga}^{A} = \frac{\gamma^{2}}{2} + \frac{|\lambda|}{(1+|\lambda|)} \frac{z_{1}^{2} }{2 \delta^{2}}
\]

\[
C_{Ga}^{B} = \frac{z_{0}}{\pi^{2} H_{0} |\lambda|} \quad \text{and} \quad \beta_{Ga}^{B} = \frac{\gamma^{2}}{2} + \frac{|\lambda|}{(1+|\lambda|)} \frac{z_{0}^{2} }{2 \delta^{2}}
\]

\[
C_{TS}^{A} = \frac{\delta \sqrt{e} \sqrt{\beta_0-1}}{\pi \beta_0} \quad \text{and} \quad \beta_{TS}^{A} = \frac{\gamma^2}{2} + \frac{\beta_{0} z_{1}^{2} }{2 \delta^2}
\]

\[
C_{TS}^{B} = \frac{2 \delta \sqrt{e} \sqrt{\beta_0-1}}{\pi^2 H_0 \beta_0} \quad \text{and} \quad \beta_{TS}^{B} = \frac{\gamma^2}{2} + \frac{\beta_{0} z_{0}^{2} }{2 \delta^2}
\]

\noindent and $\beta_0$ is a free parameter such that $\beta_0 > 1$.
\end{theorem}

Note that the lower bounds presented in Theorem \ref{thm:lower_bounds_1} are valid only for the $\lambda < 0$ setting. The additional mean and variance components present for $\lambda > 0$ are associated with a gamma process with L\'{e}vy density (\ref{gamma_process_positive_lambda}) and residual moments given by Eqs. (\ref{gamma_residual_mean}) and (\ref{gamma_residual_variance}). These terms are simply added to the lower bounds $\underline{\mu_{\varepsilon}}$ and $\underline{\sigma_{\varepsilon}^2}$ in order to obtain the final bounds on the GIG process.

\begin{remark}
For the special case of $|\lambda| = 0.5$, the GIG density $Q_{GIG}(x)$ has a functional form equivalent to a TS process with intensity function

\begin{equation*}
    \frac{e^{-x\gamma^2/2}}{x^{3/2}}\frac{\delta\Gamma(1/2)}{\sqrt{2}\pi}
\end{equation*}

\noindent Thus both the simulation algorithms in Section \ref{sec:simulation_algorithms}, and the adaptive truncation and residual approximation methods presented in this Section are significantly simplified.  It is straightforward to simulate a TS process using Alg. \ref{alg:tempered_stable} and the residual moments shown in Eqs. (\ref{mu_TS}) and (\ref{sigma_TS}) provide exact expressions for the residual moments of the equivalent GIG process. 

\noindent The true mean and variance of the residual moments replace the upper and lower bounds required for determining the adaptive truncation level $\varepsilon$ and the associated moments of the approximating Brownian term. In this case Eq. (\ref{eq:residual_series_repr}) exactly matches the first and second moments of the residual process.
\end{remark}

\section{Squeezed rejection sampling}
\label{sec:squeezing_functions}

In this section, we provide a practical extension to the sampling algorithms discussed in Section \ref{sec:simulation_algorithms} that is designed to increase the efficiency of simulation. The above methods for sampling of $N_1$ and $N_2$ in both parameter settings, given in Algs.
\ref{gen_N_1},\ref{gen_N_2} and \ref{alg:Bgen_N_1},\ref{alg:Bgen_N_2}, involve a computationally expensive pair of steps in the sampling of a truncated gamma random variate (Step 3) and a pointwise evaluation of the Hankel function (Step 4). Notice however, that Theorem \ref{theorem:1} provides both lower {\em and\/} upper bounds on the term $z|H_\nu(z)|^2$, which allows us to specify {\em squeezing\/} functions on $Q_{GIG}(x,z)$,\footnote{see \cite{devroye:1986} Section 3.6 for application of the Squeeze principle in the setting of standard random variate generation} as given in (\ref{Q_GIG_bound_A_case1})-(\ref{Q_GIG_bound_B_case2}). This allows for a labour-saving {\em retrospective sampling\/} procedure in which, for a fixed fraction of points $x_i$, we may replace the simulation of a conditional random variable $z$ and rejection sampling based on its value (Steps 3 and 4) with a simple 1-step accept/reject and no requirement to sample $z$ or evaluate $H_{|\lambda|}$. Considering first the case $|\lambda|\geq 0.5$ and the sampling of process $N_1$, see Alg. \ref{gen_N_1}, we have generated at Step 2 a single point realisation $x_i$ from the process $Q^A_{N_1}(x)$. Now consider Step 4, which accepts $x_i$ with probability

\[
    \frac{Q_{GIG}(x_i,z_i)}{Q^A_{N_1}(x_i, z_i)} \geq \frac{Q^B_{GIG}(x_i,z_i)}{Q^A_{GIG}(x_i, z_i)} = \frac{2}{\pi H_0}
\]
where we have used the squeezing inequality (\ref{eq:bound_geq_05}) and where the final equality applies when we set $z_0=z_1$, see Remark
\ref{rem:piH_0/2}. This implies that we may carry out a retrospective sampling step:  draw a uniform random variate $W_i$ on $[0,1]$, test whether it is less than or equal to $\frac{2}{\pi H_0}$, and if so, accept $x_i$ with no Steps 3 and 4 required. If $W_i>\frac{2}{\pi H_0}$, carry out steps 3 and 4, using the same realised $W_i$ to carry out the test in Step 4\footnote{The method is termed `retrospective' because we only need to generate $z$ and test the exact acceptance probability following a much simpler initial accept/reject step}. The modified version of Alg. \ref{gen_N_1} using squeezed rejection sampling is presented in Alg. \ref{gen_N_1_squeeze}.

\begin{algorithm}[H]
\caption{Squeezed generation of $N_1$ for $|\lambda| \geq 0.5$}
\label{gen_N_1_squeeze}
\begin{enumerate}
    \item $N_1={\emptyset}$,
    \item Simulate $x_i$ from the marginal point process associated with $Q^{A}_{N_1}(x)$ as given in Algorithm \ref{alg:phase1_N1},
    \item Generate $w_i\sim {\cal U}[0,1]$,
    \item If $w_i\leq \frac{2}{\pi H_0}$, accept $x_i$, i.e. set $N_1=N_1\cup x_i$,
    \item[] Else,
                \begin{enumerate}
                    \item For each rejected $x_i$, simulate a $z_i$ from a truncated square-root gamma density
                    \[
                        \frac{\Gamma(|\lambda|)\sqrt{\text{Ga}} (z||\lambda|,x_i/(2\delta^2))}{\gamma(|\lambda|,z_1^2x_i/(2\delta^2))} {\cal I}_{0<z<z_1}
                    \]
                    \item If
                    \[
                        w_i\leq \frac{Q_{GIG}(x_i,z_i)}{Q^A_{N_1}(x_i, z_i)} = \frac{2}{\pi |H_{|\lambda|}(z_i)|^2 \left( \frac{z_i^{2|\lambda|}}{z_1^{2|\lambda|-1}} \right)}
                    \]
                    \noindent accept $x_i$, i.e. set $N_1=N_1\cup x_i$; otherwise discard $x_i$.
                \end{enumerate}
\end{enumerate} 
\end{algorithm}

An exactly similar modification applies for generation of $N_2$ in Alg. \ref{gen_N_2} for $|\lambda|>0.5$. In the other parameter range $|\lambda|<0.5$, the bounds are reversed, see (\ref{eq:bound_leq_05}) and so Step 4 in the squeezed sampler is replaced with `if $w_i\leq \frac{\pi H_0}{2}$'; otherwise squeezed procedures for $N_1$ and $N_2$ in this parameter range are modified exactly as in Alg. \ref{gen_N_1_squeeze}.

\begin{figure}[!t]
\centering
\includegraphics[width=0.7\textwidth]{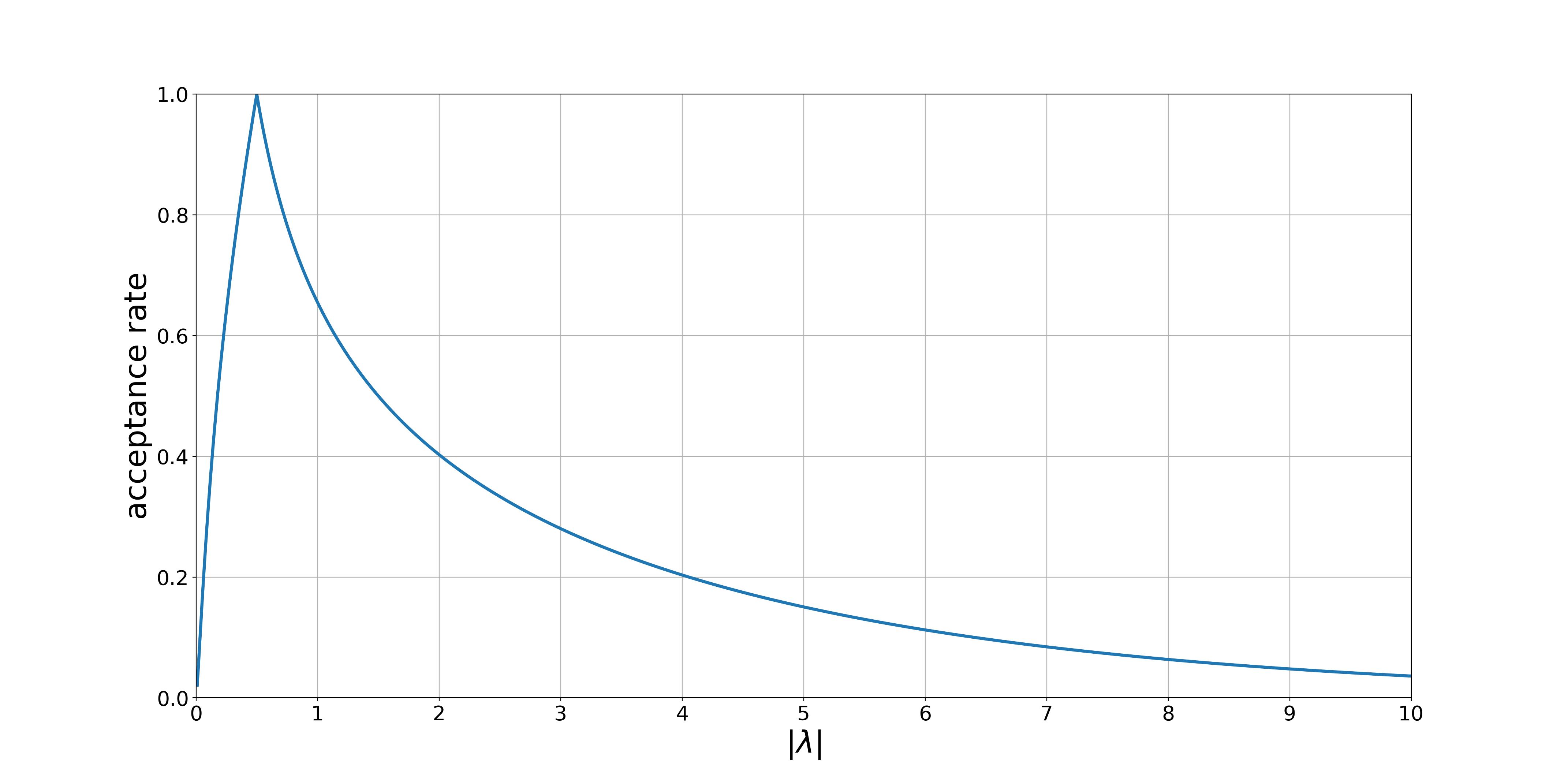}
\caption{Fractional computational saving for the retrospective sampling procedure, as a function of $|\lambda|$.}
\label{fig:squeezed_acceptance_rates}
\end{figure}

We can see that the savings arising from this method could be substantial in cases where $\frac{2}{\pi H_0}$ is close to unity, saving almost all of the heavy computations in the original Steps 3 and 4. This will occur when $|\lambda|$ is close to 0.5, and improvements will lessen as $|\lambda|$ moves away from this value. The fraction of saved computation is shown as a function of $|\lambda|$ are shown in Fig. \ref{fig:squeezed_acceptance_rates}, exhibiting a broad range of $\lambda$ values for which savings are useful.

\section{Simulations}
\label{sec:simulations}

In this section realised sample paths from the simulation algorithms in Section \ref{sec:simulation_algorithms} together with the modifications discussed in Section \ref{sec:adaptive_truncation} are presented. The distribution of the realised GH processes generated up to $T=1$ is compared against samples from the GH distribution which are generated using exact GIG random variable samplers (\cite{Devroye_2014,Statovic_2017}). While these methods are able to generate samples for a specific $t=T$, our method is able to generate the entire path of the process up to $t=T$ and hence information about the dynamics within the interval $(0,T)$ is made available. Furthermore, the accuracy and efficiency of the proposed simulation algorithm and modifications are compared with the method presented in \cite{GodsillKindap2021}.

Since there is no known closed-form cumulative distribution function (CDF) for the GH distribution, the accuracy of the simulated sample paths are measured using the two-sample Kolmogorov–Smirnov (KS) test which compares the empirical distribution functions of the sample paths at $T=1$ and samples from the GH distribution. These tests involve $10^6$ independent samples from each method and the results are shown in Tables \ref{table:ks_experiments1}, \ref{table:ks_experiments2} and \ref{table:ks_experiments3} for various parameter settings. The probability threshold $p_T=0.05$ for the adaptive truncation algorithm is found to work well and is used for all cases throughout the section. In Table \ref{table:ks_experiments1} the results are shown for the simulation procedure using only adaptive determination of the number of jumps and in Table \ref{table:ks_experiments2} the improvements in convergence by approximating the residual small jumps, as studied in \ref{subsec:lower_bounds_residual_approx}, are shown.

\begin{table}[htbp]
\caption{The results of two-sample KS tests for adaptive simulation algorithm.}
\begin{center}
\begin{tabular}{|p{1cm}|p{3cm}|p{3cm}|p{1cm}|}
\hline
$\mathbf{\lambda}$ & \textbf{KS Statistic} & \textbf{Time / sample} & $\mathbf{\tau}$ \\
\hline
\setlength{\arrayrulewidth}{1pt}
$-0.4$  & $0.00301$  & $8.27 \times 10^{-4}$ & 0.01  \\
\Xhline{0.5pt}  
 $-0.4$  & $0.00442$ & $7.77 \times 10^{-4}$ & 0.1 \\
 \Xhline{0.5pt} 
 $-0.8$  & $0.00305$  & $1.34 \times 10^{-3}$ & 0.01 \\
 \Xhline{0.5pt} 
 $-0.8$  & $0.00578$  & $9.87 \times 10^{-4}$ & 0.1 \\
 \Xhline{0.5pt} 
 $-2.5$  & $0.00369$ & $3.20 \times 10^{-3}$ & 0.01 \\
 \Xhline{0.5pt} 
 $-2.5$  & $0.00976$ & $1.27 \times 10^{-3}$ & 0.1 \\
 \Xhline{0.5pt} 
 $-10$   & $0.00485$ & $9.28 \times 10^{-3}$ & 0.01 \\
 \Xhline{0.5pt} 
 $-10$   & $0.01038$& $2.01 \times 10^{-3}$ & 0.1 \\
 \hline
\end{tabular}
\label{table:ks_experiments1}
\end{center}
\end{table}

\begin{table}[htbp]
\caption{The results of two-sample KS tests for adaptive simulation with residual approximation algorithm.}
\begin{center}
\begin{tabular}{|p{1cm}|p{3cm}|p{3cm}|p{1cm}|}
\hline
$\mathbf{\lambda}$ & \textbf{KS Statistic} & \textbf{Time / sample} & $\mathbf{\tau}$ \\
\hline
$-0.4$  & $0.00178$  & $7.98 \times 10^{-4}$ & 0.01  \\
\Xhline{0.5pt}  
 $-0.4$  & $0.00246$ & $6.57 \times 10^{-4}$ & 0.1 \\
 \Xhline{0.5pt} 
 $-0.8$  & $0.00269$  & $9.34 \times 10^{-4}$ & 0.01 \\
 \Xhline{0.5pt} 
 $-0.8$  & $0.00179$  & $7.89 \times 10^{-4}$ & 0.1 \\
 \Xhline{0.5pt} 
 $-2.5$  & $0.00325$ & $2.22 \times 10^{-3}$ & 0.01 \\
 \Xhline{0.5pt} 
 $-2.5$  & $0.00487$ & $9.71 \times 10^{-4}$ & 0.1 \\
 \Xhline{0.5pt} 
 $-10$   & $0.00276$ & $1.09 \times 10^{-2}$ & 0.01 \\
 \Xhline{0.5pt} 
 $-10$   & $0.00835$& $1.72 \times 10^{-3}$ & 0.1 \\
 \hline
\end{tabular}
\label{table:ks_experiments2}
\end{center}
\end{table}

There is a clear trade-off between the accuracy of the distribution at $T=1$ and the time required per sample for a given $\lambda$ value. This can be observed by comparing the KS statistic and time per sample for different tolerance parameters $\tau$, e.g. in Alg. \ref{alg:adaptive_simulation1}. A large tolerance results in worse convergence but allows faster simulation, and hence the tolerance parameter may be adjusted depending on the requirements of the application. Furthermore as $|\lambda|$ increases both the time required per sample and the KS statistic increases as a result of the reduced acceptance rates of the simulation algorithm as plotted in Fig. \ref{fig:acceptance_rates}. 

Comparing Tables \ref{table:ks_experiments1} and \ref{table:ks_experiments2}, it can be seen that there is no significant increase in time per sample when applying residual approximation methods and there is an improvement in the KS statistic for all parameter settings. Particularly, the results suggest that the improvement in large tolerance parameter cases where $\tau=0.1$ are substantial.

\begin{table}[htbp]
\caption{The results of two-sample KS tests for the simulation algorithm in \cite{GodsillKindap2021}.}
\begin{center}
\begin{tabular}{|p{1cm}|p{3cm}|p{3cm}|}
\hline
$\mathbf{\lambda}$ & \textbf{KS Statistic} & \textbf{Time / sample} \\
\hline
$-0.4$  & $0.00292$  & $1.72 \times 10^{-2}$   \\
\Xhline{0.5pt}  
 $-0.8$  & $0.00435$  & $8.37 \times 10^{-3}$  \\
 \Xhline{0.5pt} 
 $-2.5$  & $0.00494$ & $1.09 \times 10^{-2}$  \\
 \Xhline{0.5pt} 
 $-10$   & $0.00272$ & $2.13 \times 10^{-2}$ \\
 \hline
\end{tabular}
\label{table:ks_experiments3}
\end{center}
\end{table}

In order to compare our new methods with the method studied in \cite{GodsillKindap2021}, the adaptive truncation algorithm in Alg. \ref{alg:adaptive_simulation1} is limited to producing a certain maximum number, $M=10^{4}$, of jump magnitudes per sample path while running the experiments for Tables \ref{table:ks_experiments1} and \ref{table:ks_experiments2}. The KS statistics and the time required per sample for the same parameter settings and a fixed $M$ number of jumps from the method in \cite{GodsillKindap2021} are reviewed in Table \ref{table:ks_experiments3}. Comparing these results to Table \ref{table:ks_experiments2}, it can be seen that improved convergence results can be obtained by the novel algorithms introduced in this work while also providing significantly reduced time complexity.

To present a more intuitive understanding of the accuracy of our methods, QQ (quantile-quantile) plots of sample paths generated up to $T=1$ and samples from a random variable generator are shown in addition to histograms with the true probability density function overlaid. The parameter values in the examples are selected to reflect the different characteristic behaviour of the GH process as well as edge cases such as the normal-inverse Gaussian process $(\lambda = -1/2)$, and the Student-t process $(\gamma = 0$, $\lambda \leq 0)$. 

\begin{figure}[!h]
\centering
\includegraphics[width=0.7\textwidth]{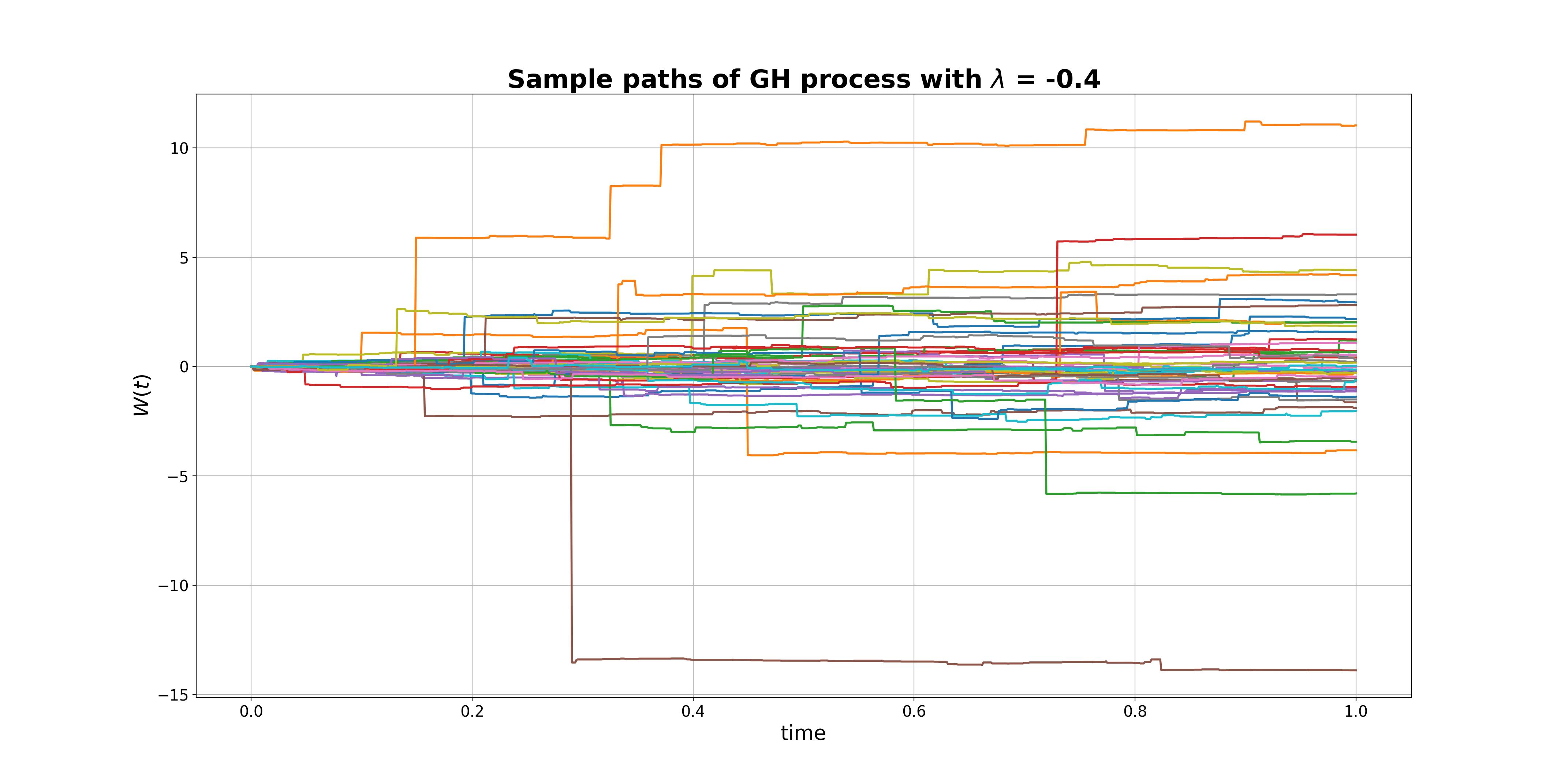}
\caption{Pathwise simulations of the GH process for $\lambda=-0.4$, $\delta=1.0$, $\gamma=0.1$ and $\beta=0$}
\label{fig:sample_path3}
\end{figure}

\begin{figure}[!h]
\centering
\includegraphics[width=0.7\textwidth]{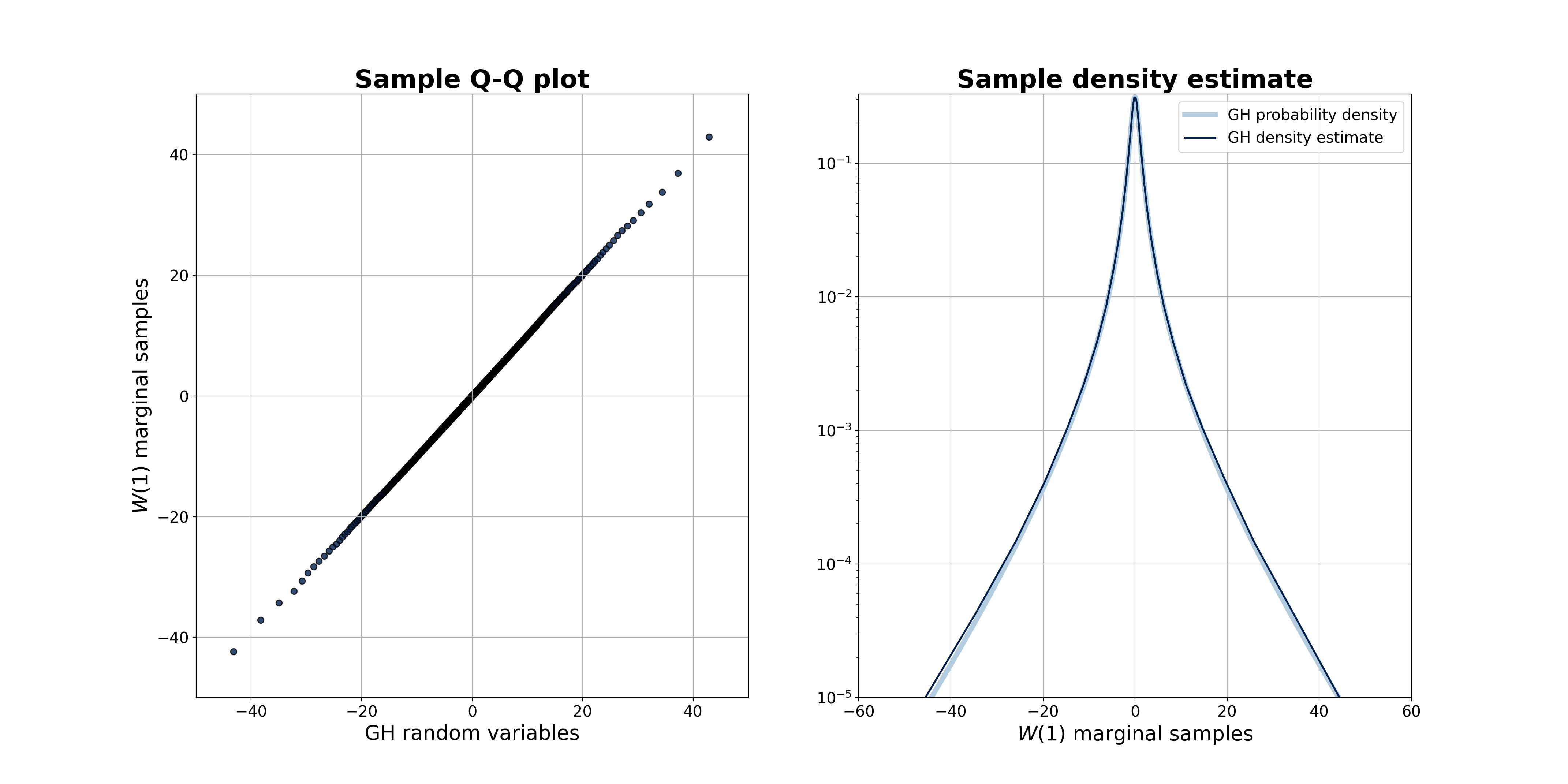}
\caption{Simulation comparison between the shot noise generated GH process and GH random variates, $\lambda=-0.4$, $\gamma=0.1$, $\delta=1$, $\beta=0$. The adaptive truncation parameters are $p_T = 0.05$ and $\tau=0.01$. Left hand panel: QQ plot comparing our shot noise method (y-axis) with random samples of the GH density generated using a random variate generator (x-axis). Right hand panel: Normalised histogram density estimate for our method compared with the true GH density function.}
\label{fig:sample_test6}
\end{figure}

\begin{figure}[!h]
\centering
\includegraphics[width=0.7\textwidth]{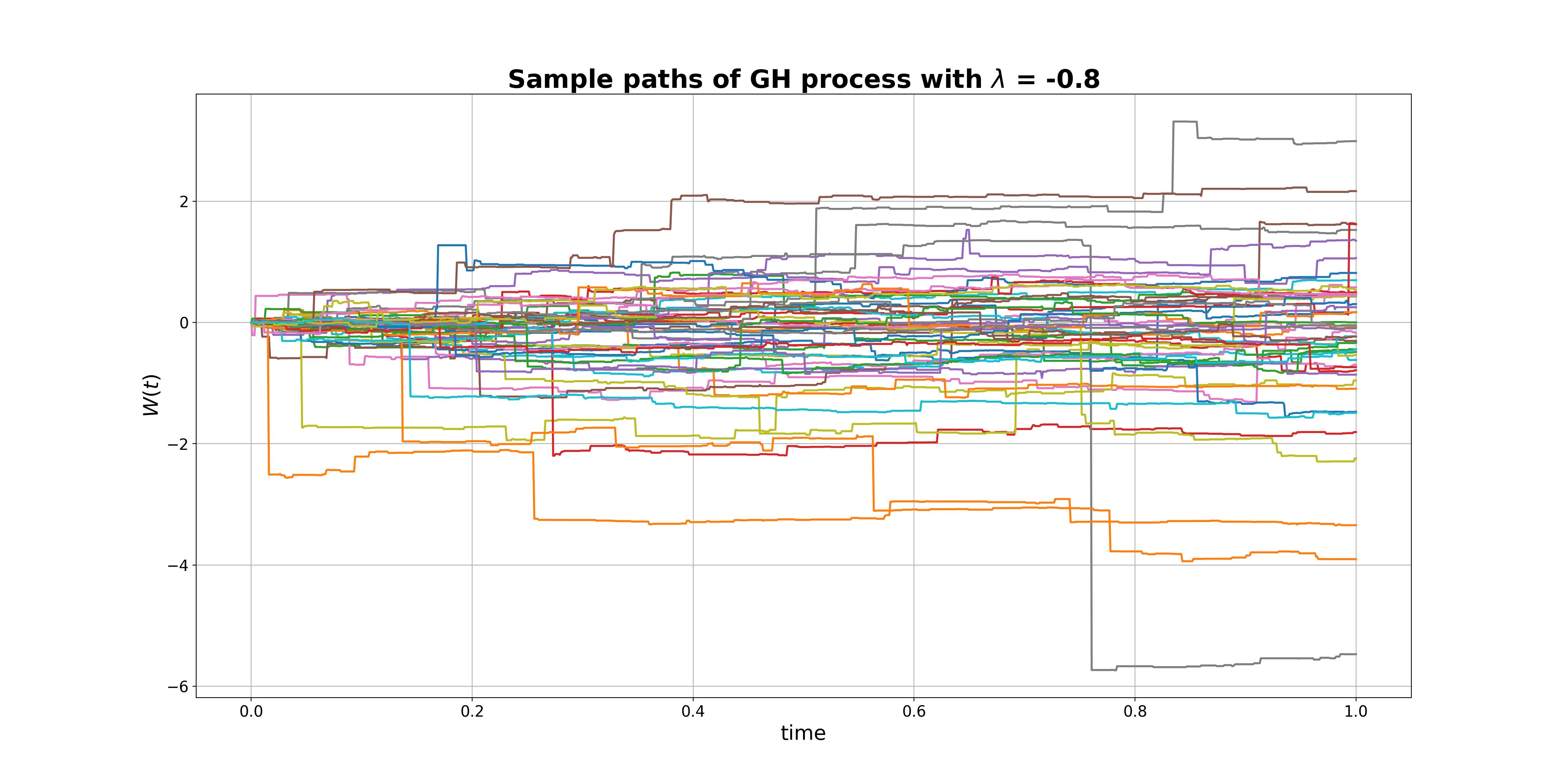}
\caption{Pathwise simulations of the GH process for $\lambda=-0.8$, $\delta=1.0$, $\gamma=0.1$ and $\beta=0$.}
\label{fig:sample_path2}
\end{figure}

\begin{figure}[!h]
\centering
\includegraphics[width=0.7\textwidth]{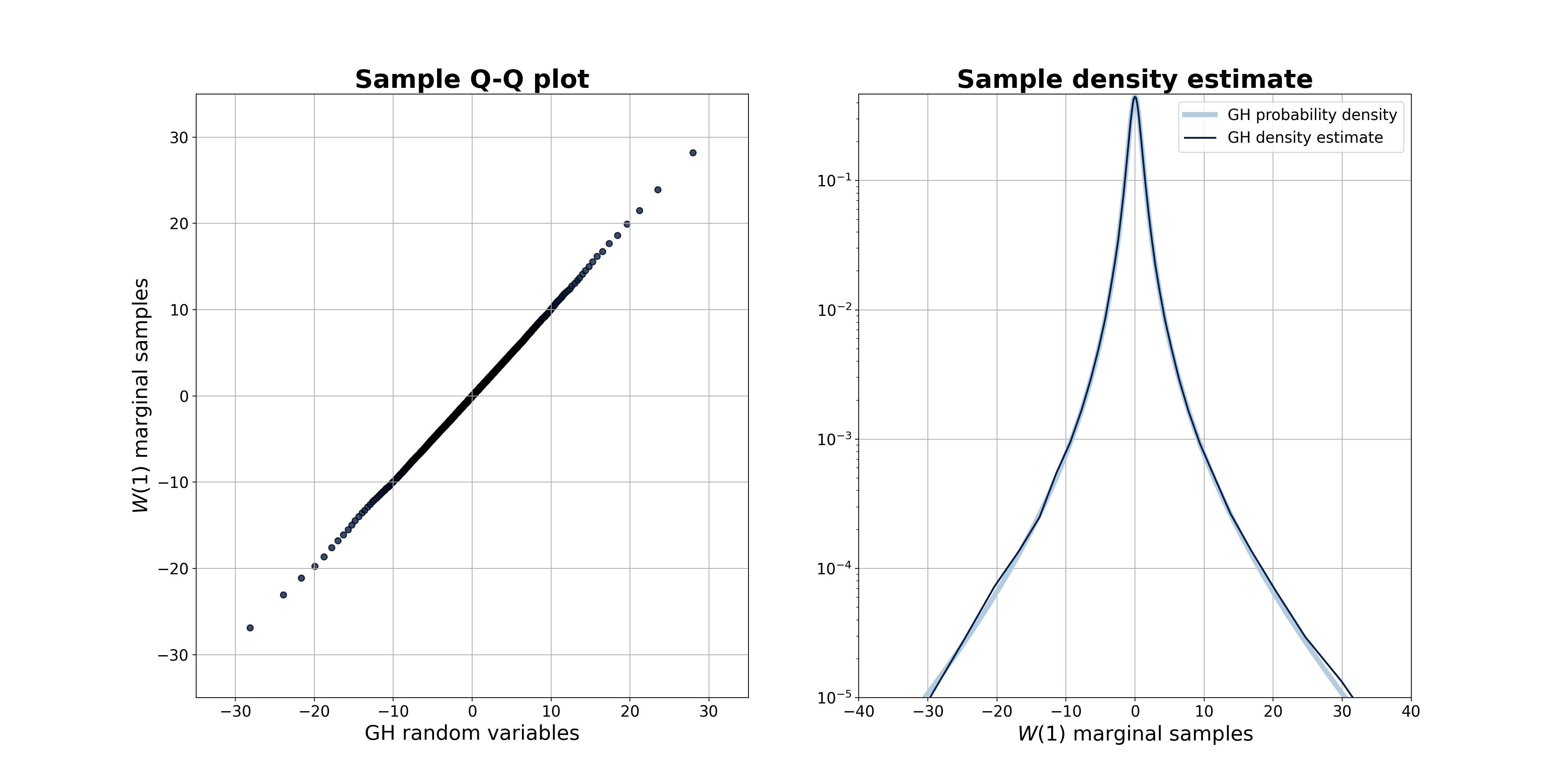}
\caption{Simulation comparison between the shot noise generated GH process and GH random variates, $\lambda=-0.8$, $\gamma=0.1$, $\delta=1$, $\beta=0$. The adaptive truncation parameters are $p_T = 0.05$ and $\tau=0.01$. Left hand panel: QQ plot comparing our shot noise method (y-axis) with random samples of the GH density generated using a random variate generator (x-axis). Right hand panel: Normalised histogram density estimate for our method compared with the true GH density function.}
\label{fig:sample_test3}
\end{figure}

\begin{figure}[!h]
\centering
\includegraphics[width=0.7\textwidth]{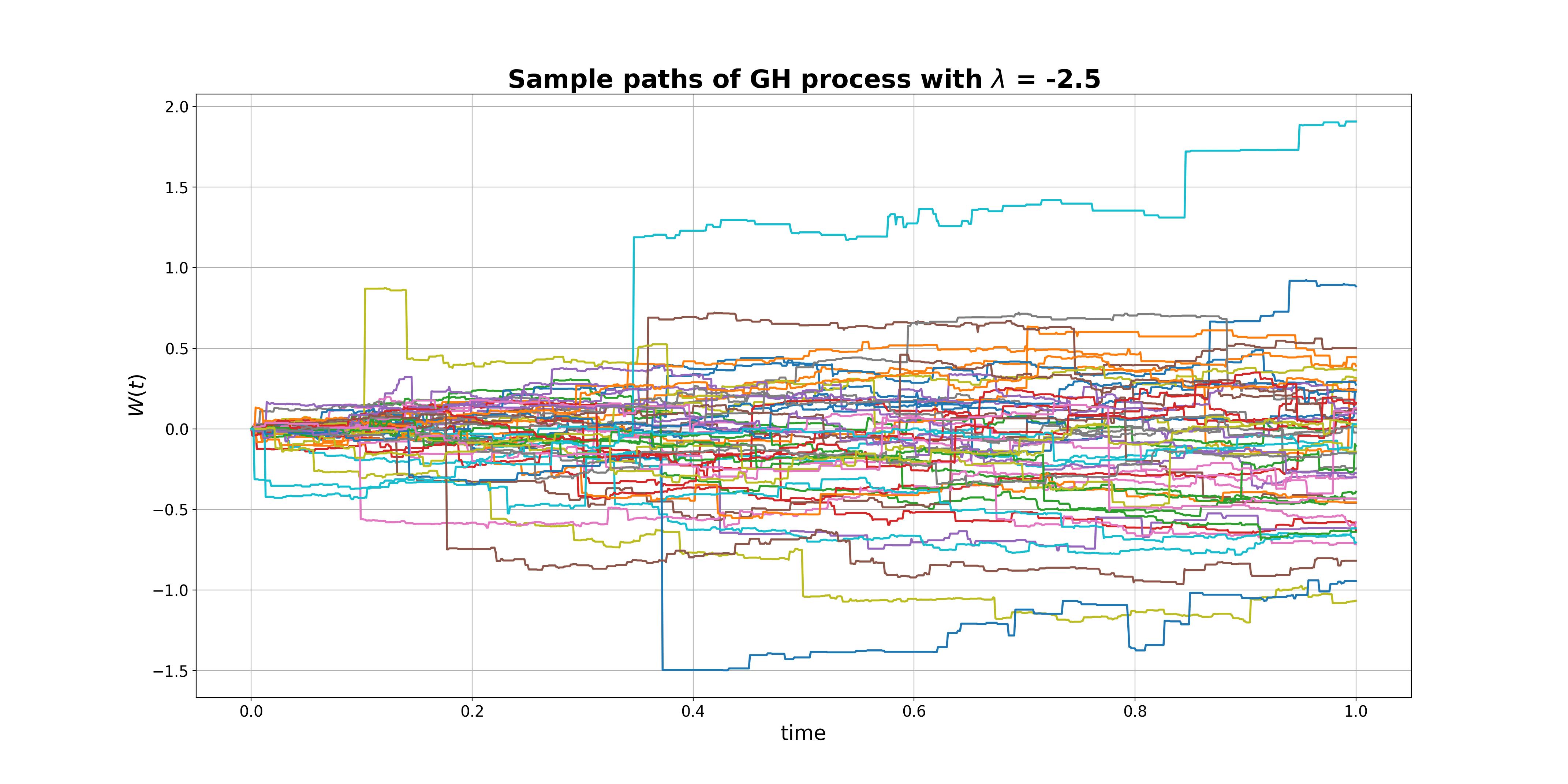}
\caption{Pathwise simulations of the GH process for $\lambda=-2.5$, $\delta=1.0$, $\gamma=0.1$ and $\beta=0$}
\label{fig:sample_path4}
\end{figure}

\begin{figure}[!h]
\centering
\includegraphics[width=0.7\textwidth]{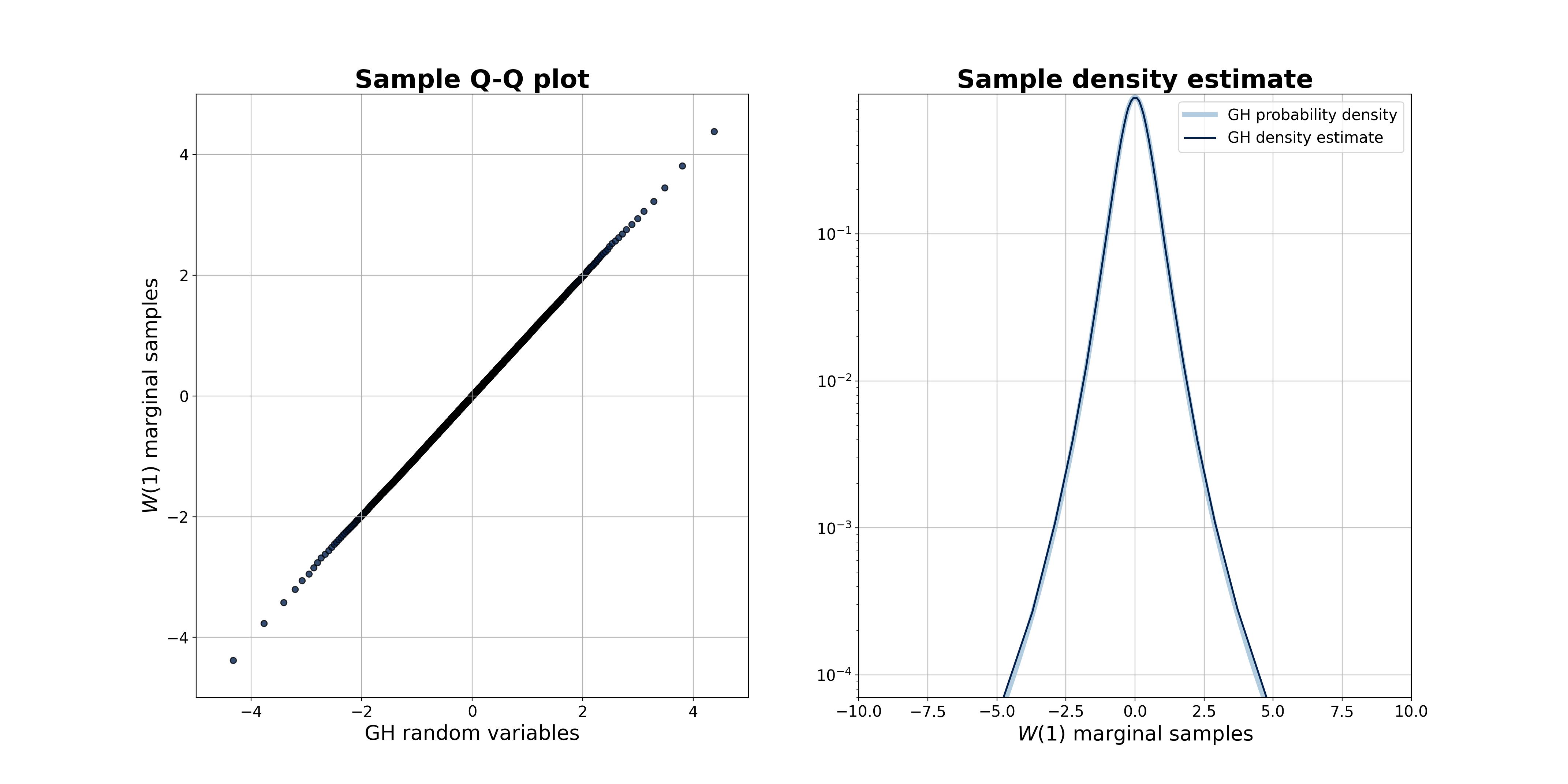}
\caption{Simulation comparison between the shot noise generated GH process and GH random variates, $\lambda=-2.5$, $\gamma=0.1$, $\delta=1$, $\beta=0$. The adaptive truncation parameters are $p_T = 0.05$ and $\tau=0.1$. Left hand panel: QQ plot comparing our shot noise method (y-axis) with random samples of the GH density generated using a random variate generator (x-axis). Right hand panel: Normalised histogram density estimate for our method compared with the true GH density function.}
\label{fig:sample_test1}
\end{figure}

\begin{figure}[!h]
\centering
\includegraphics[width=0.7\textwidth]{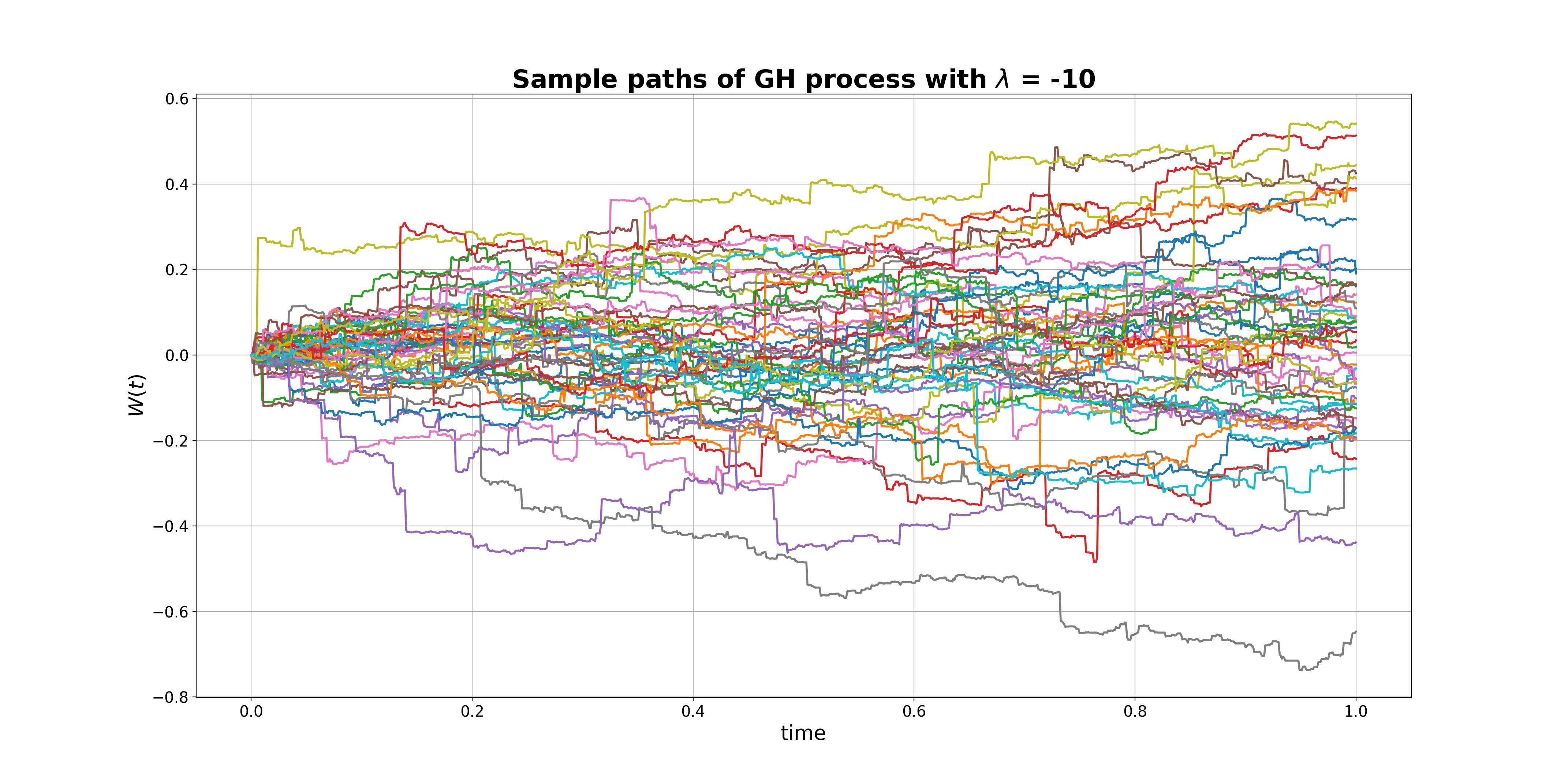}
\caption{Pathwise simulations of the GH process for $\lambda=-10$, $\delta=1.0$, $\gamma=0.1$ and $\beta=0$.}
\label{fig:sample_path5}
\end{figure}

\begin{figure}[!h]
\centering
\includegraphics[width=0.7\textwidth]{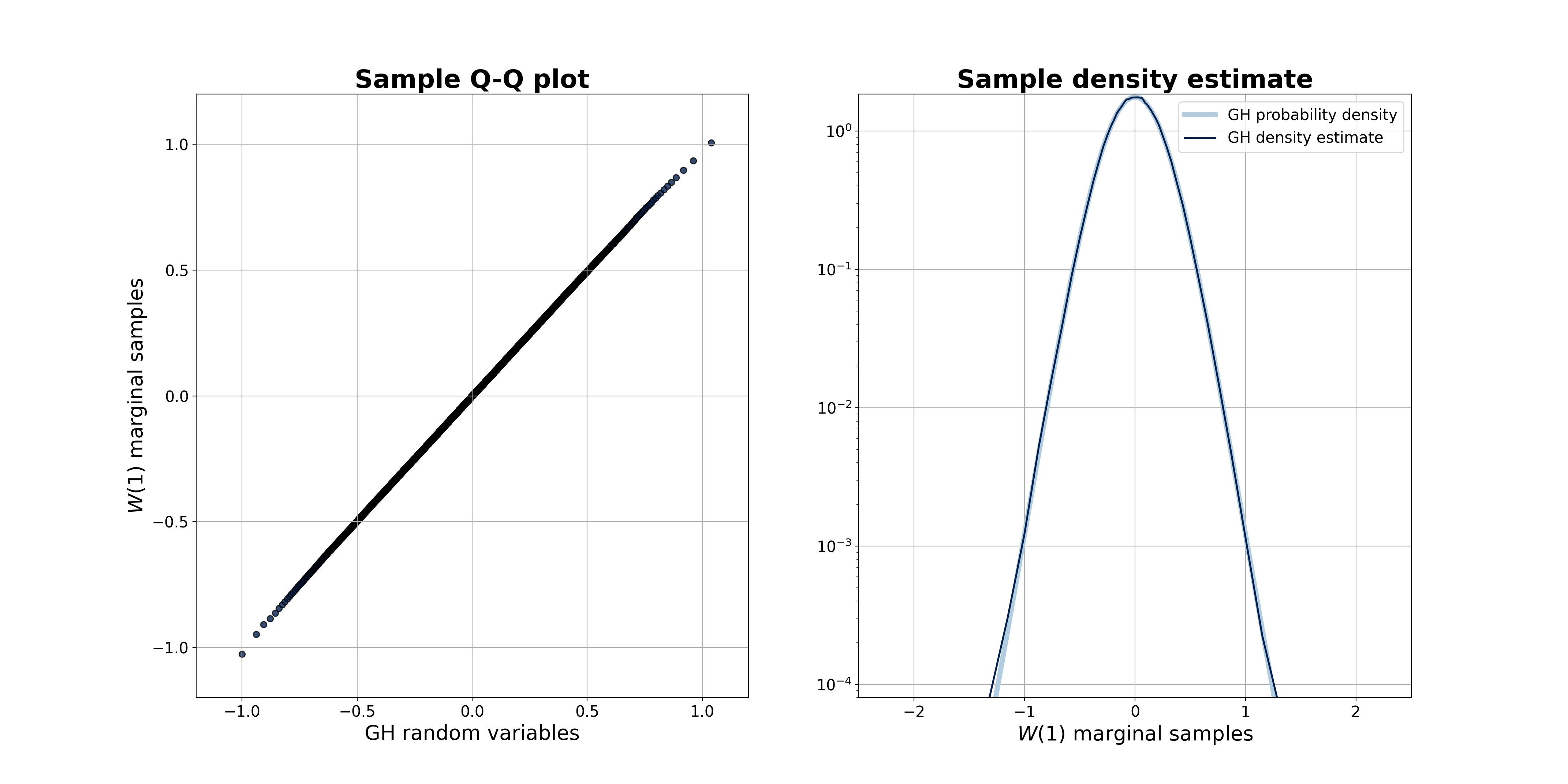}
\caption{Simulation comparison between the shot noise generated GH process and GH random variates, $\lambda=-10$, $\gamma=0.1$, $\delta=1$, $\beta=0$. The adaptive truncation parameters are $p_T = 0.05$ and $\tau=0.1$. Left hand panel: QQ plot comparing our shot noise method (y-axis) with random samples of the GH density generated using a random variate generator (x-axis). Right hand panel: Normalised histogram density estimate for our method compared with the true GH density function.}
\label{fig:sample_test2}
\end{figure}

For the case $|\lambda| > 0.5$, Figs. \ref{fig:sample_test3}, \ref{fig:sample_test1}, \ref{fig:sample_test2} shows the QQ plot and histogram for different parameter settings and Figs. \ref{fig:sample_path2},\ref{fig:sample_path4},\ref{fig:sample_path5} presents sample paths from our new adaptive simulation algorithm with residual approximation. The corresponding processes are simulated using Algorithms \ref{alg:phase1_N1}, \ref{gen_N_1}, \ref{alg:phase1_N2} and \ref{gen_N_2}. Similarly, Figs. \ref{fig:sample_path3},\ref{fig:sample_test6} presents an example in the $0 < |\lambda| < 0.5$ setting and the corresponding simulation methods are described in Algorithms \ref{alg:phase2_N1}, \ref{alg:Bgen_N_1}, \ref{alg:phase2_N2} and \ref{alg:Bgen_N_2}. The number of samples from each simulation method is $N=10^6$ and the sample path plots show randomly selected $50$ paths.

\clearpage

\clearpage

\clearpage

\clearpage

\begin{figure}[!h]
\centering
\includegraphics[width=0.7\textwidth]{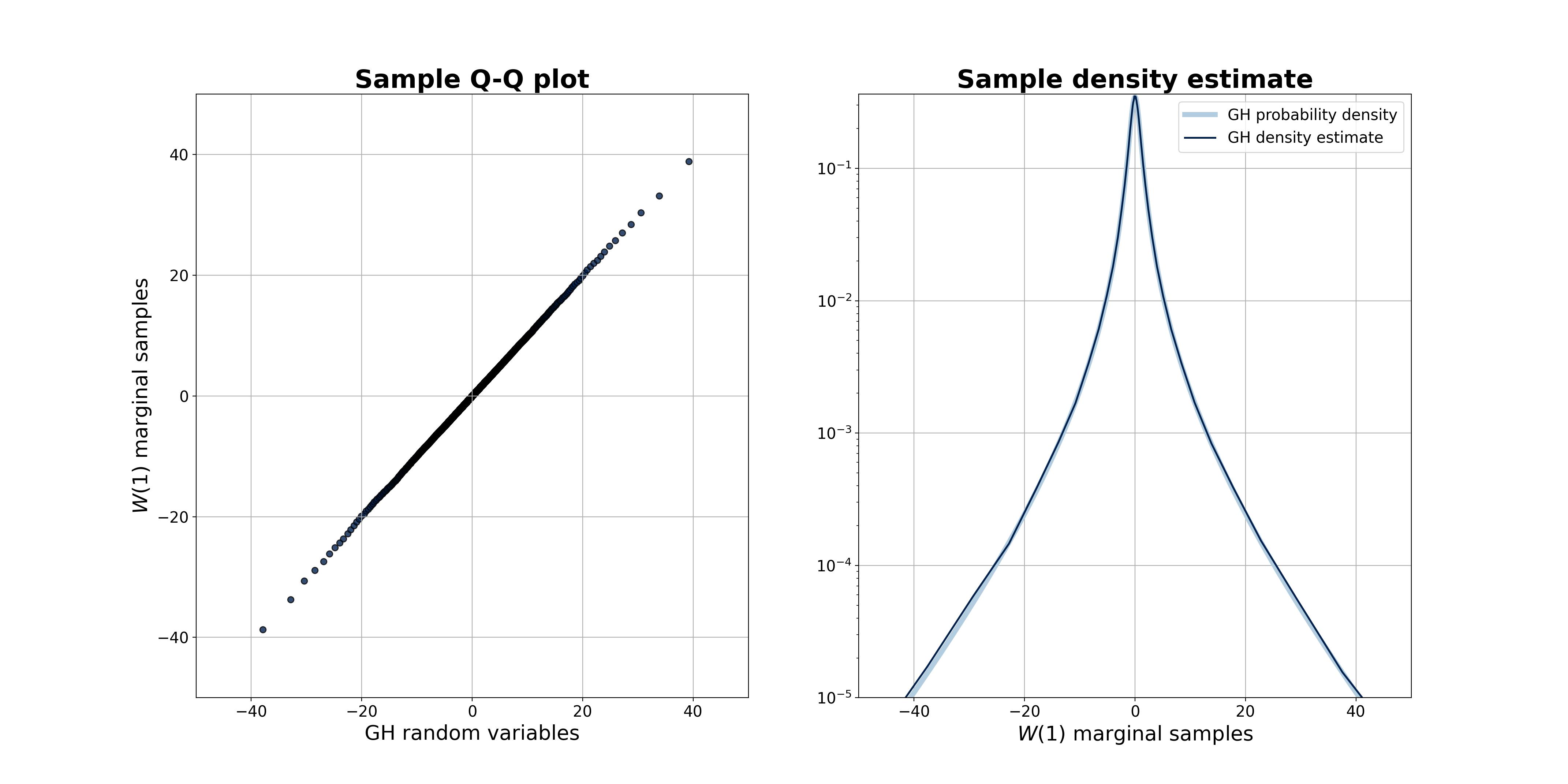}
\caption{Simulation comparison between the shot noise generated NIG process and NIG random variates, $\lambda=-0.5$, $\gamma=0.1$, $\delta=1$, $\beta=0$. The adaptive truncation parameters are $p_T = 0.05$ and $\tau=0.01$. Left hand panel: QQ plot comparing our shot noise method (y-axis) with random samples of the NIG density generated using a random variate generator (x-axis). Right hand panel: Normalised histogram density estimate for our method compared with the true NIG density function.}
\label{fig:sample_test4}
\end{figure}

\begin{figure}[!h]
\centering
\includegraphics[width=0.7\textwidth]{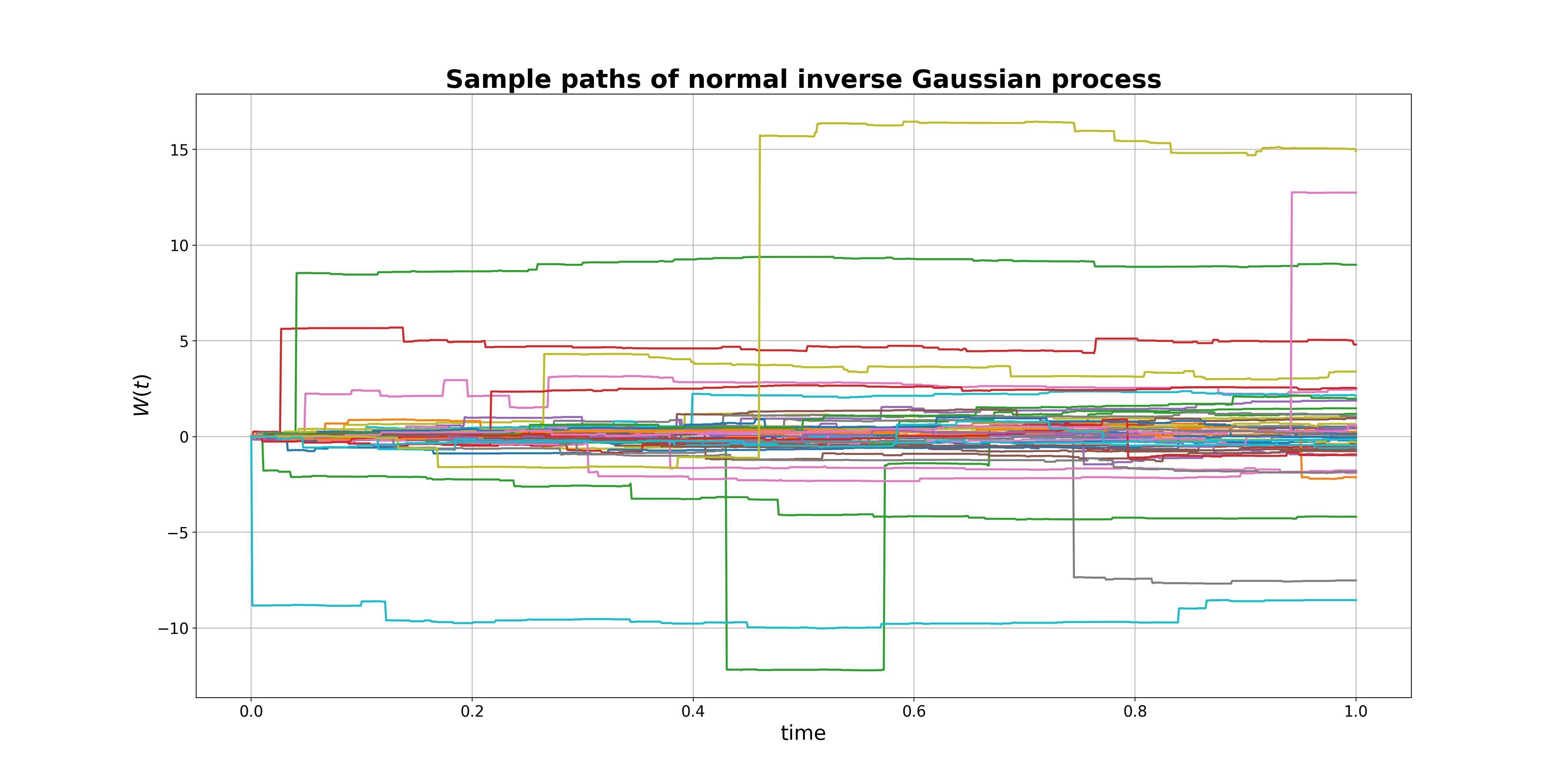}
\caption{Pathwise simulations of the NIG process for $\lambda=-0.5$, $\gamma=0.1$, $\delta=1$, $\beta=0$.}
\label{fig:sample_path6}
\end{figure}

The normal-inverse Gaussian (NIG) distribution, or the distribution of NIG processes at $T=1$, forms an exponential family and hence all of its moments have analytical expressions \cite{Barndorff-Nielsen1997}. As a result of its tractable probabilistic properties, the NIG process finds application in modelling turbulence and financial data (\cite{B_N_1997,Rydberg_1997}). The NIG process is in fact a special case of the GH process where $\lambda = -0.5$. This results in the bounds given in Corollary \ref{Q_GIG_bound_A_case2} being exactly equal to the GIG density and thus the acceptance rate of points simulated from Alg. 3 of \cite{GodsillKindap2021} is $1.0$. The QQ plot, density estimate and sample paths for this parameter setting are shown in Figs. \ref{fig:sample_test4} and \ref{fig:sample_path6}.

\clearpage

Another special case of the GH process is the Student-t process where $\lambda < 0$, $\gamma = 0$ and $\delta^2 = -2\lambda$. The Student-t distribution is parameterised using a single parameter $\nu$, called the degrees of freedom, which is a positive real number. This parameter is related to the usual parameters of a GH process such that $\lambda = -\nu/2$ and $\delta = \sqrt{\nu}$. 

\begin{figure}[!h]
\centering
\includegraphics[width=0.7\textwidth]{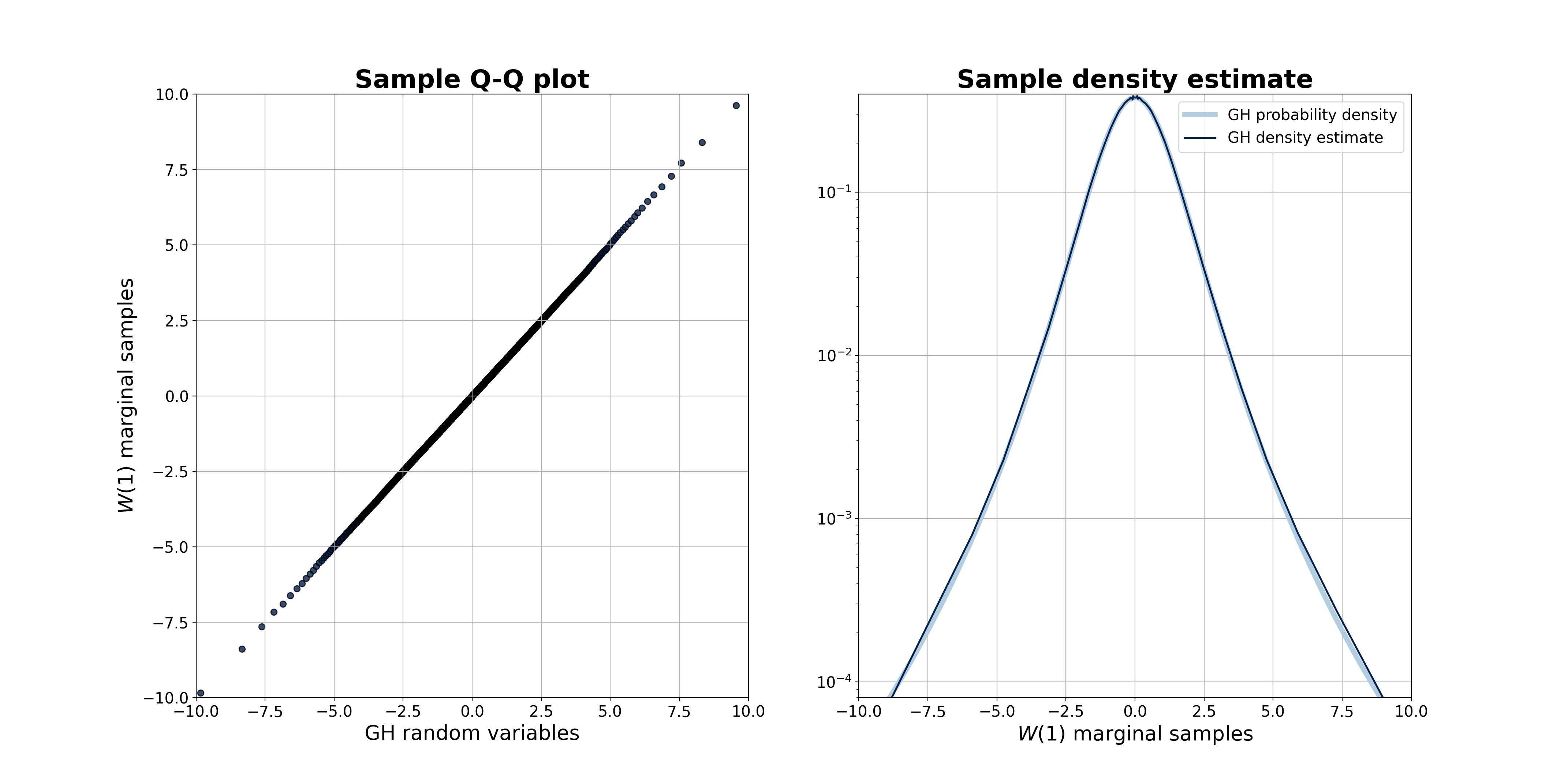}
\caption{Simulation comparison between the shot noise generated student-t process and student-t random variates, $\lambda=-2.5$, $\gamma=0$, $\delta=\sqrt{5}$, $\beta=0$. The adaptive truncation parameters are $p_T = 0.05$ and $\tau=0.1$. Left hand panel: QQ plot comparing our shot noise method (y-axis) with random samples of the student-t density generated using a random variate generator (x-axis). Right hand panel: Normalised histogram density estimate for our method compared with the true student-t density function.}
\label{fig:sample_test5}
\end{figure}

\begin{figure}[!h]
\centering
\includegraphics[width=0.7\textwidth]{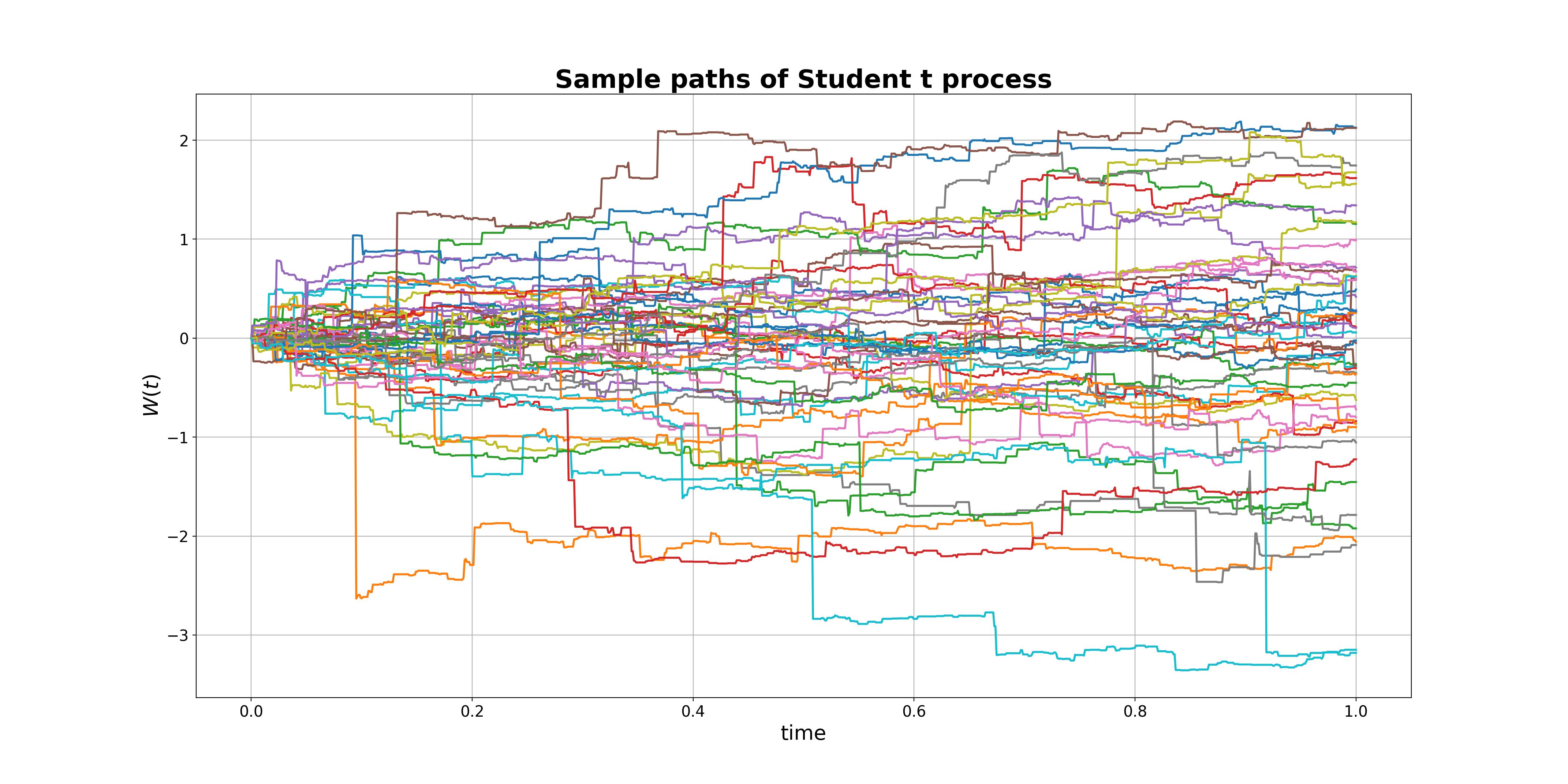}
\caption{Pathwise simulations of the Student t process for $\nu=5$ and $\beta=0$. ($\lambda=-2.5$, $\delta=\sqrt{5}$, $\gamma=0$)}
\label{fig:sample_path7}
\end{figure}

As remarked in Section \ref{sec:adaptive_truncation}, it is not possible to simulate a Student-t process using Algorithms \ref{alg:phase1_N1}, \ref{gen_N_1}, \ref{alg:phase1_N2} and \ref{gen_N_2} because the corresponding gamma processes are not well-defined in this case. Instead, for this parameter setting Algorithm 3 of \cite{GodsillKindap2021} is used together with the adaptive truncation and residual approximation methods presented in Section \ref{sec:adaptive_truncation} to produce the jumps from the subordinator GIG process. The QQ plot, density estimate and sample paths for the resulting samples from the Student t process are shown in Figs. \ref{fig:sample_test5} and \ref{fig:sample_path7}. Note that removing the condition $\delta^2 = -2\lambda$ still results in a well defined L\'{e}vy process with its marginal distribution parameterised by Eq. (3.11) in \cite{Eberlein2003}.

\begin{figure}[!h]
\centering
\includegraphics[width=0.7\textwidth]{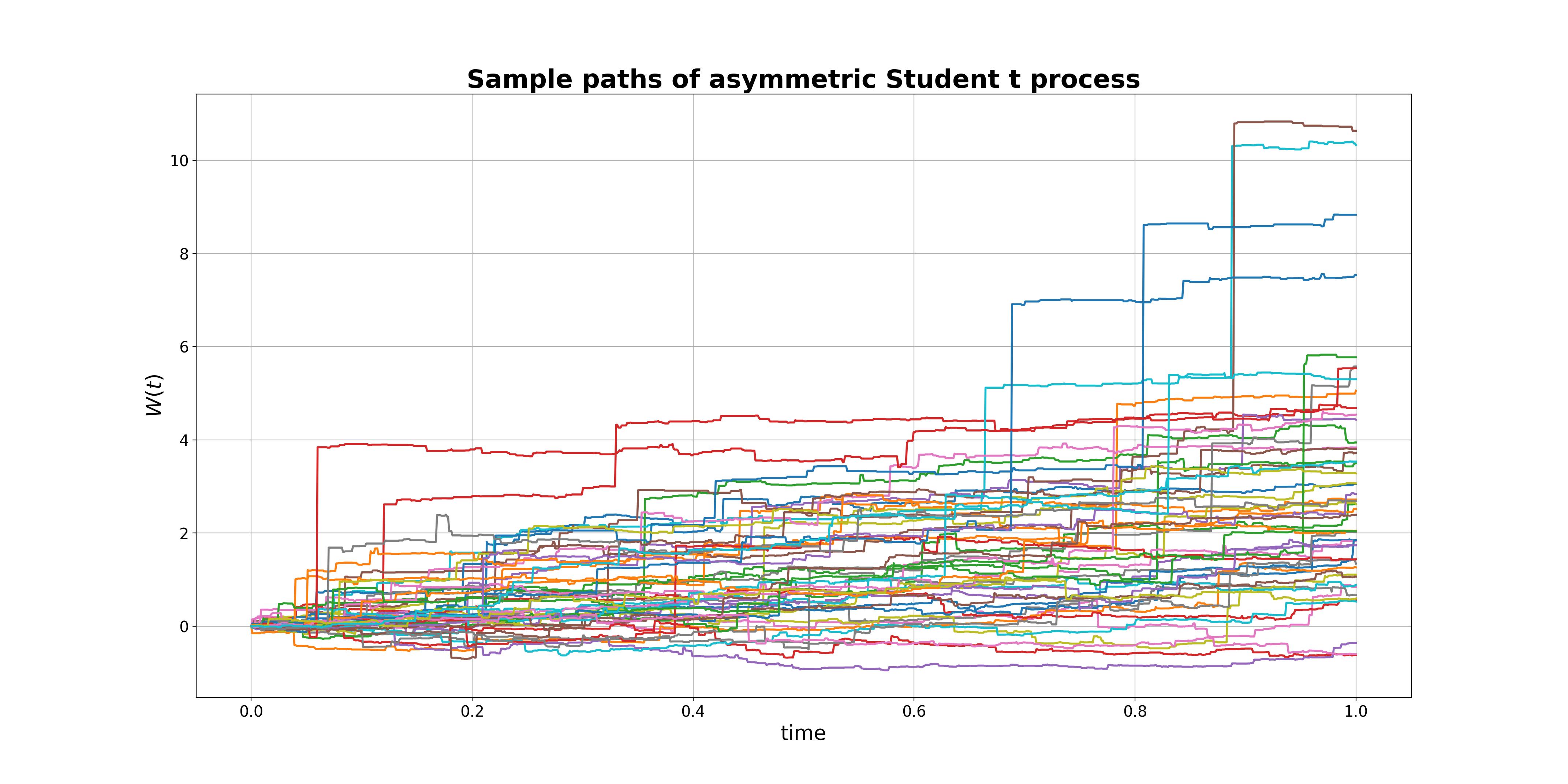}
\caption{Pathwise simulations of the asymmetric Student t process for $\nu=5$ and $\beta=2$. ($\lambda=-2.5$, $\delta=\sqrt{5}$, $\gamma=0$)}
\label{fig:sample_path1}
\end{figure}

\begin{figure}[!h]
\centering
\includegraphics[width=0.7\textwidth]{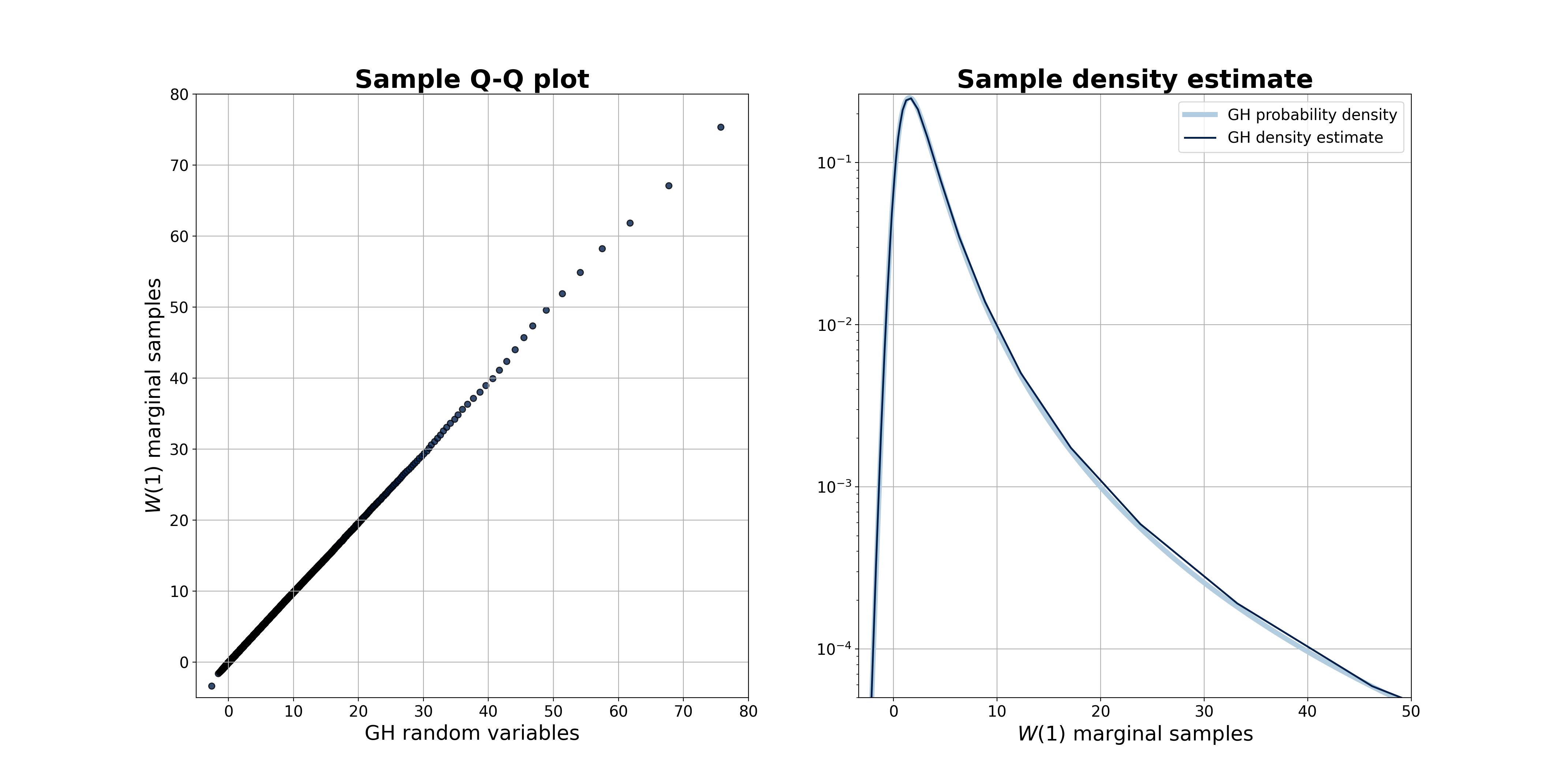}
\caption{Simulation comparison between the shot noise generated asymmetric student-t process and student-t random variates, $\lambda=-2.5$, $\gamma=0$, $\delta=\sqrt{5}$, $\beta=2$. The adaptive truncation parameters are $p_T = 0.05$ and $\tau=0.1$. Left hand panel: QQ plot comparing our shot noise method (y-axis) with random samples of the asymmetric student-t density generated using a random variate generator (x-axis). Right hand panel: Normalised histogram density estimate for our method compared with the true student-t density function.}
\label{fig:qqplot_and_histogram_asymmetric_t}
\end{figure}

It is common to apply the Student-t distribution to financial data sets as an alternative to the Gaussian distribution to account for the heavier tails observed in asset returns. There has been numerous studies that suggest that the asymmetric Student-t distribution, $\beta \neq 0$, provides a better fit to financial data sets compared to its symmetric counterpart (\cite{ZhuGalbraith2010,Alberg2008,AasHaff2006}). To the best of our knowledge, we present the sample paths of an asymmetric Student-t process for the first time in Fig. \ref{fig:sample_path1} together with the resulting marginal QQ plot and density estimate in Fig. \ref{fig:qqplot_and_histogram_asymmetric_t}. The marginal density of this limiting case of asymmetric GH processes are given in Eq. (3.9) of \cite{Eberlein2003}.

\section{Conclusions}
\label{sec:conclusions}

The point process representation of a generalised hyperbolic process and the generalised shot noise methods developed in this work provide the  the first complete methodology for simulation of generalised hyperbolic (GH) L\'{e}vy processes, giving a unified framework for a very broad  range of heavy-tailed and semi-heavy-tailed non-Gaussian processes. The continuous time formulation, simulating directly in continuous time path space, can be employed for accurate  uncertainty propagation, path visualisation, modelling and inference, especially for irregularly sampled time series datasets. 

The presented methods are based on the subordination of a Brownian motion by the generalised inverse Gaussian (GIG) process, and we have here provided novel improvements in GIG process simulation compared with our previous work \cite{GodsillKindap2021}. We have proved also that these series representations are almost surely convergent, verifying the conditions presented in \cite{Rosinski2001}. In addition to these improvements we present a novel scheme for adaptive truncation of the random shot noise representation based on probabilistic exceedance bounds, relying on new upper and lower bound expressions for the moments of truncated residual of the shot noise process; these truncations methods are shown to reduce computational burden dramatically without noticeable compromising of accuracy. Further computational savings are made through the use of squeezed rejection sampling, again based on our lower and upper moment bounds. We plan to use the new GH process simulators as a 
 fundamental building block for modelling of stochastic differential equations (SDEs) driven by GH processes, in a spirit similar to \cite{Godsill_Riabiz_Kont_2019}, where the conditionally Gaussian form of our models will be of great benefit in inference for states and parameters for GH process-driven SDEs, finding application in spatial tracking, finance and vibration data modelling, to list only a few possibilities.   These models and their applications will be further studied in future publications.

\clearpage

\printbibliography

\appendix

\section*{A: Proof of Theorem \ref{theorem4}}

\begin{proof}
For the case $|\lambda| \geq 0.5$, $B(z)$ forms an upper bound on $z|H_{|\lambda|}(z)|^2$ as stated in Eq. (\ref{bound_lam_geq_0_5}). The resulting integral obtained by replacing $z|H_{|\lambda|}(z)|^2$ with $B(z)$ is therefore a lower bound for both $\mathbb{E}\left[ \rho_1(x,z) \right]$ and $\mathbb{E}\left[ \rho_2(x,z) \right]$. As the function $B(z)$ has a piecewise form dependent on the choice of $z_0$ and the integrals are defined in terms of $z_1$, the lower bounds form a piecewise function dependent on their relative values.

Assuming $z_0 \leq z_1$, the interval $(0, z_1)$ is partitioned into two non-overlapping intervals $(0,z_0)$ and $(z_0, z_1)$. The constituent functions of $B(z)$ is used to replace $z|H_{|\lambda|}(z)|^2$ in both partitions of $\mathbb{E}\left[ \rho_1(x,z) \right]$ as follows

\begin{align*}
    \mathbb{E}\left[ \rho_1(x,z) \right] &\geq \int_{0}^{z_0} \frac{2 \,  \Gamma(|\lambda|)\sqrt{\text{Ga}}
(z||\lambda|, \frac{x}{2\delta^2})}{\pi H_0 \left( \frac{z_0}{z} \right)^{2|\lambda|-1} \left( \frac{z}{z_1} \right)^{2|\lambda|-1}  \gamma(|\lambda|, \frac{z_1^2x}{2\delta^2})} dz + \int_{z_0}^{z_1} \frac{2 \, \Gamma(|\lambda|)\sqrt{\text{Ga}}
(z||\lambda|, \frac{x}{2\delta^2}) }{\pi H_0 \left( \frac{z}{z_1} \right)^{2|\lambda|-1} \gamma(|\lambda|, \frac{z_1^2x}{2\delta^2})} dz \\
 &= \frac{2^2 \left( \frac{x}{2 \delta^2} \right)^{|\lambda|}}{\pi H_0} \frac{\left( \frac{z_1}{z_0} \right)^{2|\lambda|-1}}{\gamma(|\lambda|, \frac{z_1^2 x}{2\delta^2})} \int_{0}^{z_0} z^{2|\lambda|-1} e^{-\frac{z^2 x}{2 \delta^2}} dz + \frac{2^2 \left( \frac{x}{2 \delta^2} \right)^{|\lambda|}}{\pi H_0} \frac{z_1^{2|\lambda|-1}}{\gamma(|\lambda|, \frac{z_1^2 x}{2\delta^2})} \int_{z_0}^{z_1} e^{-\frac{z^2 x}{2 \delta^2}} dz \\
&= \frac{2}{\pi H_0} \Bigg( \left( \frac{z_1}{z_0} \right)^{2|\lambda|-1} \frac{\gamma(|\lambda|, \frac{z_0^2 x}{2 \delta^2})}{\gamma(|\lambda|, \frac{z_1^2 x}{2 \delta^2})} + \left( \frac{z_1^2 x}{2 \delta^2} \right)^{|\lambda|-0.5} \frac{\left[ \gamma(0.5, \frac{z_1^2 x}{2 \delta^2}) - \gamma(0.5, \frac{z_0^2 x}{2 \delta^2}) \right]}{\gamma(|\lambda|, \frac{z_1^2 x}{2 \delta^2})} \Bigg)
\end{align*}

For $\mathbb{E}\left[ \rho_2(x,z) \right]$, the same assumption that $z_0 \leq z_1$ leads to a simple lower bound as

\begin{align*}
&\mathbb{E}\left[ \rho_2(x,z) \right] \geq \frac{2}{\pi H_0} \int_{z_1}^{\infty} \frac{\Gamma(0.5)\sqrt{\text{Ga}}
(z|0.5, \frac{x}{2\delta^2})}{\Gamma(0.5,\frac{z_1^2x}{2\delta^2})} dz \\
& \quad \quad \quad \, \, \, \, = \frac{2^2}{\pi H_0} \frac{\left( \frac{x}{2 \delta^2} \right)^{0.5}}{\Gamma(0.5, \frac{z_1^2 x}{2 \delta^2})} \int_{z_1}^{\infty} e^{-\frac{z^2 x}{2 \delta^2}} dz \\
& \quad \quad \quad \, \, \, \, = \frac{2}{\pi H_0}
\end{align*}

For the case of $z_0 \geq z_1$, replacing $z|H_{|\lambda|}(z)|^2$ with $B(z)$ again leads to a simple lower bound on $\mathbb{E}\left[ \rho_1(x,z) \right]$ as

\begin{align*}
&\mathbb{E}\left[ \rho_1(x,z) \right] \geq \int_{0}^{z_1} \frac{2 \, \Gamma(|\lambda|)\sqrt{\text{Ga}}
(z||\lambda|, \frac{x}{2\delta^2})}{\pi H_0 \left(\frac{z_0}{z}\right)^{2|\lambda|-1} \left(\frac{z}{z_1}\right)^{2|\lambda|-1} \gamma(|\lambda|, \frac{z_1^2x}{2\delta^2})} dz \\
& \quad \quad \quad \, \, \, \, = \frac{2^2 \left(\frac{x}{2\delta^2}\right)^{|\lambda|} z_{1}^{2|\lambda|-1}}{\pi H_0 \gamma(|\lambda|, \frac{z_1^2 x}{2\delta^2}) z_{0}^{2|\lambda|-1}}  \int_{0}^{z_1} z^{2|\lambda|-1} e^{-\frac{z^2 x}{2 \delta^2}} dz \\
& \quad \quad \quad \, \, \, \, = \frac{2}{\pi H_0} \left( \frac{z_1}{z_0} \right)^{2|\lambda|-1}
\end{align*}

Lastly, the lower bound on $\mathbb{E}\left[ \rho_2(x,z) \right]$ for $z_0 \geq z_1$ is obtained by partitioning $(z_1, \infty)$ into two non-overlapping intervals $(z_1, z_0)$, $(z_0, \infty)$, and the constituent functions of $B(z)$ is used to replace $z|H_{|\lambda|}(z)|^2$ as

\begin{align*}
\mathbb{E}\left[ \rho_2(x,z) \right] &\geq \int_{z_1}^{z_0} \frac{2 \, \Gamma(0.5)\sqrt{\text{Ga}}
(z|0.5, \frac{x}{2\delta^2})  }{\pi H_0 \left( \frac{z_0}{z} \right)^{2|\lambda|-1} \Gamma(0.5,\frac{z_1^2x}{2\delta^2})} dz + \int_{z_0}^{\infty} \frac{2}{\pi H_0} \frac{\Gamma(0.5)\sqrt{\text{Ga}}
(z|0.5, \frac{x}{2\delta^2})}{\Gamma(0.5,\frac{z_1^2x}{2\delta^2})} dz \\
&= \frac{2^2 \left( \frac{x}{2\delta^2} \right)^{0.5}}{\pi H_0 z_0^{2|\lambda|-1} \Gamma(0.5, \frac{z_1^2 x}{2\delta^2})} \int_{z_1}^{z_0} z^{2|\lambda|-1} e^{-\frac{z^2 x}{2 \delta^2}} dz + \frac{2^2 \left( \frac{x}{2\delta^2} \right)^{0.5}}{\pi H_0 \Gamma(0.5, \frac{z_1^2 x}{2\delta^2})} \int_{z_0}^{\infty} e^{-\frac{z^2 x}{2 \delta^2}} dz \\
&= \frac{2}{\pi H_0} \Bigg( \frac{\Gamma(0.5, \frac{z_0^2 x}{2 \delta^2})}{\Gamma(0.5, \frac{z_1^2 x}{2 \delta^2})} + \left( \frac{z_0^2 x}{2 \delta^2} \right)^{0.5-|\lambda|} \frac{\left[ \gamma(|\lambda|, \frac{z_0^2 x}{2 \delta^2}) - \gamma(|\lambda|, \frac{z_1^2 x}{2 \delta^2}) \right]}{\Gamma(0.5, \frac{z_1^2 x}{2 \delta^2})} \Bigg)
\end{align*}

\end{proof}

\section*{B: Proof of Theorem \ref{thm:lower_bounds_1}}

\begin{proof}
In principle the residual moments of a L\'{e}vy process can be obtained directly from Eqs. (\ref{mu_c}) and (\ref{sigma_c}) as before. However, once again these cannot be evaluated exactly for most parameter settings. Instead we provide lower bounds for the integrand term $Q_{GIG}(x)$ which are obtained from the marginal point processes $Q^{A/B}_{N_1/N_2}(x)$ defined in Eqs.~(\ref{eq:QAN1}), (\ref{eq:QAN2}), (\ref{eq:QBN1}) and (\ref{eq:QBN2}).

The starting point is the bivariate L\'{e}vy density $Q_{GIG}(x,z)$, and the lower bounds on this function given in Corollary \ref{cor:theorem1}. For the case $|\lambda|\geq 0.5$, the marginal $Q_{GIG}(x)$ is bounded as

\begin{equation*}
    Q^{B}_{N_1}(x) + Q^{B}_{N_2}(x) \leq Q_{GIG}(x)
\end{equation*}

Hence for $n\in\{1,2\}$, bounds on the residual moments may be found by studying each term in

\begin{align*}
\int_{0}^{\varepsilon}x^n Q^B_{N_1}(x) & dx + \int_{0}^{\varepsilon}x^n Q^B_{N_2}(x)dx \nonumber\\ &\leq \int_{0}^{\varepsilon}x^n Q_{GIG}(x)dx 
\end{align*}

The first integral over $Q_{N_1}^{B}(x)$ is still intractable and hence has to be further lower bounded. A lower bound on the incomplete gamma function is defined in (\cite{neuman:13}, Th. 4.1)\footnote{Note that an alternative bound is given by $(1-e^{-x})/x\leq a/x^a\gamma(a,x)$ for $0\leq a \leq 1$} as

\[
\frac{(z_0^2x/(2\delta^2))^{|\lambda|}}{|\lambda|}\exp\left(\frac{-|\lambda|z_0^2x/(2\delta^2)}{|\lambda|+1}\right) \leq \gamma(|\lambda|,z_0^2x/(2\delta^2))
\]

\noindent Therefore a lower bounding density is obtained as

\begin{equation*}
    \frac{z_{0}}{\pi^{2} H_{0} |\lambda|} x^{-1} \exp \left( - \left[ \frac{\gamma^{2}}{2} + \frac{|\lambda|}{(1+|\lambda|)} \frac{z_{0}^{2} }{2 \delta^{2}} \right] x \right) \leq Q_{N_1}^{B} (x)
\end{equation*}

\noindent which is the density of a gamma process. The associated residual moments have already been given in Theorem \ref{thm:probabilistic_bound} and shown in Eqs. (\ref{gamma_residual_mean}) and (\ref{gamma_residual_variance}). The associated mean and variance of the residual of  $Q^B_{N_1}$ for a truncation level $\varepsilon$ are then given directly as

\begin{align*}
    \mu_{N_1}^{B}(\varepsilon) &= \frac{C_{Ga}^{B}}{\beta_{Ga}^{B}} \gamma \left( 1, \beta_{Ga}^{B} \varepsilon \right) \\
    {\sigma_{N_1}^{B}}^2(\varepsilon) &= \frac{C_{Ga}^{B}}{{\beta_{Ga}^{B}}^2} \gamma \left( 2, \beta_{Ga}^{B} \varepsilon \right)
\end{align*}

\noindent where

\begin{equation*}
    C_{Ga}^{B} = \frac{z_{0}}{\pi^{2} H_{0} |\lambda|} \quad \text{and} \quad \beta_{Ga}^{B} = \frac{\gamma^{2}}{2} + \frac{|\lambda|}{(1+|\lambda|)} \frac{z_{0}^{2} }{2 \delta^{2}}
\end{equation*}

The density $Q^B_{N_2}$ is a modified tempered stable characterised by the upper incomplete gamma function, which can be lower bounded as (\cite{Changetal2011}, Th. 2)

\begin{equation*}
    \sqrt{2e} \frac{\sqrt{\beta_{0}-1}}{\beta_{0}} \exp \left( - \beta_{0} \frac{z_{0}^{2} x}{2 \delta^2} \right)  \leq \Gamma(0.5, \frac{z_{0}^{2} x}{2 \delta^2})
\end{equation*}

\noindent where $\beta_0 > 1$ and $x \geq 0$. The free parameter $\beta_0$ can be adaptively optimised across the whole domain. The bound can be derived using the well-known equivalence $\Gamma(0.5, x) = \sqrt{\pi} \, \mathrm{erfc}(\sqrt{x})$.

Hence a lower bound on $Q^B_{N_2}$ is obtained as

\begin{equation*}
    \frac{2 \delta \sqrt{e} \sqrt{\beta_0-1}}{\pi^2 H_0 \beta_0} x^{-3/2} \exp \left( - \left[ \frac{\gamma^2}{2} + \frac{\beta_{0} z_{0}^{2} }{2 \delta^2} \right] x \right) \leq Q^B_{N_2}(x)
\end{equation*}

\noindent which is the density of a tempered stable process and the associated residual moments are previously derived in the proof of Theorem \ref{thm:probabilistic_bound} and shown in Eq. (\ref{mu_TS}) and (\ref{sigma_TS}). The lower bound on the residual mean and variance of $Q^B_{N_2}$ for a truncation level $\varepsilon$ are then obtained directly as 

\begin{align*}
    \mu_{N_2}^{B}(\varepsilon) &= C_{TS}^{B} {\beta_{TS}^{B}}^{-0.5} \gamma \left( 0.5, \beta_{TS}^{B} \varepsilon \right) \\
    {\sigma_{N_2}^{B}}^2(\varepsilon) &= C_{TS}^{B} {\beta_{TS}^{B}}^{-1.5} \gamma \left( 1.5, \beta_{TS}^{B} \varepsilon \right)
\end{align*}

\noindent where

\begin{equation*}
    C_{TS}^{B} = \frac{2 \delta \sqrt{e} \sqrt{\beta_0-1}}{\pi^2 H_0 \beta_0} \quad \text{and} \quad \beta_{TS}^{B} = \frac{\gamma^2}{2} + \frac{\beta_{0} z_{0}^{2} }{2 \delta^2}
\end{equation*}

Finally the lower bound on the residual mean and variance of the GIG density can be expressed in terms of the residual means and variances of the lower bounding gamma and TS processes as $\underline{\mu_{\varepsilon}} = \mu_{N_1}^{B}(\varepsilon) + \mu_{N_2}^{B}(\varepsilon)$ and $\underline{\sigma_{\varepsilon}^2} = {\sigma_{N_1}^{B}}^2(\varepsilon) + {\sigma_{N_2}^{B}}^2(\varepsilon)$.

For the case $0 < |\lambda| \leq 0.5$, the marginal $Q_{GIG}(x)$ is bounded as

\begin{equation*}
    Q^{A}_{N_1}(x) + Q^{A}_{N_2}(x) \leq Q_{GIG}(x)
\end{equation*}

Similarly for $n\in\{1,2\}$, lower bounds on the moments may be found as

\begin{align*}
\int_{0}^{\varepsilon}x^n Q^A_{N_1}(x) & dx + \int_{0}^{\varepsilon}x^n Q^A_{N_2}(x)dx \leq \int_{0}^{\varepsilon}x^n Q_{GIG}(x)dx 
\end{align*}

The lower bounds on the incomplete gamma functions established for the previous parameter setting can be directly substituted into $Q^A_{N_1}(x)$ and $Q^A_{N_2}$. The resulting densities are again characterised as a gamma process and a tempered stable process respectively such that

\begin{equation*}
    \frac{z_{1}}{2 \pi |\lambda|} x^{-1} \exp \left( - \left[ \frac{\gamma^{2}}{2} + \frac{|\lambda|}{(1+|\lambda|)} \frac{z_{1}^{2} }{2 \delta^{2}} \right] x \right) \leq Q_{N_1}^{A} (x)
\end{equation*}

\begin{equation*}
    \frac{\delta \sqrt{e} \sqrt{\beta_0-1}}{\pi \beta_0} x^{-3/2} \exp \left( - \left[ \frac{\gamma^2}{2} + \frac{\beta_{0} z_{1}^{2} }{2 \delta^2} \right] x \right) \leq Q^B_{N_2}(x)
\end{equation*}

Hence the associated residual moments of these processes can be directly found again using Eqs. (\ref{gamma_residual_mean}), (\ref{gamma_residual_variance}) and (\ref{mu_TS}), (\ref{sigma_TS}). The corresponding lower bound on the residual mean and variance of the GIG process for the case $0 < |\lambda| \leq 0.5$ and truncation level $\varepsilon$ are then found as 

\begin{equation*}
    \underline{\mu_{\varepsilon}} = \frac{C_{Ga}^{A} \gamma \left( 1, \beta_{Ga}^{A} \varepsilon \right)}{\beta_{Ga}^{A}} + \frac{C_{TS}^{A} \gamma \left( 0.5, \beta_{TS}^{A} \varepsilon \right)}{{\beta_{TS}^{A}}^{0.5}}
\end{equation*}
\begin{equation*}
     \underline{\sigma_{\varepsilon}^2} = \frac{C_{Ga}^{A} \gamma \left( 2, \beta_{Ga}^{A} \varepsilon \right)}{{\beta_{Ga}^{A}}^2} + \frac{ C_{TS}^{A} \gamma \left( 1.5, \beta_{TS}^{A} \varepsilon \right) }{{\beta_{TS}^{A}}^{1.5}}
\end{equation*}

\end{proof}

\end{document}